\documentclass[12pt]{article}
\pdfoutput=1
\usepackage{jheppub}
\usepackage{amsmath,bbm,array,amsfonts,graphicx,wrapfig,lscape,float,mathtools,multirow,longtable,rotating,subfigure,ulem,cancel}
\usepackage[section]{placeins}

\newcommand{\be}{\begin{equation}}
\newcommand{\ee}{\end{equation}}
\newcommand{\beq}{\begin{equation}}
\newcommand{\beql}[1]{\begin{equation}\label{#1}}
\newcommand{\eeq}{\end{equation}}
\newcommand{\ba}{\begin{array}}
\newcommand{\ea}{\end{array}}
\newcommand{\bea}{\begin{eqnarray}}
\newcommand{\beal}[1]{\begin{eqnarray}\label{#1}}
\newcommand{\eea}{\end{eqnarray}}
\newcommand{\ben}{\begin{enumerate}}
\newcommand{\een}{\end{enumerate}}
\newcommand{\bean}{\begin{eqnarray*}}
\newcommand{\eean}{\end{eqnarray*}}

\newcommand{\btab}[1]{\begin{tabular}{#1}}
\newcommand{\etab}{\end{tabular}}

\def \Black {\pdfsetcolor{0 0 0 1}}

\newcommand{\comment}[1]{}

         \let\q=\theta

   \let\Q=\Theta

\newcommand{\qed}{\nobreak \ifvmode \relax \else
      \ifdim\lastskip<1.5em \hskip-\lastskip
      \hskip1.5em plus0em minus0.5em \fi \nobreak
      \vrule height0.75em width0.5em depth0.25em\fi}

\numberwithin{equation}{section}

\title{Construction and Deconstruction of Single Instanton Hilbert Series}

\author{Amihay Hanany, }
\author{Rudolph Kalveks}

\affiliation{
Theoretical Physics Group, The Blackett Laboratory,
Imperial College London, \\
Prince Consort Road, London SW7 2AZ, United Kingdom
}

\emailAdd{a.hanany@imperial.ac.uk, rudolph.kalveks09@imperial.ac.uk}

\preprint{Imperial/TP/15/AH/04}

\abstract{Many methods exist for the construction of the Hilbert series describing the moduli spaces of instantons. We explore some of the underlying group theoretic relationships between these various constructions, including those based on the Coulomb branches and Higgs branches of SUSY quiver gauge theories, as well as those based on generating functions derivable from the Weyl Character Formula. We show how the character description of the reduced single instanton moduli space (``RSIMS") of any Classical or Exceptional group can be deconstructed faithfully in terms of characters or modified Hall-Littlewood polynomials of its regular semi-simple subgroups. We derive and utilise Highest Weight Generating (``HWG") functions, both for the characters of Classical or Exceptional groups and for the Hall-Littlewood polynomials of unitary groups. We illustrate how the root space data encoded in extended Dynkin diagrams corresponds to relationships between the Coulomb branches of quiver gauge theories for RSIMS and those for $T(SU(N))$ moduli spaces.
\\

~\today}

\begin{document}

\maketitle

\listoftables

\listoffigures
\section{Introduction}
\label{sec:intro}

The moduli spaces of instantons remain the subject of much research and new constructions continue to be presented in the literature. What is perhaps most remarkable is the wide variety of different approaches that can be deployed to construct the complicated Hilbert series describing these moduli spaces, along with the possibility of generating their expansions from the combinatorics of a few relatively simple building blocks. The construction methods range from those that are purely group theoretic in nature, through methods associated with semi-simple subgroup decompositions, to those that draw upon the Higgs or Coulomb branches of supersymmetric (``SUSY") quiver gauge theories.

The aim of this paper is to examine a number of these approaches, to try to elucidate the manner in which they are related by common group theoretic constructs, and to develop methods for extending the range of possible constructions. Furthermore, while some of these constructions, such as Coulomb branch quiver theories, are essentially reductive in nature, so that it is difficult to recover the design of the construction from the resulting Hilbert series, other constructions, such as those involving Hall-Littlewood polynomials, are reversible, so that the specification for the construction can be recovered from any generating function for the (refined) Hilbert series. We refer to this reversible process as {\it deconstruction}. We emphasise that we focus on the analysis of character decompositions of Hilbert series of instanton moduli spaces; we do not analyse the underlying instanton theories, such as the ADHM construction, which are already well covered in the literature \cite{Nekrasov:2004vw}.

As discussed in \cite{Benvenuti:2010pq}, the moduli space of single $G$-instantons over ${\mathbb C^2}$ decouples into the $SU(2)$ component associated with the ${\mathbb C^2}$ and a reduced moduli space associated with the Yang-Mills group $G$. Our principal focus will be on the reduced moduli spaces of single instantons (``RSIMS"). These possess the simplest group theoretic descriptions and are therefore good candidates for study. It may in due course be possible to bring a similar approach to bear on the more intricate subject of multiple instanton moduli spaces.

A construction of the Hilbert series for any number of instantons with $G$ taken as $SU(N)$ was given in \cite{Nakajima:2003pg, Nakajima:2005fg}. It was subsequently shown \cite{Benvenuti:2010pq, Hanany:2012dm} how such character expansions of instanton moduli spaces can be constructed on the Higgs branches of particular $N=2$ SUSY quiver gauge theories, not just for $SU(N)$, but for any Classical  symmetry group. In all cases, the RSIMS correspond to fields transforming as highest weight symmetrisations of the adjoint representation of $G$. The details of the quiver theory constructions required to yield these character expansions differ according to the symmetry group. We give a brief review of these in Section \ref{sec:Higgs}.

We follow the literature \cite{Benvenuti:2010pq, Hanany:2012dm, Keller:2012da} in taking the property of transforming in a symmetrisation of the adjoint as the defining characteristic of the reduced moduli space of a single instanton. Working with this definition, it is in principle a relatively straightforward exercise to construct a refined Hilbert series (``HS") for the RSIMS of any group using plethystic character generating functions. We do this by following a group theoretic analysis that starts from the Weyl Character Formula \cite{Fuchs:1997bb, Keller:2012da}. In Section \ref{sec:plethystics}, we set out the general methodology and give the plethystic character generating functions for low rank Classical groups and for $G_2$ and $F_4$. The results correspond to those obtained by following \cite{Keller:2011ek}. This approach is naturally agnostic with respect to any explicit field construction for the instanton moduli spaces, but provides useful insight into their group-theoretic structure.

More recently, a completely different approach to the construction of instanton moduli spaces has been developed. This draws upon early work on the GNO lattice \cite{Goddard:1976qe}, as well as more recent developments in $N=4$ quiver theory \cite{Borokhov:2002cg, Gaiotto:2008ak, Bashkirov:2010kz, Bashkirov:2010hj}. Specifically, the approach in \cite{Cremonesi:2013lqa} uses the Dynkin diagram of the extended (or untwisted affine) Lie algebra corresponding to the instanton symmetry group $G$ to specify a Coulomb branch quiver theory. Initially formulated for instanton moduli spaces of simply laced ADE symmetry groups \cite{Intriligator:1996ex}, the construction has been extended to non simply laced BCFG groups \cite{Cremonesi:2014xha}. There are interesting relationships between these Coulomb branch quiver theories and those for $T(SU(N))$, as will be discussed.

In Section \ref{sec:coulomb}, as a useful preliminary, we summarise the relationship between Lie algebras and their affine counterparts. We also set out the Coulomb branch quiver theory methodology for constructing RSIMS by mapping monopole charges from the GNO lattice of the affine Dynkin diagram of $G$ to the root lattice of $G$. For $SU(2)$, $SU(3)$, $SO(5)$ and $G_2$, we demonstrate the analytic equivalence of this Coulomb branch monopole construction to the RSIMS obtained from the character generating functions set out in Section \ref{sec:plethystics}. It follows that these Coulomb branch constructions are also equivalent to the Higgs branch constructions set out in Section \ref{sec:Higgs}. These are examples of {\it mirror symmetry} between the Coulomb branches of one class of SUSY gauge theories and the Higgs branches of a different class of SUSY gauge theories \cite{Intriligator:1996ex}.

While the Higgs branch constructions draw upon the basic irreducible representations (``irreps"), such as fundamentals, vectors and spinors, of $G$, the Coulomb branch constructions draw directly upon the root system of $G$. Further types of RSIMS construction draw upon the characters or modified Hall-Littlewood (``mHL") polynomials of semi-simple subgroups of $G$, which can be identified from extended Dynkin diagrams.  We list in Table \ref{tab1} these different approaches to RSIMS construction, indicating the groups for which the various constructions are known. The notable gap is the absence of a Higgs branch construction for Exceptional groups.

The mHL polynomials for $SU(N)$ are equivalent (up to a normalisation factor) to the Hilbert series generated by $T(SU(N))$ quiver theories. Constructions out of mHL polynomials, guided by a string theoretic analysis of M5 branes wrapping spheres with three punctures, are known for $E_6$, $E_7$ and $E_8$ instantons \cite{Gadde:2011uv}. We show in Section \ref{sec:mHL} how to use the orthogonality and completeness properties of these mHL polynomials to deconstruct the RSIMS of any Classical or Exceptional group into a sum of mHL polynomials, or equivalently, how to construct any RSIMS out of some combination of $T(SU(N))$ quiver theories.

\begin{table}
\caption{Types of RSIMS Construction}
\begin{center}
\begin{tabular}{|c|c|c|}
\hline
 Type of RSIMS Construction & Section & Groups\\
\hline
\hline
Character Generating Function &  \ref{sec:plethystics} & $ABCDEFG$ \\
\hline
Higgs Branch &  \ref{sec:Higgs} &$ABCD$\\
\hline
Coulomb Branch &  \ref{sec:coulomb} & $ABCDEFG$ \\
\hline
Subgroup Representations &  \ref{sec:weylbranch} & $ABCDEFG$\\
\hline
modified Hall-Littlewood Polynomials &  \ref{sec:mHL} & $ABCDEFG$ \\
\hline
\end{tabular}
\end{center}
\label{tab1}
\end{table}
We do not analyse the moduli spaces of multiple instanton theories herein. While progress has been made on these moduli spaces \cite{Keller:2012da, Hanany:2012dm, Cremonesi:2014xha}, they do not have an equally simple description in terms of the representation theory of their constituent groups due to mixing effects between the instanton and global symmetry groups. We comment on the dualities and other relationships between the various types of RSIMS construction in the concluding Section.

\paragraph{Notation and Terminology}
We freely use the terminology and concepts of the Plethystics Program, including the Plethystic Exponential (``PE"), its inverse, the Plethystic Logarithm (``PL"), the Fermionic Plethystic Exponential (``PEF") and, its inverse, the Fermionic Plethystic Logarithm(``PFL"). The reader is referred to \cite{Benvenuti:2006qr} or \cite{Hanany:2014dia} for a summary. Where no ambiguity arises, we may refer to RSIMS simply as instantons. 

We present the characters of groups either in the generic form ${\cal X}_{Group}$ or, more specifically, using Dynkin labels such as ${[ {{n_1}, \ldots , {n_r}}]_{Group}}$, where $r$ is the rank of the group (dropping subscripts if no ambiguities arise). We may refer to \textit{series}, such as $1 + f + {f^2} +  \ldots $, by their \textit{generating functions} $1/\left( {1 - f} \right)$. We rely on the use of distinct coordinates/variables to help distinguish the different types of generating function, as indicated in Table \ref{tab2}.
%
\begin{table}[htp]
\caption{Types of Generating Function}
\begin{center}
\begin{tabular}{c}
$\begin{array}{*{20}{c}}
{Generating~Function}&\vline& {{g^{Group}}\left( {coordinates} \right)}\\
\hline
{HWG}&\vline& {{g^{Group}}\left( {{t_A},{m_i}} \right)}\\
{Character}&\vline& {{g^{Group}_{\cal X}}\left( {{m_i},{x_i}} \right)}\\
{Refined~HS~(CSA~coordinates)}&\vline& {{g^{Group}}\left( {{t_A},{x_i}} \right)}\\
{Refined~HS~(roots)}&\vline& {{g^{Group}}\left( {{t_A},{z_i}} \right)}\\
{Unrefined~HS~(distinct~counting)}&\vline& {{g^{Group}}\left( {{t_A}} \right)}\\
{Unrefined~HS}&\vline& {{g^{Group}}\left( t \right)}
\end{array}$
\end{tabular}
\end{center}
\label{tab2}
\end{table}

These different types of generating function are related and can be considered as a hierarchy in which the highest weight generating functions (``HWG"), character and refined HS generating functions fully encode the group theoretic information. We label unimodular Cartan subalgebra (``CSA") coordinates for weights within characters by $x$ or $y$, using subscripts when necessary. We label simple root coordinates by $z_i$, where $i$ ranges from $1$ to rank $r$. We generally label field counting variables with $t$. Depending on the constructions used for RSIMS, they appear enumerated either by $t$ or $t^2$ - the moduli spaces are the same. Finally, we often deploy highest weight notation \cite{Hanany:2014dia}, which uses fugacities to track highest weight Dynkin labels, and describes the structure of a Hilbert series in terms of the highest weights of its consituent irreps. We typically denote such Dynkin label counting variables for representations based on characters with $m_i$ and those for Hall-Littlewood polynomials with $h_i$, although we may also use other letters, where this is helpful. We define these counting variable to have a complex modulus of less than unity and follow established practice in referring to them as ``fugacities", along with the monomials formed from the products of CSA coordinates.


\section{RSIMS from Character Generating Functions}
\label{sec:plethystics}

A reduced single instanton moduli space consists of highest weight symmetrisations of the adjoint representation \cite{Benvenuti:2010pq}. This comprises the subsequence of irreps generated by symmetrisations of the adjoint, whose highest weights have the longest root length. They are distinguished by having Dynkin labels that are a non-negative integer multiple of those of the adjoint.  We are therefore seeking to construct class functions, whether expressed as infinite sums, or as rational quotients of polynomials, that generate the series, expanded in terms of CSA coordinates:
\begin{equation} 
\label{eq:pleth2.1}
\begin{aligned}
{g^{G}_{instanton}}\left( {t,{x_i}} \right) &\equiv \sum\limits_{n = 0}^\infty  {{{\left[ n \theta_1,\ldots,n \theta_r \right]}}{t^n}},
 \end{aligned}
\end{equation}
where $[\theta_1,\ldots, \theta_r]$ are the Dynkin labels for the highest weight of the adjoint representation $\theta$. We can express such a series using HWG notation \cite{Hanany:2014dia}, which results from mapping the characters in the series to fugacities for highest weight Dynkin labels. Using the Dynkin label fugacities $\{ m_1,\ldots, m_r \}$ and taking the Dynkin labels of the highest weight of the adjoint representation as $[\theta_1,\ldots, \theta_r]$, the instanton series can equivalently be expressed in terms of monomials as:
\begin{equation} \label{eq:pleth2.2}
\begin{aligned}
{g^{G}_{instanton}}\left( {t,{x_i}} \right) \Leftrightarrow g_{instanton}^G\left( {t,{m_1}, \ldots, {m_r}} \right) &\equiv \sum\limits_{n = 0}^\infty  {m_1^{n{\theta _1}} \ldots m_r^{n{\theta _r}}{t^n}} \\
 &  = PE\left[ {m_1^{{\theta _1}} \ldots m_r^{{\theta _r}}t} \right].
 \end{aligned}
\end{equation}
Obtaining a generating function for \ref{eq:pleth2.1} is not straightforward, since a symmetrisations of the adjoint just using the PE function invariably give rise to many representations besides the required series:
\begin{equation} \label{eq:pleth2.3}
\begin{aligned}
PE\left[ {[\theta_1,\ldots, \theta_r]~t} \right] &= \sum\limits_{n = 0}^\infty  {Sy{m^n}\left[ \theta_1,\ldots, \theta_r \right]{t^n}}\\
&= \sum\limits_{n = 0}^\infty  {\left[ n \theta_1,\ldots, n \theta_r \right]{t^n}}  +  \ldots [other~irreps].
 \end{aligned}
\end{equation}
Thus, for $SU(2)$, character expansion yields the result:
\begin{equation} \label{eq:pleth2.4}
\begin{aligned}
PE\left[ {[2] t} \right] &= 1 + \left[ 2 \right]t + {t^2} + \left[ 4 \right]{t^2} + \left[ 2 \right]{t^3} + \left[ 6 \right]{t^3} +  \ldots \\
 &  = \sum\limits_{{n_1},{n_2} = 0}^\infty  {\left[ {2{n_1}} \right]{t^{{n_1} + 2{n_2}}}} \\
 &  = \frac{1}{{\left( {1 - {t^2}} \right)}}\sum\limits_{n = 0}^\infty  {\left[ {2n} \right]{t^n}} \\
 &  = \frac{1}{{\left( {1 - {t^2}} \right)}}g_{instanton}^{SU(2)}\left( {t,{{x_i}}} \right).\\
 \end{aligned}
\end{equation}
We can summarise this series most efficiently using HWG notation:
\begin{equation} \label{eq:pleth2.4a}
\begin{aligned}
PE\left[ {[2] t} \right] & \Leftrightarrow   1 +m^2 t + {t^2} +m^4{t^2} + m^2 {t^3} + m^6 {t^3} +  \ldots \\
& = 1/(1-t^2)/(1-m^2 t)\\
& = PE[t^2 + m^2 t].\\
 \end{aligned}
\end{equation}
Using HWG notation, we set out in Table \ref{table3} the results of such a symmetrisation exercise for a selection of low rank groups.
\begin{table}
\caption{HWGs for PE of Adjoint for Low Rank Classical Groups}
\begin{center}
\begin{tabular}{|c|c|c|c|c|c|}
\hline
 $Group$& $Adjoint$&$ PL[g_{PE[adj~t]}^G\left( {t,{m_i}} \right)]$\\
\hline
$ {{A_1} \cong {B_1} \cong {C_1}} $ &$ {\left[ 2 \right]}$&$ {{t^2} + {m^2}t} $\\
\hline
$ {{A_2}} $&$ {\left[ {1,1} \right]} $&$\begin{array}{c}{t^2} + {t^3} + {m_1}^3{t^3} + {m_2}^3{t^3} \\+ {m_1}{m_2}t + {m_1}{m_2}{t^2} - {m_1}^3{m_2}^3{t^6} \end{array}$\\
\hline
$ {{B_2}}$&$ {\left[ {0,2} \right]}$&$ \begin{array}{c}{t^2} + {t^4} + {m_1}{t^2} + {m_1}^2{t^2} + {m_2}^2t\\ + {m_2}^2{t^3} + {m_1}{m_2}^2{t^4} 
- {m_1}^2{m_2}^4{t^8}\end{array}$\\
\hline
$ {{C_2}}$&$ {\left[ {2,0} \right]}$&$ {As~{B_2}~with~{m_1} \Leftrightarrow {m_2}}$\\
\hline
$ {{D_2}}$&$ {\left[ {2,0} \right] \oplus [0,2]}$&$ {2{t^2} + m_1^2t + m_2^2t}$\\
\hline
\end{tabular}
\end{center}
\label{table3}
\end{table}
Returning to our $SU(2)$ example, we can read off the relations:
\begin{equation} 
\label{eq:pleth2.5}
\begin{aligned}
PE\left[ {[2] t} \right] & \Leftrightarrow PE\left[ {{t^2} + {m^2}t} \right]\\
 &  = \frac{1}{{\left( {1 - {t^2}} \right)}}PE\left[ {{m^2}t} \right]\\
 &  = \frac{1}{{\left( {1 - {t^2}} \right)}}g_{instanton}^{SU(2)}\left( {t,m} \right).\\
\end{aligned}
\end{equation}
In this case, a simple rearrangement of \ref{eq:pleth2.4} or \ref{eq:pleth2.5} gives us the generating function we seek, so that:
\begin{equation}
\label{eq:pleth2.6}
\begin{aligned}
g_{instanton}^{SU(2)}\left( {t,{{x_i}}} \right) = \left( {1 - {t^2}} \right)PE\left[ {[2] t} \right].
 \end{aligned}
\end{equation}
This ansatz generalises to RSIMS series associated with any group, with the important proviso that the pre-factor to the PE term is generally a non-trivial class function transforming in some combination of irreps, rather than just a polynomial in the fugacity $t$:
\begin{equation} 
\label{eq:pleth2.7}
\begin{aligned}
g_{instanton}^{G}\left( {t,{{x_i}}} \right) =P^G_{instanton}[{\cal X}(x_i),t]~PE\left[ {\theta t} \right].
 \end{aligned}
\end{equation}
The class function $P^G_{instanton}[{\cal X}(x_i),t]$ can be found by a variety of routes, including from the quiver gauge theory constructions described in later sections. For groups where the adjoint combines one or more basic irreps (i.e. has Dynkin labels equal to one or zero only), the RSIMS generating function can also be obtained by simplifying a character generating function; this is the route that has been taken here for $G_2$\footnote{The $G_2$ instanton generating function has also been calculated using dimensional analysis \cite{Benvenuti:2010pq}.} and for $F_4$.

To elaborate on this method, we can, as shown in \cite{Hanany:2014dia}, obtain a generating function for the characters ${{\cal X}{(x_i)}}$ of any representation of a group $G$ from the Weyl Character Formula as:
\begin{equation}
\label{eq:pleth2.8}
{g^G_{\cal X}}\left( {{m_i},x_i} \right) = \frac{1}{{x\prod\limits_{\alpha  \in \Phi  + }^{} {\left( {1 - {x^{ - \alpha }}} \right)} }}\sum\limits_{w \in W}^{} {\det \left[ w \right]\prod\limits_j {\frac{{\prod\limits_i {{x_i}^{{w_{ij}}}} }}{{1 - {m_j}\prod\limits_i {{x_i}^{{w_{ij}}}} }}} },
\end{equation}
where the Weyl vector is ${x } \equiv \prod\limits_i {x_i} = \prod\limits_{\alpha  \in \Phi  + } {x^{\alpha /2}} $ and the $m_i$ are Dynkin label fugacities. This generating function specialises to the RSIMS series as:
\begin{equation}
\label{eq:pleth2.9}
g_{instanton}^G\left( {{t},x_i} \right) = \frac{1}{{x\prod\limits_{\alpha  \in \Phi  + } {\left( {1 - {x^{ - \alpha }}} \right)} }}\sum\limits_{w \in W}^{} {\det \left[ w \right]\frac{{\prod\limits_{i,j} {{x_i}^{{w_{ij}}}} }}{{1 - t\prod\limits_i {{x_i}^{{w_{ij}}{\theta _j}}} }}}.
\end{equation}
The Weyl group matrices required for calculations can be obtained from {\it Mathematica} add-on programs such as LieArt \cite{Feger:2012bs}.

An equivalent formula for the generating functions of RSIMS is provided by \cite{Keller:2011ek, Keller:2012da}. This method expresses \ref{eq:pleth2.9} purely in terms of roots and their inner products, thereby avoiding the need for explicit determination of the full Weyl group of matrices.

Since the highest weight $\theta$ of the adjoint representation is a longest root, and since root length is invariant under Weyl group reflections, the action of elements $w \in W$ of the Weyl group, $\theta  \to w\theta$, can be used to decompose the Weyl group into a subgroup ${W_0} \equiv {\left\{w_0 \right\}} \equiv \left\{ {w: w\theta  = \theta } \right\}$, which leaves $\theta$ invariant, and its cosets $\left\{ {{w_\gamma }} \right\} \equiv \left\{ {w:{w_\gamma }\theta  = \gamma } \right\}$, where ${\gamma}$ are the long roots. By choosing a representative ${w_\gamma }$ from each coset, we can write any Weyl group element as $w={w_\gamma }{w_0 }$, for some element of ${W_0}$.

Under such a decomposition, the subgroup ${W_0}$ is the Weyl group of the Lie algebra ${G_0} \subset G$, that is determined by the maximal subset of the simple roots of $G$ that are not linked to the extended (or affine) node of the Dynkin diagram for $G$ (see later). The simple roots of $G_0$ have the property of being orthogonal to the (highest weight of the) adjoint of $G$. These Weyl group decompositions are described in Table \ref{table4}\footnote{based on \cite{Keller:2011ek}, corrected for the C series}.

\begin{table}
\caption{Weyl Group Decomposition by Action on Adjoint}
\begin{center}
\begin{tabular}{|c|c|c|c|c|}
\hline
$ G$&$ {\left| W \right|}$&$ {\left| {{\Phi _{long}}} \right|}$&$ {\left| {{W_0}} \right| = \left| W \right|/ {\left| {{\Phi _{long}}} \right|} }$&$ {{G_0}}$ \\
\hline
$ {{A_n}}$&$ {\left( {n + 1} \right)!}$&$ {n\left( {n + 1} \right)}$&$ {\left( {n - 1} \right)!}$&$ {{A_{n - 2}}}$ \\
$ {{B_{n\ge2}}}$&$ {{2^n}n!}$&$ {2n\left( {n - 1} \right)}$&$ {{2^{n - 1}}\left( {n - 2} \right)!}$&$ {{A_1} \times {B_{n - 2}}}$ \\
$ {{C_n}}$&$ {{2^n}n!}$&$ {2n}$&$ {{2^{n - 1}}\left( {n - 1} \right)!}$&$ {{C_{n - 1}}}$ \\
$ {{D_{n \ge 4}}}$&$ {{2^{n - 1}}n!}$&$ {2n\left( {n - 1} \right)}$&$ {{2^{n - 2}}\left( {n - 2} \right)!}$&$ {{A_1} \times {D_{n - 2}}}$ \\
$ {{E_6}}$&$ {72.6!}$&$ {72}$&$ {6!}$&$ {{A_5}}$ \\
$ {{E_7}}$&$ {72.8!}$&$ {126}$&$ {{2^5}.6!}$&$ {{D_6}}$ \\
$ {{E_8}}$&$ {192.10!}$&$ {240}$&$ {72.8!}$&$ {{E_7}}$ \\
$ {{F_4}}$&$ {1152}$&$ {24}$&$ {{2^3}.3!}$&$ {{C_3}}$ \\
$ {{G_2}}$&$ {12}$&$ 6$&$ {2!}$&$ {{A_1}}$ \\
\hline
\end{tabular}
\label{table4}
\end{center}
\end{table}
Using such a decomposition, we can rewrite \ref{eq:pleth2.9} as:
\begin{equation}
\label{eq:pleth2.10}
g_{instanton}^G\left( {t,x_i} \right) =\sum\limits_{\gamma  \in {\Phi _{long}}}^{} {\frac{{\det \left[ {{w_\gamma }} \right]}}{{\left( {1 - {x^\gamma }t} \right)x\prod\limits_{\alpha  \in \Phi  + } {\left( {1 - {x^{ - \alpha }}} \right)} }}} \sum\limits_{{w_0} \in {W_0}}^{} {\det \left[ {{w_0}} \right]\prod\limits_{i,j} {{x_i}^{{{\left( {{w_\gamma }{w_0}} \right)}_{ij}}}} }.
\end{equation}
By drawing on Weyl's identity \cite{Macdonald:1995fk}, which can be applied equally to $W$ and to $W_0$:
\begin{equation}
\label{eq:pleth2.11}
\sum\limits_{w \in W}^{} {\det[w]}\prod\limits_{i,j} {x_i^{{w_{ij}}}} = \prod\limits_{\alpha  \in \Phi  + } {\left( {{x^{\alpha /2}} - {x^{ - \alpha /2}}} \right)},
\end{equation}
and by following the group theoretic calculations in \cite{Keller:2011ek, Keller:2012da}, we can reduce \ref{eq:pleth2.10} to a general result for an RSIMS. This can be written most concisely as:
\begin{equation}
\label{eq:pleth2.12}
{g_{instanton}^G\left( {t,x_i} \right) = \sum\limits_{\gamma  \in {\Phi _{long}}} {\frac{1}{{\left( {1 - {x^\gamma }t} \right)\left( {1 - {x^{ - \gamma }}} \right)\prod\limits_{{\alpha \in \Phi}: (\alpha, \gamma)  = 1} {\left( {1 - {x^{ - \alpha }}} \right)} }}} }.
\end{equation}
As previously, the terms $x^{ \gamma }$ and $x^{ \alpha}$ in \ref{eq:pleth2.12} represent monomials in CSA coordinates and $(\alpha, \gamma)$ is the inner product that selects the required subsets of the roots. It follows from the form of \ref{eq:pleth2.12}, in which $t$ appears coupled to long roots only, that the dimension of the {\it refined} RSIMS is given by the number of long roots. 

The class functions $P^G_{instanton}$ can be separated out, once the various generating functions have been calculated using either \ref{eq:pleth2.9} or \ref{eq:pleth2.12}, and we tabulate these in Tables \ref{table5} and \ref{table6} for low rank Classical and Exceptional groups.
\begin{table}
\caption{RSIMS Generating Functions for Low Rank Classical Groups}
\begin{tabular}{|c|c|c c|}
\hline
$ {Series}$
&$\begin{array}{c}RSIMS\\HWG\end{array}$&$P^G_{instanton}$&${PE}$ \\
\hline
$ {{A_1}} $&
${{m^2}t}$&${\left( 1 - {t^2} \right)[0,0]}$&${PE\left[ {\left[ 2 \right]t} \right]}$ \\
\hline
$ {{A_2}} $&
${{m_1}{m_2}t}$&
$\left( {\begin{array}{*{20}{c}}
{1 - {t^2} - {t^4} + {t^6}}&\left[ {0,0} \right]\\
 - {t^2} + 2{t^3} - {t^4}&{\left[ {1,1} \right]}
 \end{array}} \right)$&
${PE\left[ {\left[ {1,1} \right]t} \right]}$\\
\hline
$ {{A_3}} $&
${{m_1}{m_3}t}$&
${P_{instanton}^{A_3}}$&
${PE\left[ {\left[ {1,0,1} \right]t} \right]}$ \\
\hline
$ {{B_2}} $&
${m_2^2t}$&
${\left( {\begin{array}{*{20}{c}}
{1 - {t^2} - {t^6} + {t^8}}&{\left[ {0,0} \right]}\\
{{t^3} - 2{t^4} + {t^5}}&{\left[ {0,2} \right]}\\
{ - {t^2} + {t^3} + {t^5} - {t^6}}&{\left[ {1,0} \right]}\\
{{t^3} - 2{t^4} + {t^5}}&{\left[ {1,2} \right]}\\
{- {t^2} + {t^3} + {t^5} - {t^6}}&{\left[ {2,0} \right]}
\end{array}} \right)}$&
${PE\left[ {\left[ {0,2} \right]t} \right]}$ \\
\hline
$ {{B_3}} $&
${{m_2}t}$&${P_{instanton}^{B_3}}$&${PE\left[ {\left[ {0,1,0} \right]t} \right]}$ \\
\hline
$ {{C_2}} $&
${{m_1}^2t}$&
${\left( {\begin{array}{*{20}{c}}
{1 - {t^2} - {t^6} + {t^8}}&{\left[ {0,0} \right]}\\
{ - {t^2} + {t^3} + {t^5} - {t^6}}&{\left[ {0,1} \right]}\\
{- {t^2} + {t^3} + {t^5} - {t^6}}&{\left[ {0,2} \right]}\\
{{t^3} - 2{t^4} + {t^5}}&{\left[ {2,0} \right]}\\
{{t^3} - 2{t^4} + {t^5}}&{\left[ {2,1} \right]}
\end{array}} \right)}$&
${PE\left[ {\left[ {2,0} \right]t} \right]}$ \\
\hline
$ {{C_3}} $&
${m_1^2t}$&${P_{instanton}^{C_3}}$&${PE\left[ {\left[ {2,0,0} \right]t} \right]}$ \\
\hline
$ {{D_2}} $&
${m_1^2t + m_2^2t}$&
${\left( {\begin{array}{*{20}{c}}
{2 - 2{t^2} - 2{t^3} +2 {t^5}}&{\left[ {0,0} \right]}\\
{-{t}+{t^2}+{t^3} -{t^4} }&{\left[ {0,2} \right]}\\
{-{t}+{t^2}+{t^3} -{t^4} }&{\left[ {2,0} \right]}
\end{array}} \right)}$&
${PE\left[ {\left[ {2,0} \right]t + \left[ {0,2} \right]t} \right]}$ \\
\hline
$ {{D_3}} $&
${{m_2}{m_3}t}$&${P_{instanton}^{D_3}}$&${PE\left[ {\left[ {0,1,1} \right]t} \right]}$ \\
\hline
$ {{D_4}} $&
${{m_2}t}$&${P_{instanton}^{D_4}}$&${PE\left[ {\left[ {0,1,0,0} \right]t} \right]}$ \\
 \hline
\end{tabular}
\label{table5}
\end{table}

\begin{table}
\caption{RSIMS Generating Functions for Exceptional Groups}
\begin{tabular}{|c|c|c c|}
\hline
$ {Series}$&
$\begin{array}{c}
RSIMS\\
HWG
\end{array}$&$P^G_{instanton}$&${PE}$ \\
\hline
$ {{E_6}}$&
${{m_6}t}$&${to~be~calculated}$&${PE\left[ {\left[ 0,0,0,0,0,1 \right]t} \right]}$ \\
$ {{E_7}}$&
${{m_1}t}$&${to~be~calculated}$&${PE\left[ {\left[ {1,0,0,0,0,0,0} \right]t} \right]}$ \\
$ {{E_8}}$&
${{m_7}t}$&${to~be~calculated}$&${PE\left[ {\left[ {0,0,0,0,0,0,1,0} \right]t} \right]}$ \\
\hline
$ {{F_4}}$&
${{m_1}t}$&${P_{instanton}^{F_4}}$&${PE\left[ {\left[ {1,0,0,0} \right]t} \right]}$ \\
\hline
$ {{G_2}}$&
${{m_1}t}$&${\left( {\begin{array}{*{20}{c}}
{1 - {t^2} - {t^9} + {t^{11}}}&{\left[ {0,0} \right]}\\
{ - {t^4} + {t^5} + {t^6} - {t^7}}&{\left[ {0,1} \right]}\\
{ - {t^2} + {t^3} + {t^8} - {t^9}}&{\left[ {0,2} \right]}\\
{ - {t^4} + {t^5} + {t^6} - {t^7}}&{\left[ {0,3} \right]}\\
{{t^3} - {t^4} - {t^7} + {t^8}}&{\left[ {1,0} \right]}\\
{{t^3} - {t^4} - {t^7} + {t^8}}&{\left[ {1,1} \right]}
\end{array}} \right)}$&${PE\left[ {\left[ {1,0} \right]t} \right]}$ \\
\hline
\end{tabular}
\label{table6}
\end{table}

The numerator class functions $P^G_{instanton}$ for $A_3$, $B_3$, $C_3$, $D_3$, $D_4$ and $F_4$ are somewhat lengthy and are given in Appendix 1. The isomorphisms $B_2  \cong C_2$ and $A_3  \cong D_3$ are apparent under interchange of Dynkin labels and their fugacities. All the $P^G_{instanton}$ class functions take a particular form when viewed as polynomials in $t$, being palindromic (or anti-palindromic) with some maximum degree $d$ and with the absolute values of the coefficients of $t^k$ and $t^{d-k}$ being equal. Also, for simple groups, the coefficient of $t^0$ is always equal to unity and that of $t$ always vanishes.

The class functions $P^G_{instanton}$ for the E series groups have yet to be calculated explicitly \footnote{Owing to memory constraints in {\it Mathematica}}, although it remains feasible to use \ref{eq:pleth2.12} in unfactored form. It is a straightforward matter to verify that the Taylor expansions of all these generating functions in powers of $t$ yield the characters of the reduced single instanton moduli spaces in accordance with \ref{eq:pleth2.1}.
\FloatBarrier
\section{RSIMS from Coulomb Branches of Extended Dynkin Diagrams}
\label{sec:coulomb}
\subsection{Introduction}

The monopole construction of RSIMS in \cite{Cremonesi:2013lqa,Cremonesi:2014xha}, also referred to as a Coulomb branch construction of RSIMS, draws upon a lattice determined by the simple roots and dual Coxeter labels of $G$.\footnote{This lattice is often referred to as a GNO lattice.\cite{Goddard:1976qe}} It exploits an intriguing and highly non-trivial relationship between $G$ and a unitary product group defined by the dual Coxeter labels of $G$, and inherits further structure from the extended Dynkin diagram of $G$.\footnote{For simply laced groups, the extended Dynkin diagram of $G$ differs from the extended Dynkin diagram of the GNO dual of $G$.}

The monopole construction is built directly upon the root space of the Lie algebra and assembles RSIMS out of sums of monomials in simple roots. For A series constructions, the simple roots are each associated with a $U(1)$ symmetry group. The algorithm used in the general monopole construction works, however,  with $U(N)$ rather than $U(1)$ symmetry groups. A $U(N)$ group is assigned to each simple root of an algebra, having its rank set by the {\it dual} Coxeter label ${\mathord{\buildrel{\lower3pt\hbox{$\scriptscriptstyle\smile$}} 
\over a}_i}$ of the simple root . Thus, any Lie group of rank $r$ is associated with a unitary product group $\prod\limits_{i = 1}^r {\otimes U({\mathord{\buildrel{\lower3pt\hbox{$\scriptscriptstyle\smile$}} 
\over a}_i})}$. The monopole construction also counts root monomials according to a precise definition of {\it conformal dimension} that depends upon the linking pattern of the {\it extended} (or  {\it untwisted affine}) Dynkin diagram \cite{Dynkin:1957um} of the Lie algebra, as well as upon root length information encoded in the Cartan matrix, as will be elaborated.

\subsection{Affine Lie Algebras}
\begin{figure}[htbp]
\begin{center}
\includegraphics[scale=0.35]{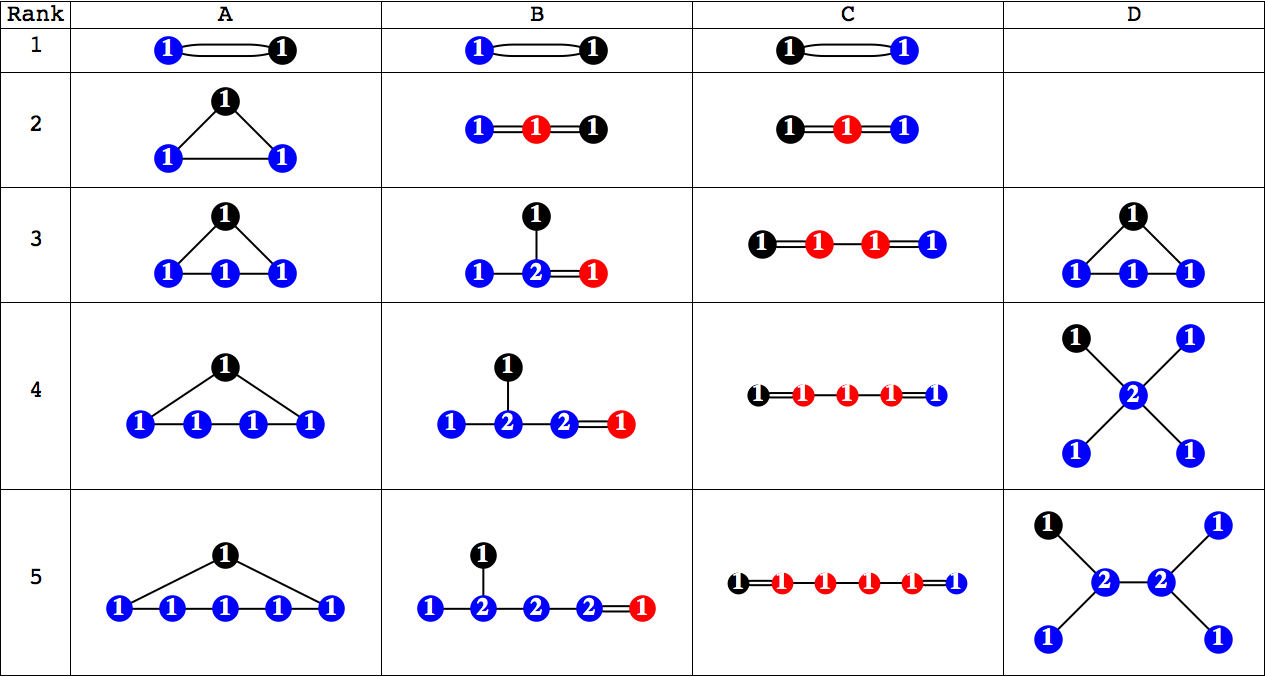}\\
\caption[Extended Dynkin diagrams for Classical Groups.]{Extended Dynkin diagrams for simple Classical Groups up to rank 5. Blue nodes denote long roots with length 2. Red nodes denote short roots. A black node denotes the long root added in the affine construction. The dual Coxeter labels giving the $U(N)$ symmetry for each node are also shown.}
\label{figABCD}
\end{center}
\end{figure}

\begin{figure}[htbp]
\begin{center}
\includegraphics[scale=0.35]{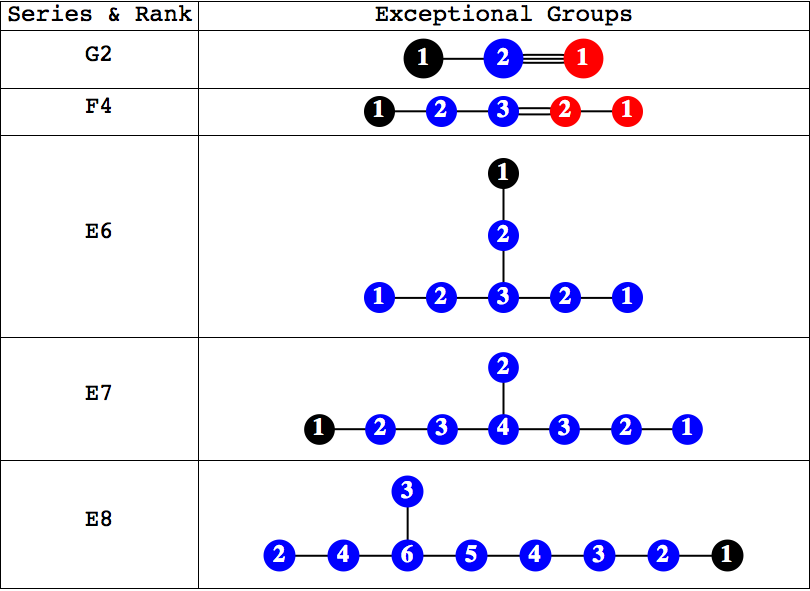}\\
\caption[Extended Dynkin diagrams for Exceptional Groups.]{Extended Dynkin diagrams for Exceptional Groups. Blue nodes denote long roots with length 2. Red nodes denote short roots. A black node denotes the long root added in the affine construction. The dual Coxeter labels giving the $U(N)$ symmetry for each node are also shown.}
\label{figEFG}
\end{center}
\end{figure}

It is useful to give a brief summary of the relationship between a simple Lie algebra and its related untwisted affine Lie algebra. For further detail the reader is referred to \cite{Fuchs:1997bb}. An affine Lie algebra is formed by generalising a Cartan matrix $A^{ij}$ through the addition of an extra row and column, corresponding to an extra simple root and an extra eigenvalue operator, or, equivalently, to adding an extra node to the Dynkin diagram. Specifically, a Cartan matrix $A^{ij}$, with entries $2$ on the diagonal, is modified to form an untwisted affine Cartan matrix according to the schema:
\begin{equation}
\label{eq:mon3.2}
affine~A^{ij} = \left( {\begin{array}{*{20}{c}}
A^{ij}&{\left[ {col} \right]}\\
{ - \left[ {adjoint} \right]}&2
\end{array}} \right),
\end{equation}
where the column vector $[col]$ is obtained by transposing the Dynkin labels of the adjoint representation and replacing all non-zero entries with $-1$ or $-2$ (in the case of $A_1$, for example), such that the affine Cartan matrix acquires a zero determinant and becomes degenerate.\footnote{There is also a class of twisted affine Lie algebras \cite{Fuchs:1997bb}, whose Cartan matrices are similarly degenerate, but in which the extra node is connected to a representation other than the adjoint. The Dynkin diagrams of these twisted affine Lie algebras do not correspond to canonical {\it extended} Dynkin diagrams \cite{Dynkin:1957um} and are not studied herein.} For reference, we tabulate in Figures \ref{figABCD} and \ref{figEFG} the extended (or untwisted affine) Dynkin diagrams for the simple Classical and Exceptional groups respectively \cite{Fuchs:1997bb}.

The defining feature of an affine Lie algebra is that the affine Cartan matrix is degenerate positive semidefinite, having a zero determinant and one zero eigenvalue; this in turn means that the additional root and eigenvalue operators are linear combinations of the other operators. Naturally, the rank is unchanged.  The linear relationship between the operators is encapsulated in the Coxeter labels ${a_j}$ and dual Coxeter labels ${\mathord{\buildrel{\lower3pt\hbox{$\scriptscriptstyle\smile$}} \over a} _i}$ of each node. These labels are respectively the left and right eigenvectors with zero eigenvalue of the affine Cartan matrix:
\begin{equation}
\label{eq:mon3.3}
\sum\limits_{i = 0}^r {{a_i}} {A^{ij}}= 0 =\sum\limits_{j = 0}^r {{A^{ij}}{{\mathord{\buildrel{\lower3pt\hbox{$\scriptscriptstyle\smile$}} \over a} }_j}}.
\end{equation}
The two types of Coxeter label differ according to the length of the simple root to which they refer: the ratio between the dual Coxeter label and the Coxeter label of a root is equal to the ratio of its length to the length of the longest root \cite{Fuchs:1997bb}. 

The Cartan matrix for an affine Lie algebra can be reduced to that for a regular Lie algebra by the elimination of a row and its corresponding column (not necessarily the row and column that were added to form the affine Cartan matrix). An important feature of the construction is that both the dual Coxeter and Coxeter labels of other nodes are invariant under the addition or subtraction of untwisted affine nodes.

The eigenvector of the Cartan matrix with zero eigenvalue given by the dual Coxeter labels has important properties: it defines a linear relationship between the eigenvalue operators $H^j$ and a central charge C, which is invariant under the action of the root (i.e. raising/lowering) operators $E^i$ \cite{Fuchs:1997bb}:
\begin{equation}
\label{eq:mon3.4}
C = \sum\limits_{j = 0}^r {{H^j}} {{\mathord{\buildrel{\lower3pt\hbox{$\scriptscriptstyle\smile$}} \over a} }_j},
\end{equation}
\begin{equation}
\label{eq:mon3.5}
\left[ {C,E_ \pm ^i} \right] = 0.
\end{equation}
In the case where the central charge C is zero, the untwisted affine Lie algebra is equivalent to the original Lie algebra, with some degeneracy/redundancy amongst operators and Dynkin labels of irreps.\footnote{Other constructions are also studied, such the addition of derivations to the affine Lie algebra \cite{Fuchs:1997bb} to realise an algebra with a non-zero central charge}. For the purpose of the monopole construction of RSIMS, we work with a central charge of zero and simply make use of the linking pattern of the extended Dynkin diagram, as encoded in the untwisted affine Cartan matrix, and its dual Coxeter labels.
\FloatBarrier
\subsection{Coulomb Branch or Monopole Construction}
Having covered some preliminaries, we can now give the general monopole construction of RSIMS, which is valid for all simple Classical and Exceptional Lie groups. This follows the schema, refined from \cite{Cremonesi:2013lqa}:
\begin{equation}
\label{eq:mon3.6}
g_{instanton}^G\left( {t,z} \right) \equiv  \sum\limits_{n = 0}^\infty  {\left[ {adj~n} \right]{t^n}}   =\sum\limits_q^{} {{z^q}} {t^{\Delta \left( q \right)}} \prod\limits_{i,j} {\frac{1}{{\left( {1 - {t^{{d_{i,j}(q)}}}} \right)}}}. 
\end{equation}
The formula makes use of simplifying notation, which requires explication to give an unambiguous construction:
\begin{enumerate}

\item The variable t is a fugacity for the Dynkin labels of the adjoint representation.

\item The label $z$ is a collective coordinate for a monomial $z_1 \ldots z_r$ in the simple roots of the Lie algebra.

\item The rank of the $U(N_i)$ symmetry group of a simple root is given by the dual Coxeter label ${\mathord{\buildrel{\lower3pt\hbox{$\scriptscriptstyle\smile$}} \over a} _i}$ of its node on the Dynkin diagram.

\item The label $q$ is a collective coordinate for the monopole fluxes (or ``GNO charges") $\left\{ {\left\{ {{q_{1,1}}, \ldots ,{q_{1,{N_1}}}} \right\}, \ldots ,\left\{ {{q_{r,1}}, \ldots ,{q_{r,{N_r}}}} \right\}} \right\}$, arranged into subsets for each $U(N_i)$ symmetry group. 

\item The term $z^q$ combines the collective $z$ and $q$ coordinates into overall charges for each monomial in the roots and is expanded as ${z^q} \equiv \prod\limits_{i = 1}^r {z_i^{\sum\limits_{j = 1}^{{N_i}} {{q_{i,j}}} }}$.

\item The limits of summation for the monopole charges are $\infty  \ge {q_{i,1}} \ge \ldots {q_{i,j}}  \ge \ldots {q_{i,{N_i}}} \ge  - \infty $ for $i=1,\ldots r$. (In the case of $U(1)$ symmetry it is convenient to drop the redundant second index on $q_{i,j}$.)

\item The terms $d_{i,j}$ give the degrees of the Casimirs of the residual $U(N_i)$ symmetries that remain for each root under each assignment of $q$ charges (explained below).

\item The term $\Delta \left( q \right)$ gives the conformal dimension (explained below) associated with each assignment of $q$ charges.
\end{enumerate}
The determination of residual symmetries for each root under each assignment of monopole charges follows \cite{Cremonesi:2013lqa}. We construct a partition of $N_i$ for each root, which counts how many of the charges $q_{i,j}$ are equal, such that $\lambda(q_i)=(\lambda_{i,1},\ldots,\lambda_{i,N_i})$, where $\sum\limits_{j = 1}^{{N_i}} {{\lambda _{i,j}}}  = {N_i}$ and ${\lambda _{i,j}} \ge {\lambda _{i,j + 1}}$. The terms $\lambda_{i,j}$ in the partition give the ranks of the residual $U(N)$ symmetries associated with each root, so that it is a straightforward matter to compound the terms in the degrees of Casimirs, recalling that a $U(N)$ group has Casimirs of degrees 1 through $N$:
\begin{equation}
\label{eq:mon3.7}
\prod\limits_{i,j} {\frac{1}{{\left( {1 - {t^{{d_{i,j}(q)}}}} \right)}}}  \equiv \prod\limits_{i = 1\hfill \atop j = 1\hfill}^{i = r\hfill \atop
j = {N_i}\hfill} {\frac{1}{{\prod\limits_{k = 1}^{{\lambda _{ij}}\left( q_i \right)} {\left( {1 - {t^k}} \right)} }}}.
\end{equation}
So, for example, if $q_{i,j}= q_{i,k}$ for all $j ,k$, then $\{d_{i,1},\ldots d_{i,N_i}\}=\{1,\ldots N_i$\} and if $q_{i,j}\neq q_{i,k}$ for all $j ,k$, then $\{d_{i,1},\ldots d_{i,N_i}\}=\{1,\ldots 1$\}.

Thus far, all the group theoretic parameters involved in the monopole construction of the reduced moduli spaces of single instantons have simply been those of the Classical or Exceptional Lie group. The calculation of conformal dimension also draws upon the linking pattern of the {\it extended} Dynkin diagram, or, equivalently, the {\it extended} Cartan matrix ${A_{ij}}$. The conformal dimension is given by the formula:
\begin{equation}
\label{eq:mon3.8}
\Delta \left( q \right) = \frac{1}{2}\left( {\sum\limits_{\begin{array}{*{20}{c}}
{j> i \ge 0\atop Affine~{A_{ij}} \ne 0}
\end{array}}^r {\sum\limits_{m,n} {\left| {{q_{i,m}}{A_{ij}} - {q_{j,n}}{A_{ji}}} \right|} } } \right) - \sum\limits_{i = 1}^r {\sum\limits_{m > n}^{} {\left| {{q_{i,m}} - {q_{i,n}}} \right|} }.
\end{equation}
In the conformal dimension formula, the extra affine root, labelled by $i=0$ is typically assigned a $U(1)$ monopole charge $q_0$ of zero. Nonetheless, it still plays a role in the first term in \ref{eq:mon3.8} in accordance with the linking pattern in the extended $A_{ij}$ Cartan matrix. There are other possible gauge choices, as will be discussed.

The above procedure gives an algorithm for the monopole construction of RSIMS for any simple Classical or Exceptional group, including those of the non-simply laced BCFG series, in addition to the ADE series. However, in order for the formulae to be valid for non-simply laced groups, it is essential to use the {\it dual} Coxeter labels associated with the nodes of the Dynkin diagram \cite{Cremonesi:2014xha}; it is also essential that differences in root lengths are treated using the extended Cartan matrix, as implemented in \ref{eq:mon3.8}.

The character of the adjoint representation of any group is given by the sum of its roots, which have as their basis a set of simple roots, plus the rank of the group. The root space is in turn spanned by the monomials $z^q \equiv {z_1}^{q_1} \dots {z_r}^{q_r}$, used in \ref{eq:mon3.6}. An RSIMS construction using the root space therefore requires the collection of the root monomials into representations, at the correct positive and negative integer powers and multiplicities. As set out above, central roles are played by the fugacity $t$, in conjunction with its exponent, the conformal dimension $\Delta (q)$, and the $U(N)$ symmetry groups associated with the dual Coxeter labels of the roots.

We can obtain further insight into the mechanisms behind the workings of the monopole construction by studying the structure of the root space of the adjoint representation and its symmetrisations, and we do this in the following sections.

The conformal dimension, as defined, has a number of important properties. Firstly, as we illustrate below, conformal dimension is invariant under the Weyl group of reflections of the root space and so effects a foliation of a root space into sets of dominant weights and their associated orbits. Secondly, this foliation requires that the conformal dimension is a non-negative integer.\footnote{We note in passing that \cite{Gaiotto:2008ak} classifies theories as ``good", ``ugly" or  ``bad", depending on whether conformal dimension is 1, 1/2 or 0. In the case of the RSIMS construction, conformal dimension ranges over all non-negative integer values.} This requirement of integer shifts around the root space driven by the $q$ charges is satisfied as a result of the {\it balanced} property of all the extended Dynkin diagrams, shown in Figures \ref{figABCD} and  \ref{figEFG}. A quiver is defined as {\it balanced} \cite{Gaiotto:2008ak}, if the $U(N)$ charge on each node obeys the rule:
\begin{equation}
\label{eq:mon3.9a}
{N_i} ={\frac{1}{2}}  \sum\limits_{j \in \left\{ {\scriptstyle adjacent \atop
\scriptstyle nodes} \right\}}^{}  \left| A_{ij} \right| {{N_j}},
\end{equation}
where the weighting factors $| A_{ij}|$ are taken from the Cartan matrix as before. Under this condition, the unit displacement of any one of the $q$ charges, taking account of all the links in \ref{eq:mon3.8}, always leads to a unit (or zero) shift in conformal dimension.

We can obtain an expression for the unrefined moduli space or Hilbert series associated with the monopole construction $g_{instanton}^G\left( {1,t} \right)$ by the simple expedient of setting the root space coordinates to unity. Then, since the number of poles contributed by each $U(N_i)$ group depends only on rank $N_i$, and is invariant under the gauge group breaking by the monopole flux $q$, the dimension of this moduli space can be expanded as:
\begin{equation}
\label{eq:mon3.9}
Dim[g_{instanton}^G\left( {1,t} \right)] = Dim[\sum\limits_{{q}}^{} {t^{\Delta \left( q \right)}}] + Dim[\prod\limits_{i,j} {\frac{1}{{\left(1 - t \right)}}} ].
\end{equation}
The dimension of each of the RHS terms is determined by the sum of the ranks of the $U(N_i)$ symmetry groups associated with the nodes of the Dynkin diagram, that is to say, the sum of the dual Coxeter labels, in both cases. Hence, the dimension of the {\it unrefined} moduli space generated by the monopole construction is equal to twice the sum of the dual Coxter labels of the group. As noted in \cite{Hanany:2014dia}, the dimension of an RSIMS is equal to twice the sum of the dual Coxeter labels of the group.\footnote{This corresponds to the relationship between the dimension of a reduced single instanton moduli space and the quaternionic dual Coxter number established in \cite{Benvenuti:2010pq}, recalling that the (dual) Coxeter number of a Lie algebra is given by the sum of the (dual) Coxeter labels plus $1$.} This provides a non-trivial consistency check on the monopole construction.

While we cannot, at this time, present a general analytic proof of the equivalence between monopole constructions of RSIMS and those based on character generating functions, we can, in principle, demonstrate the analytic equivalence on a case by case basis; we do this below for $A_1, A_2$ and $B_2$. We can also check that expansion of each monopole construction generates the RSIMS series of characters (which we have done to as high an order as is practicable for all the Classical and Exceptional groups).

\subsection{Construction for Simply Laced Groups}
We now set out how the ADE series RSIMS constructions emerge from the general construction given by \ref{eq:mon3.6}, \ref{eq:mon3.7} and \ref{eq:mon3.8}. The treatment largely follows \cite{Cremonesi:2014xha}.  We then analyse the A series, showing the formal equivalence of monopole instanton constructions for $A_1$ and $A_2$ to ones based on character generating functions, and using the root structure of $A_2$ to illustrate the group theoretic properties of the conformal dimension construct.

\subsubsection{A Series}
The monopole construction for A series instantons of rank 2 and above is based on the extended Cartan matrix, defined in accordance with the schema \ref{eq:mon3.2}, and the dual Coxter labels of the simple roots (shown as a column vector), where we have labelled the affine root by $z_0$:
\begin{equation}
\label{eq:mon3.10}
\begin{tabular}{c|c c c c c c|c|}
${{z_1}}$&$ 2$&${ - 1}$&$ \ldots $&$0$&$0$&${ - 1}$&$ 1$ \\
${{z_2}}$&$ { - 1}$&$2$&$ \ldots $&$0$&$0$&$0$&$ 1$ \\
 $\ldots $&$  \ldots $&$ \ldots $&$ \ldots $&$ \ldots $&$ \ldots $&$ \ldots $&$  1 $ \\
${{z_{r - 1}}}$&$ 0$&$0$&$ \ldots $&$2$&${ - 1}$&$0$&$ 1$ \\
${{z_r}}$ & $0$ & $0$ & $ \ldots $ & ${-1}$ & $2$ & ${-1}$ & $1$ \\
${{z_0}}$&$ { - 1}$&$0$&$ \ldots $&$0$&${ - 1}$&$2$&$ 1$ \\
\end{tabular}.
\end{equation}
For $A_1$, the extended Cartan matrix and dual Coxeter labels are:
\begin{equation}
\label{eq:mon3.11}
\begin{tabular}{c|c c|c|}
$ z_1 $&$ 2 $&$ -2 $&$ 1$ \\
$ z_0 $&$ -2 $&$ 2 $&$ 1$ \\
\end{tabular}.
\end{equation}
Applying the prescription set out in \ref{eq:mon3.6}, \ref{eq:mon3.7} and \ref{eq:mon3.8}, we obtain the equation for an A series RSIMS:
\begin{equation}
\label{eq:mon3.12}
\begin{aligned}
g_{instanton}^{{A_r}} = {\frac{1}{{{{\left( {1 - t} \right)}^r}}}\sum\limits_{{q_1}, \ldots {{{q}}_r} =  - \infty }^\infty  {{{{z}}_1}^{{{{q}}_1}}} {{{z}}_2}^{{{{q}}_2}} \ldots {{{z}}_r}^{{{{q}}_r}}~{t^{\Delta \left( q \right)}}},\\
\end{aligned}
\end{equation}
where
\begin{equation}
\label{eq:mon3.13}
\begin{aligned}
\Delta \left( q \right) &= \frac{1}{2}\left( {\left| {{{{q}}_1}} \right| + \sum\limits_{i = 1}^{r - 1} {\left| {{{{q}}_i} - {{{q}}_{i + 1}}} \right|}  + \left| {{{{q}}_r}} \right|} \right).
\end{aligned}
\end{equation}
The resulting monopole constructions for $A_1$ and $A_2$ can be rearranged into the equivalent character generating functions. For $A_1$, where we are working with root space vectors expressed as $z_1$ in the basis of simple roots, rather than as $x^2$ in the basis of CSA coordinates, we have:
\begin{equation}
\label{eq:mon3.14}
\begin{aligned}
g_{instanton}^{{A_1}} &= \frac{1}{{\left( {1 - t} \right)}}\sum\limits_{{q_1} =  - \infty }^\infty  {z_1^{{q_1}}{t^{\left| {{q_1}} \right|}}} \\
 &  = \frac{1}{{\left( {1 - t} \right)}}\left( {\sum \limits_{{q_1} = 0}^\infty  {z_1^{{q_1}}{t^{{q_1}}}}  + \sum\limits_{{q_1} = 0}^\infty  {z_1^{ - {q_1}}{t^{{q_1}}} - 1} } \right)\\
 &  = \frac{{1 - {t^2}}}{{\left( {1 - t} \right)\left( {1 - {z_1}t} \right)\left( {1 - t/{z_1}} \right)}}\\
 &  = \left( {1 - {t^2}} \right)PE\left[ {\left[ 2 \right] t} \right].
\end{aligned}
\end{equation}
This yields the instanton character generating function for $A_1$ given in Table \ref{table5}.

For $A_2$ the rearrangement of the series, which follows the boundaries of the Weyl chambers of the group, is more intricate:
\begin{equation}
\label{eq:mon3.15}
\begin{aligned}
g_{instanton}^{{A_2}} &= \frac{1}{{{{\left( {1 - t} \right)}^2}}}\sum\limits_{{q_1},{q_2} =  - \infty }^\infty  {z_1^{{q_1}}z_2^{{q_2}}{t^{\frac{1}{2}\left( {\left| {{q_1}} \right| + \left| {{q_1} - {q_2}} \right| + \left| {{q_2}} \right|} \right)}}} \\
 &  = \frac{1}{{{{\left( {1 - t} \right)}^2}}}\left( \begin{array}{l}
\sum\limits_{{q_2} = 0}^\infty  {\sum\limits_{{q_1} = 0}^{{q_2}} {\left( {z_1^{{q_1}}z_2^{{q_2}} + z_1^{ - {q_1}}z_2^{ - {q_2}} + z_1^{{q_2}}z_2^{{q_1}} + z_1^{ - {q_2}}z_2^{ - {q_1}}} \right){t^{{q_2}}}} } \\
 - \sum\limits_{{q_1} = 0}^\infty  {\left( {z_1^{{q_1}}z_2^{{q_1}} + z_1^{ - {q_1}}z_2^{ - {q_1}}} \right){t^{{q_1}}}} \\
 - \sum\limits_{{q_1} = 0}^\infty  {\left( {z_1^{{q_1}} + z_1^{ - {q_1}} + z_2^{{q_1}} + z_2^{ - {q_1}}} \right){t^{{q_1}}}} \\
 + \sum\limits_{{q_1} = 0}^\infty  {\sum\limits_{{q_2} = 0}^\infty  {\left( {z_1^{ - {q_1}}z_2^{{q_2}} + z_1^{{q_1}}z_2^{ - {q_2}}} \right){t^{\left( {{q_1} + {q_2}} \right)}}} }  + 1\end{array} \right)\\
 &= \frac{{(1 - {t^2} - {t^4} + {t^6}) - \left( {{t^2} - 2{t^3} + {t^4}} \right)\left( {{z_1} + {z_2} + {z_1}{z_2} + z_1^{ - 1} + z_2^{ - 1} + z_1^{ - 1}z_2^{ - 1} + 2} \right)}}{{\left( {1 - {z_1}t} \right)\left( {1 - {z_2}t} \right)\left( {1 - {z_1}{z_2}t} \right)\left( {1 - z_1^{ - 1}t} \right)\left( {1 - z_2^{ - 1}t} \right)\left( {1 - z_1^{ - 1}z_2^{ - 1}t} \right){{\left( {1 - t} \right)}^2}}}\\
& = \left( {(1 - {t^2} - {t^4} + {t^6})\left[ {0,0} \right] - \left( {{t^2} - 2{t^3} + {t^4}} \right)\left[ {1,1} \right]} \right)PE[\left[ {1,1} \right] t].
\end{aligned}
\end{equation}
Once again, we obtain the instanton character generating function for $A_2$ as given in Table \ref{table5}.

Some insight into the structure of the monopole formula can be obtained by reversing the above procedure and seeking to derive the monopole constructions from the plethystic generating functions for RSIMS identified in Section \ref{sec:plethystics}. For $A_1$, summing  \ref{eq:pleth2.12}, the derivation proceeds as below:
\begin{equation}
\label{eq:mon3.15a}
\begin{aligned}
g_{instanton}^{{A_1}} &= \frac{1}{(1-z) \left(1-\frac{t}{z}\right)}+\frac{1}{\left(1-\frac{1}{z}\right) (1-t z)}\\
&=\sum _{a=0}^{\infty } \sum _{b=0}^{\infty } t^a z^{b-a}+\sum _{a=0}^{\infty } \sum _{b=0}^{\infty } t^a z^{a-b}\\
&=\sum _{q=-\infty }^{\infty } \sum _{b=\max (0,-q)}^{\infty }  t^{b+q}z^q +\sum _{q=-\infty }^{\infty } \sum _{b=\max (0,q)}^{\infty } t^{b-q} z^q \\
& = \frac{1}{{1 - t}}\sum\limits_{q =  - \infty }^\infty  {\left( {{t^{\max (0, - q) + q}} + {t^{\max (0,q) - q}}} \right){z^q}}\\
& = \frac{1}{{1 - t}}\left( {\sum\limits_{q =  - \infty }^\infty  {{t^{\left| q \right|}}}{z^q}  + \sum\limits_{q =  - \infty }^\infty  {z^q} } \right)\\
&=\frac{1}{1-t} {\sum _{q=-\infty }^{\infty }  t^{\left| q\right| }z^q}.
\end{aligned}
\end{equation}
The key steps in the derivation include (i) Taylor expansion of the summand associated with each long root, (ii) rearrangement of the limits of summation, such that the summands share the same simple root fugacities $z^q$ and the charges $q$ range from $-\infty$ to $\infty$, (iii) implementation of sums with the respect to the charges that are not carried by the simple roots and (iv) simplification of the resulting piecewise functions. When boiling down the latter it is useful to draw on identities that follow from the complex unimodular nature of the root space coordinates.

\begin{figure}[htbp]
\begin{center}
\includegraphics[scale=0.5]{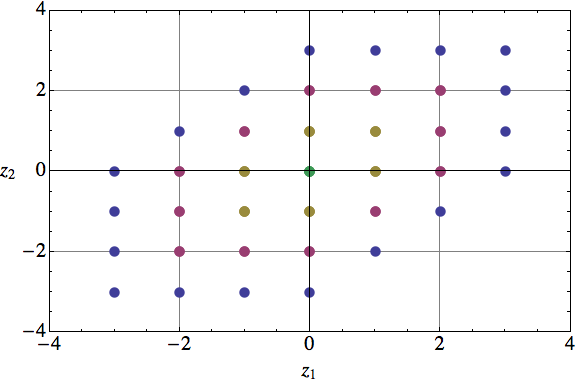}\\
\caption[Root Space of $A_2$ Foliated by Conformal Dimension.]{Root Space of $A_2$ Foliated by Conformal Dimension. The colour sequence corresponds to conformal dimensions of 0 for (0,0), 1 for the Weyl orbit of (1,1), 2 for the Weyl orbits of (2,2), (1,2) and (2,1), and 3 for the Weyl orbits of (3,3), (2,3) and (3,2). The adjoint representation is given by the orbit of (1,1) with conformal dimension 1 plus 2 orbits with conformal dimension 0.}
\label{figA2roots}
\end{center}
\end{figure}

While we should in principle be able to find such derivations for higher rank groups, the simplification of the piecewise functions becomes increasingly non-trivial. Thus, for $SU(3)$, we have:
\begin{equation}
\label{eq:mon3.15c}
\begin{aligned}
g_{instanton}^{{A_2}} &= \sum _{\text{Weyl}} \frac{1}{\left(1-\frac{1}{z_1}\right) \left(1-\frac{1}{z_1 z_2}\right) \left(1-z_2\right) \left(1-t z_1\right)}\\
&=\sum _{\text{Weyl}} \sum _{a=0}^{\infty } \sum _{b=0}^{\infty } \sum _{c=0}^{\infty } \sum _{d=0}^{\infty } t^a  z_1^{a-b-c} z_2^{d-c}\\
&=\sum _{\text{Weyl}} \sum _{q_1=-\infty }^{\infty } \sum _{q_2=-\infty }^{\infty } \sum _{c=\max \left(0,-q_2\right)}^{\infty } \sum _{b=\max \left(0,-c-q_1\right)}^{\infty } z_1^{q_1} z_2^{q_2} t^{b+c+q_1}\\
&=\frac{1} {(1-t)} {\sum _{\text{Weyl}} \sum _{q_1=-\infty }^{\infty } \sum _{q_2=-\infty }^{\infty } \sum _{c=\max \left(0,-q_2\right)}^{\infty } z_1^{q_1} z_2^{q_2} t^{\max \left(0,c+q_1\right)}}\\
&=\frac {1}{(1-t)} {\sum _{\text{Weyl}} \sum _{q_1=-\infty }^{\infty } \sum _{q_2=-\infty }^{\infty } \sum _{a=0}^{\infty } z_1^{q_1} z_2^{q_2} t^{\max \left(0,a+q_1,a+q_1-q_2\right)}}\\
&=\frac {1} {(1-t)^2}{\sum _{\text{Weyl}} \sum _{q_1=-\infty }^{\infty } \sum _{q_2=-\infty }^{\infty } z_1^{q_1} z_2^{q_2} \left(t^{\max \left(0,q_1,q_1-q_2\right)}-(1-t) \min \left(0,\max \left(q_1,q_1-q_2\right)\right)\right)}\\
&=\frac{1}{(1-t)^2} {\sum _{\text{Weyl}} \sum _{q_1=-\infty }^{\infty } \sum _{q_2=-\infty }^{\infty } z_1^{q_1} z_2^{q_2} t^{\max \left(0,q_1,q_1-q_2\right)}},
\end{aligned}
\end{equation}
where we have used an identity, which is valid for the root coordinates:
\begin{equation}
\label{eq:mon3.15d}
\begin{aligned}
\sum _{q_1=-\infty }^{\infty } \sum _{q_2=-\infty }^{\infty } z_1^{q_1} z_2^{q_2} \min \left(0,\max \left(q_1,q_1-q_2\right)\right)=0.
\end{aligned}
\end{equation}
We continue by carrying out the Weyl reflections to obtain:
\begin{equation}
\label{eq:mon3.15e}
\begin{aligned}
g_{instanton}^{A_2} &= \frac{1}{(1-t)^2}  {\sum _{q_1=-\infty }^{\infty } \sum _{q_2=-\infty }^{\infty } z_1^{q_1} z_2^{q_2} } (t^{\max \left(0,-q_1,q_2-q_1\right)}+t^{\max \left(0,q_1,q_1-q_2\right)}+t^{\max \left(0,-q_2,-q_1\right)} \\
& ~~~~~~~~~~~~~~~~~~~~~~~~ +t^{\max \left(0,-q_2,q_1-q_2\right)}+t^{\max \left(0,q_2,q_2-q_1\right)}+t^{\max \left(0,q_2,q_1\right)})\\
&=\frac{1} {(1-t)^2} {\sum _{q_1=-\infty }^{\infty } \sum _{q_2=-\infty }^{\infty } z_1^{q_1} z_2^{q_2} } 
\left( {2 + {t^{\left| {q_1} \right|}} + {t^{\left| {q_2} \right|}} + {t^{\left| {q_1 - q_2} \right|}} + {t^{\frac{1}{2}\left| {q_1 - q_2} \right| + \frac{1}{2}\left| {q_1} \right| + \frac{1}{2}\left| {q_2} \right|}}} \right)\\
&=\frac{1} {(1-t)^2} {\sum _{q_1=-\infty }^{\infty } \sum _{q_2=-\infty }^{\infty } z_1^{q_1} z_2^{q_2} } 
\left( { {t^{\frac{1}{2}\left| {q_1 - q_2} \right| + \frac{1}{2}\left| {q_1} \right| + \frac{1}{2}\left| {q_2} \right|}}} \right),
\end{aligned}
\end{equation}
where we have rearranged the parts of the six piecewise functions and then used unimodular coordinate identities to eliminate five of the resulting functions:
\begin{equation}
\label{eq:mon3.15f}
\begin{aligned}
\sum\limits_{{q_1} =  - \infty }^\infty  {z_1^{{q_1}}}  =0= \sum\limits_{{q_1} =  - \infty }^\infty  {\sum\limits_{{q_2} =  - \infty }^\infty  {z_1^{{q_1}}} } z_2^{{q_2}}{t^{\left| {{q_1} - {q_2}} \right|}}.
\end{aligned}
\end{equation}
A key feature of the monopole construction is the manner in which conformal dimension foliates the root space into sets of Weyl group orbits that correspond to the adjoint and its symmetrisations. This is shown in Figure \ref{figA2roots} for the first few orbits of $A_2$, where we label states in terms of their root space coordinates, rather than their weight space coordinates (Dynkin labels).
\FloatBarrier
\subsubsection{D Series}
The monopole construction for D series RSIMS of rank 4 and above is based on the extended Cartan matrix, defined in accordance with the schema \ref{eq:mon3.2}, and the dual Coxter labels of the simple roots (shown as a column vector), where we have labelled the affine simple root by $z_0$:
\begin{equation}
\label{eq:mon3.16}
\begin{tabular}{c|c c c c c c c|c|}
${{z_1}}$&$ 2$&${ - 1}$&$ \ldots $&$0$&$0$&$0$&$0$&$ 1$ \\
${{z_2}}$&$ { - 1}$&$2$&$ \ldots $&$0$&$0$&$0$&${ - 1}$&$ 2$ \\
 $\ldots $&$  \ldots $&$ \ldots $&$ \ldots $&$ \ldots $&$ \ldots $&$ \ldots $&$ \ldots $&$ 2$ \\
${{z_{r - 2}}}$&$ 0$&$0$&$ \ldots $&$2$&${ - 1}$&${ - 1}$&$0$&$ 2$ \\
${{z_{r - 1}}}$&$ 0$&$0$&$ \ldots $&${ - 1}$&$2$&$0$&$0$&$ 1$ \\
${{z_r}}$&$ 0$&$0$&$ \ldots $&${ - 1}$&$0$&$2$&$0$&$ 1$ \\
${{z_0}}$&$ 0$&${ - 1}$&$ \ldots $&$0$&$0$&$0$&$2$&$ 1$\\
\end{tabular}.
\end{equation}
Applying the prescription set out in \ref{eq:mon3.6}, \ref{eq:mon3.7} and \ref{eq:mon3.8}, we obtain the equation for a D series RSIMS of rank 4 or greater:
\begin{equation}
\label{eq:mon3.17}
\begin{aligned}
g_{instanton}^{{D_r}}  &=\sum\limits_{{q_1},{q_{r - 1}},{q_r} =  - \infty }^\infty  {\sum\limits_{{q_{j,1}} \ge {q_{j,2}} \ge  - \infty \atop
r - 2 \ge j \ge 2}^\infty  {{z_1}^{{q_1}}{z_2}^{{q_{2,1}} + {q_{2,2}}} \ldots {z_{r - 2}}^{{q_{r - 2,1}} + {q_{r - 2,2}}}{z_{r - 1}}^{{q_{r - 1}}}{z_r}^{{q_r}}}}\\
 &~~~~~~~~~~~~~~~~~~~~~~~~~~~~~~~~~~~~\times P_{U(N)}^{{D_r}}\left( {q,t} \right){t^{\Delta (q)}},
\end{aligned}
\end{equation}
where
\begin{equation}
\label{eq:mon3.18}
\begin{aligned}
P_{U(N)}^{{D_r}}\left( {q,t} \right) &= \frac{1}{{{{\left( {1 - t} \right)}^r}}}\prod\limits_{j = 2}^{r - 2} {{
\left\{ \begin{array}{l}
{q_{j,1}} = {q_{j,2}}:1/(1 - {t^2})\\
{q_{j,1}} \ne {q_{j,2}}:1/(1 - t)
\end{array} \right.}
}
\end{aligned}
\end{equation}
and
\begin{equation}
\label{eq:mon3.19}
\begin{aligned}
\Delta (q) &= \frac{1}{2}\left( {\sum\limits_{i = 1}^2 {\left| {{q_{2,i}}} \right|}  + \sum\limits_{i = 1}^2 {\left| {{q_1} - {q_{2,i}}} \right|}  + \sum\limits_{k = 2}^{r - 3} {\sum\limits_{i,j = 1}^2 {\left| {{q_{k,i}} - {q_{k + 1,j}}} \right|} }  + \sum\limits_{i = 1}^2 {\left| {{q_{r - 2,i}} - {q_{r - 1}}} \right|}  + \sum\limits_{i = 1}^2 {\left| {{q_{r - 2,i}} - {q_r}} \right|} } \right)\\
& ~~~~~~~~~~~~~~~~~~~~~ - \sum\limits_{k = 2}^{r - 2} {\left| {{q_{k,1}} - {q_{k,2}}} \right|}
\end{aligned}
\end{equation}
The construction can, in principle, be rearranged into the character generating functions shown in Table \ref{table5}, similarly to the cases of the A series constructions shown above.

As in the case of the A series, the conformal dimension measure has the effect of foliating the root system into orbits of dominant weights associated with successive multiples of the adjoint representation.

Also, the gauge choice $q_0 =0$ has alternatives and, indeed, any one of the monopole charges can be defined to zero, providing the summand is modified to include both $z_0$ and ${q_0}$ and care is taken over the $P_{U(N)}$ symmetry factors.  For star shaped quivers, such as $D_4$, a particularly convenient choice of gauge is $q_{2,2}=0$, and this leads directly to a decomposition into a symmetric sum over all the representations of four T(SU(2)) quiver theories, as discussed in Section \ref{sec:mHL}.
\subsubsection{E Series}
The monopole construction for $E_6$ instantons is based on an extended Cartan matrix and dual Coxter labels of the form:
\begin{equation}
\label{eq:mon3.20}
\begin{tabular}{c|c c c c c c c|c|}
${{z_1}}$&$ 2$&${ - 1}$&$0$&$0$&$0$&$0$&$0$&$ 1$ \\
${{z_2}}$&$ { - 1}$&$2$&${ - 1}$&$0$&$0$&$0$&${ - 1}$&$ 2$ \\
${{z_3}}$&$ 0$&${ - 1}$&$2$&${ - 1}$&$0$&${ - 1}$&$0$&$ 3$ \\
${{z_4}}$&$ 0$&$0$&${ - 1}$&$2$&${ - 1}$&$0$&$0$&$ 2$ \\
${{z_5}}$&$ 0$&$0$&$0$&${ - 1}$&$2$&$0$&$0$&$ 1$ \\
${{z_6}}$&$ 0$&$0$&${ - 1}$&$0$&$0$&$2$&${ - 1}$&$ 2$ \\
${{z_0}}$&$ 0$&$0$&$0$&$0$&$0$&${ - 1}$&$2$&$ 1$ \\
\end{tabular}.
\end{equation}
Applying the prescription set out in \ref{eq:mon3.6}, \ref{eq:mon3.7} and \ref{eq:mon3.8}, we obtain the monopole equation for an $E_6$ instanton:
\begin{equation}
\label{eq:mon3.21}
\begin{aligned}
g_{instanton}^{{E_6}} &=\sum\limits_{{q_1},{q_5} =  - \infty }^\infty  {\sum\limits_{{q_{j,1}} \ge {q_{j,2}} \ge  - \infty \atop j = 2,4,6}^\infty  {\sum\limits_{{q_{3,1}} \ge {q_{3,2}} \ge {q_{3,3}} \ge  - \infty }^\infty  {{z_1}^{{q_1}}{z_2}^{{q_{2,1}} + {q_{2,2}}}{z_3}^{{q_{3,1}} + {q_{3,2}} + {q_{3,3}}}{z_4}^{{q_{4,1}} + {q_{4,2}}}{z_5}^{{q_5}}{z_6}^{{q_{6,1}} + {q_{6,2}}}} } }\\
 &~~~~~~~~~~~~~~~~~~~~~~~~~~~~~~~~~~~~~~~~~~~~~~~~~~\times P_{U(N)}^{{E_6}}\left( {q,t} \right){t^{\Delta (q)}},
\end{aligned}
\end{equation}
where
\begin{equation}
\label{eq:mon3.22}
\begin{aligned}
P_{U(N)}^{{E_6}}\left( {q,t} \right) &= \frac{1}{{{{\left( {1 - t} \right)}^6}{{\left( {1 - {t^2}} \right)}^4}\left( {1 - {t^3}} \right)}} \\
&\times If\left[ {{q_{3,1}} \ne {q_{3,2}} \vee {q_{3,1}} \ne {q_{3,3}} \vee {q_{3,2}} \ne {q_{3,3}},\left( {1 + t + {t^2}} \right)} \right] \\
& \times If\left[ {{q_{3,1}} \ne {q_{3,2}} \wedge {q_{3,1}} \ne {q_{3,3}} \wedge {q_{3,2}} \ne {q_{3,3}}),\left( {1 + t} \right)} \right] \\
&\times \prod\limits_{j = 2,4,6}^{} {If\left[ {{q_{j,1}} \ne {q_{j,2}},\left( {1 + t} \right)} \right]} \\
\end{aligned}
\end{equation}
and
\begin{equation}
\label{eq:mon3.23}
\begin{aligned}
\Delta (q) &= \frac{1}{2}\left( {\sum\limits_{i = 1}^2 {\left| {{q_1} - {q_{2,i}}} \right|}  + \sum\limits_{k = 2,4,6}^{} {\sum\limits_{i,j = 1}^{i = 2,j = 3} {\left| {{q_{3,j}} - {q_{k,i}}} \right|} }  + \sum\limits_{i = 1}^2 {\left| {{q_{4,i}} - {q_5}} \right|}  + \sum\limits_{i = 1}^2 {\left| {{q_{6,i}}} \right|} } \right)\\
& - \sum\limits_{k = 2,4,6}^{} {\sum\limits_{ i > j \ge 1}^{2} {\left| {{q_{k,i}} - {q_{k,j}}} \right| - \sum\limits_{ i > j \ge 1}^{3} {\left| {{q_{3,i}} - {q_{3,j}}} \right|} }}.
\end{aligned}
\end{equation}
We do not give the explicit instanton constructions for $E_7$ and $E_8$ groups; however, these follow a similar pattern to the $E_6$ construction. The constructions for the ADE series given above are equivalent to those in \cite{Cremonesi:2013lqa}. Again, the gauge choice $q_0=0$ has alternatives.  For star shaped quivers, such as $E_6$, a particularly convenient choice of gauge is $q_{3,3}=0$, and this leads directly to a decomposition into a symmetric sum over all the representations of three T(SU(3)) quiver theories, as discussed in Section \ref{sec:mHL}.
\FloatBarrier
\subsection{Construction for Non-Simply Laced Groups}
\subsubsection{B Series}
The monopole construction for B series instantons is based on the extended Cartan matrix, defined in accordance with the schema \ref{eq:mon3.2}, and its dual Coxeter labels, where we have labelled the affine simple root $z_0$:
\begin{equation}
\label{eq:mon3.24}
\begin{tabular}{c|c c c c c c|c|}
${{z_1}}$&$ 2$&${ - 1}$&$ \ldots $&$0$&$0$&$0$&$ 1$ \\
${{z_2}}$&$ { - 1}$&$2$&$ \ldots $&$0$&$0$&${ - 1}$&$ 2$ \\
 $\ldots $&$  \ldots $&$ \ldots $&$ \ldots $&$ \ldots $&$ \ldots $&$ \ldots $&$ 2$ \\
${{z_{r - 1}}}$&$ 0$&$0$&$ \ldots $&$2$&${ - 2}$&$0$&$ 2$ \\
${{z_r}}$&$ 0$&$0$&$ \ldots $&${ - 1}$&$2$&$0$&$ 1$ \\
${{z_0}}$&$ 0$&${ - 1}$&$ \ldots $&$0$&$0$&$2$&$ 1$ 
\end{tabular}.
\end{equation}
Applying the prescription set out in \ref{eq:mon3.6}, \ref{eq:mon3.7} and \ref{eq:mon3.8}, we obtain the monopole equation for a B series instanton of rank 2 and above:
\begin{equation}
\label{eq:mon3.25}
\begin{aligned}
g_{instanton}^{{B_r}} =\sum\limits_{{q_1},{q_r} =  - \infty }^\infty  {\sum\limits_{{q_{j,1}} \ge {q_{j,2}} \ge  - \infty \atop r - 1 \ge j \ge 2}^\infty  {{z_1}^{{q_1}}{z_2}^{{q_{2,1}} + {q_{2,2}}} \ldots {z_{r - 1}}^{{q_{r - 1,1}} + {q_{r - 1,2}}}{z_r}^{{q_r}}} } P_{U(N)}^{{B_r}}\left( {q,t} \right){t^{\Delta(q)}},
\end{aligned}
\end{equation}
where
\begin{equation}
\label{eq:mon3.26}
\begin{aligned}
P_{U(N)}^{{B_r}}\left( {q,t} \right)&=\frac{{\prod\limits_{j = 2}^{r - 1} {If\left[ {{q_{j,1}} \ne {q_{j,2}},(1 + t)} \right]} }}{{{{\left( {1 - t} \right)}^r}{{\left( {1 - {t^2}} \right)}^{r - 2}}}}
\end{aligned}
\end{equation}
and
\begin{equation}
\label{eq:mon3.27}
\begin{aligned}
\Delta(q)&=\frac{1}{2}\left( {\sum\limits_{i = 1}^2 {\left(  \left| {{q_1} - {q_{2,i}}} \right| +{\left| {{q_{2,i}}} \right| + \left| {2{q_{r - 1,i}} - {q_r}} \right|} \right)}  + \sum\limits_{k = 2}^{r - 2} {\sum\limits_{i,j = 1}^2 {\left| {{q_{k,i}} - {q_{k + 1,j}}} \right|} } } \right) \\
&- \sum\limits_{k = 2}^{r - 1} {\sum\limits_{i > j}^{} {\left| {{q_{k,i}} - {q_{k,j}}} \right|} }.
\end{aligned}
\end{equation}
We can extract the monopole construction for $B_2$ from \ref{eq:mon3.25}, \ref{eq:mon3.26} and \ref{eq:mon3.27} and rearrange it into the character generating function for $B_2$ in Table \ref{table5}:
\begin{equation}
\label{eq:mon3.28}
\begin{aligned}
g_{instanton}^{{B_2}} &= \frac{1}{{{{\left( {1 - t} \right)}^2}}}\sum\limits_{{q_1},{q_2} =  - \infty }^\infty  {{{{z}}_1}^{{{{q}}_1}}} {{{z}}_2}^{{{{q}}_2}}{t^{\frac{1}{2}\left( {\left| {2{{{q}}_1} - {{{q}}_2}} \right| + \left| {{{{q}}_2}} \right|} \right)}}\\
& = \frac{1}{{{{\left( {1 - t} \right)}^2}}}\left( \begin{array}{l}
\sum\limits_{q_1 = 0}^\infty  {\sum\limits_{q_2 = 0}^{2{q_1}} {\left( {{{{z}}_1}^{{{{q}}_1}}{{{z}}_2}^{{{{q}}_2}} + {{{z}}_1}^{ - {{{q}}_1}}{{{z}}_2}^{ - {{{q}}_2}}} \right){t^{{{{q}}_1}}}} } \\
 + \sum\limits_{q_1 = 0}^\infty  {\sum\limits_{q_2 = 2{q_1}}^\infty  {\left( {{{{z}}_1}^{{{{q}}_1}}{{{z}}_2}^{{{{q}}_2}} + {{{z}}_1}^{ - {{{q}}_1}}{{{z}}_2}^{ - {{{q}}_2}}} \right){t^{\left( {{{{q}}_2} - {{{q}}_1}} \right)}}} } \\
 - \sum\limits_{q_1 = 0}^\infty  {\left( {{{{z}}_1}^{{{{q}}_1}}{{{z}}_2}^{2{{{q}}_1}} + {{{z}}_1}^{ - {{{q}}_1}}{{{z}}_2}^{ - 2{{{q}}_1}}} \right){t^{{{{q}}_1}}}} \\
 + \sum\limits_{q_1 = 0}^\infty  {\sum\limits_{q_2 = 0}^\infty  {\left( {{{{z}}_1}^{{{{q}}_1}}{{{z}}_2}^{ - {{{q}}_2}} + {{{z}}_1}^{ - {{{q}}_1}}{{{z}}_2}^{{{{q}}_2}}} \right){t^{\left( {{{{q}}_1} + {{{q}}_2}} \right)}}} } \\
 - \sum\limits_{q_1 = 0}^\infty  {\left( {{{{z}}_1}^{{{{q}}_1}} + {{{z}}_1}^{ - {{{q}}_1}}} \right){t^{{{{q}}_1}}}} \\
 - \sum\limits_{q_2 = 0}^\infty  {\left( {{{{z}}_2}^{{{{q}}_2}} + {{{z}}_2}^{ - {{{q}}_2}}} \right){t^{{{{q}}_2}}}} \\
 + 1
\end{array} \right)\\
\ldots\\
& =\left( {\begin{array}{*{20}{c}}
{1 - {t^2} - {t^6} + {t^8}}&{\left[ {0,0} \right]}\\
{{t^3} - 2{t^4} + {t^5}}&{\left[ {0,2} \right]}\\
{ - {t^2} + {t^3} + {t^5} - {t^6}}&{\left[ {1,0} \right]}\\
{{t^3} - 2{t^4} + {t^5}}&{\left[ {1,2} \right]}\\
{ - {t^2} + {t^3} + {t^5} - {t^6}}&{\left[ {2,0} \right]}
\end{array}} \right)PE\left[ {\left[ {0,2} \right] t} \right].
\end{aligned}
\end{equation}
Similarly to the A series, we can also derive the monopole expression for $B_2$ RSIMS from the plethystic formula \ref{eq:pleth2.12}. The long roots are given by $\{ z_1,z_1 {z_2}^2, {z_1}^{-1}, {z_1}^{-1} {z_2}^{-2} \}$ and, selecting those Weyl reflections that transform between $z_1$ and the other long roots, we obtain:
\begin{equation}
\label{eq:mon3.29a}
\begin{aligned}
g_{instanton}^{{B_2}}{\rm{ }} &= \sum\limits_{Weyl: \Phi  \in long} {\frac{1}{{\left( {1 - t{z_1}} \right)\left( {1 - 1/{z_1}} \right)\left( {1 - {{{z}}_{{2}}}} \right)\left( {1 - 1/{z_1}{z_2}} \right)}}} \\
&= \sum\limits_{Weyl:\Phi  \in long} \sum _{a=0}^{\infty } \sum _{b=0}^{\infty } \sum _{c=0}^{\infty } \sum _{d=0}^{\infty } t^a  z_1^{a-b-d} z_2^{c-d}\\
&= \sum\limits_{Weyl: \Phi  \in long} \sum _{{q_1}=-\infty }^{\infty } \sum _{{q_2}=-\infty }^{\infty } \sum _{d=\max (0,-{q_2})}^{\infty } \sum _{b=\max (0,-d-{q_1})}^{\infty } {z_1}^{{q_1}} {z_2}^{{q_2}} t^{b+d+{q_1}}\\
&=\frac{1} {(1-t)} \sum\limits_{Weyl: \Phi  \in long}     \sum _{{q_1}=-\infty }^{\infty } \sum _{{q_2}=-\infty }^{\infty } \sum _{d=0}^{\infty } {z_1}^{{q_1}} {z_2}^{{q_2}} t^{\max (0,d+{q_1},d+{q_1}-{q_2})}\\
&=\frac {1} {(1-t)^2}  \sum\limits_{Weyl: \Phi  \in long}    \sum _{q_1=-\infty }^{\infty } \sum _{q_2=-\infty }^{\infty } z_1^{q_1} z_2^{q_2} (t^{\max (0,{q_1},{q_1}-{q_2})}-(1-t) \min (0,\max ({q_1},{q_1}-{q_2})))\\
&=\frac {1} {(1-t)^2}  \sum\limits_{Weyl: \Phi  \in long}    \sum _{q_1=-\infty }^{\infty } \sum _{q_2=-\infty }^{\infty } z_1^{q_1} z_2^{q_2} {t^{\max (0,{q_1},{q_1}-{q_2})}},
\end{aligned}
\end{equation}
where we have used an identity, that is valid for unimodular coordinates:
\begin{equation}
\label{eq:mon3.29b}
\begin{aligned}
\sum _{q_1=-\infty }^{\infty } \sum _{q_2=-\infty }^{\infty } z_1^{q_1} z_2^{q_2} \min (0,\max ({q_1},{q_1}-{q_2}))=0.
\end{aligned}
\end{equation}
We continue by carrying out the relevant Weyl reflections and rearranging the piecewise functions to obtain the RSIMS:
\begin{equation}
\label{eq:mon3.29c}
\begin{aligned}
g_{instanton}^{B_2} &= \frac{1}{(1-t)^2}  {\sum _{q_1=-\infty }^{\infty } \sum _{q_2=-\infty }^{\infty } z_1^{q_1} z_2^{q_2} } (t^{\max (0,-{q_1},{q_2}-{q_1})}+t^{\max (0,-{q_1},{q_1}-{q_2})}\\
&~~~~~~~~~~~~~~~~~~~~~~~~~~~~~~~~~~+t^{\max (0,{q_1},{q_2}-{q_1})}+t^{\max (0,{q_1},{q_1}-{q_2})})\\
&=\frac{1} {(1-t)^2} {\sum _{q_1=-\infty }^{\infty } \sum _{q_2=-\infty }^{\infty } z_1^{q_1} z_2^{q_2} } 
(t^{\frac{1}{2}{\left| 2 {q_1}-{q_2}\right| }+\frac{1}{2} {\left| {q_2}\right| }}+t^{\left| {q_1}-{q_2}\right| }+t^{\left| {q_1}\right| }+1)\\
&=\frac{1} {(1-t)^2} {\sum _{q_1=-\infty }^{\infty } \sum _{q_2=-\infty }^{\infty } z_1^{q_1} z_2^{q_2} } {t^{\frac{1}{2}{\left| 2 {q_1}-{q_2}\right| }+\frac{1}{2} {\left| {q_2}\right| }}},
\end{aligned}
\end{equation}
where we have eliminated piecewise terms using root identities, as before.
\subsubsection{C Series}
The monopole construction for C series instantons is based on the extended Cartan matrix, defined in accordance with the schema \ref{eq:mon3.2}, and its dual Coxeter labels, where we have labelled the affine simple root $z_0$:
\begin{equation}
\label{eq:mon3.30}
\begin{tabular}{c|c c c c c c|c|}
${{z_1}}$&$ 2$&${ - 1}$&$ \ldots $&$0$&$0$&${ - 1}$&$ 1$ \\
${{z_2}}$&$ { - 1}$&$2$&$ \ldots $&$0$&$0$&$0$&$ 1$ \\
 $\ldots $&$  \ldots $&$ \ldots $&$ \ldots $&$ \ldots $&$ \ldots $&$ \ldots $&$ 1$ \\
${{z_{r - 1}}}$&$ 0$&$0$&$ \ldots $&$2$&${ - 1}$&$0$&$ 1$ \\
${{z_r}}$&$ 0$&$0$&$ \ldots $&${ - 2}$&$2$&$0$&$ 1$ \\
${{z_0}}$&$ { - 2}$&$0$&$ \ldots $&$0$&$0$&$2$&$ 1$ \\
\end{tabular}.
\end{equation}
Applying the prescription set out in \ref{eq:mon3.6}, \ref{eq:mon3.7} and \ref{eq:mon3.8}, we obtain the monopole equation for a C series instanton:
\begin{equation}
\label{eq:mon3.31}
\begin{aligned}
g_{instanton}^{{C_r}}={\frac{1}{{{{\left( {1 - t} \right)}^r}}}\sum\limits_{{{{q}}_i} =  - \infty }^\infty  {{{{z}}_1}^{{{{q}}_1}}} {{{z}}_2}^{{{{q}}_2}} \ldots {{{z}}_r}^{{{{q}}_r}}{t^{\Delta(q)} }},
\end{aligned}
\end{equation}
where
\begin{equation}
\label{eq:mon3.32}
\begin{aligned}
\Delta(q)&={\frac{1}{2}\left( {\left| {{{{q}}_1}} \right| + \sum\limits_{i = 1}^{r- 2} {\left| {{{{q}}_i} - {{{q}}_{i + 1}}} \right|}  + \left| {{{{q}}_{r - 1}} - 2{{{q}}_r}} \right|} \right)}.
\end{aligned}
\end{equation}
It follows from \ref{eq:mon3.28}, \ref{eq:mon3.31} and \ref{eq:mon3.32} that the constructions for $B_2$ and $C_2$ are isomorphic under interchange of root labels, as required by consistency.
\subsubsection{$F_4$ and $G_2$}
\begin{figure}[htbp]
\begin{center}
\includegraphics[scale=0.5]{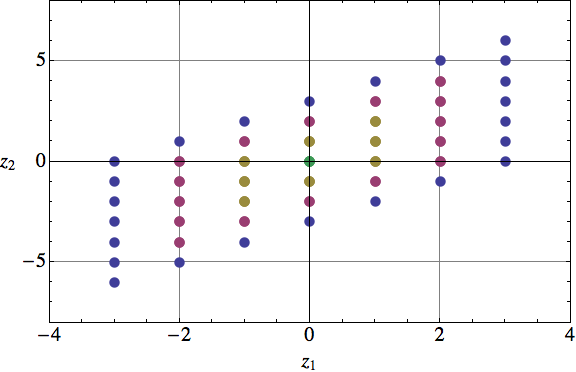}\\
\caption[Root Space of $B_2$ Foliated by Conformal Dimension.] {Root Space of $B_2$ Foliated by Conformal Dimension. The colour sequence corresponds to conformal dimensions of 0 for (0,0), 1 for the Weyl orbits of (1,2) and (1,1), 2 for the Weyl orbits of (2,4), (2,3) and (2,2), and 3 for the Weyl orbits of (3,6), (3,5), (3,4) and (3,3). The long root of the adjoint representation is (1,2).}
\label{figB2roots}
\end{center}
\end{figure}

\begin{figure}[htbp]
\begin{center}
\includegraphics[scale=0.5]{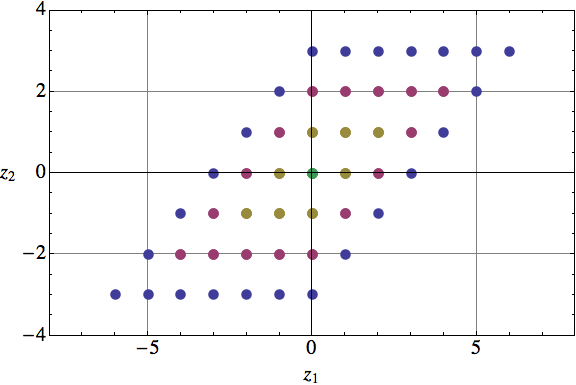}\\
\caption[Root Space of $C_2$ Foliated by Conformal Dimension.]{Root Space of $C_2$ Foliated by Conformal Dimension.The colour sequence corresponds to conformal dimensions of 0 for (0,0), 1 for the Weyl orbits of (2,1) and (1,1), 2 for the Weyl orbits of (4,2), (3,2) and (2,2), and 3 for the Weyl orbits of (6,3), (5,3), (4,3) and (3,3). The long root of the adjoint representation is (2,1).}
\label{figC2roots}
\end{center}
\end{figure}

\begin{figure}[htbp]
\begin{center}
\includegraphics[scale=0.5]{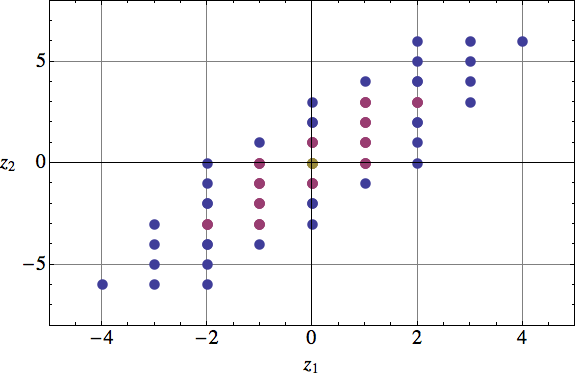}\\
\caption[Root Space of $G_2$ Foliated by Conformal Dimension.]{Root Space of $G_2$ Foliated by Conformal Dimension. The colour sequence corresponds to conformal dimensions of 0 for (0,0), 1 for the Weyl orbits of (2,3) and (1,2), and 2 for the Weyl orbits of (4,6), (3,6), (3,5) and (2,4). The long root of the adjoint representation is (2,3).}
\label{figG2roots}
\end{center}
\end{figure}

The monopole construction for the $F_4$ instanton is based on the extended Cartan matrix, defined in accordance with the schema \ref{eq:mon3.2}, and its dual Coxeter labels, where we have labelled the affine simple root $z_0$:
\begin{equation}
\label{eq:mon3.33}
\begin{tabular}{c|c c c c c|c|}
${{z_1}}$&$ 2$&${ - 1}$&$0$&$0$&${ - 1}$&$ 2$ \\
${{z_2}}$&$ { - 1}$&$2$&${ - 2}$&$0$&$0$&$ 3$ \\
${{z_3}}$&$ 0$&${ - 1}$&$2$&${ - 1}$&$0$&$ 2$ \\
${{z_4}}$&$ 0$&$0$&${ - 1}$&$2$&$0$&$ 1$ \\
${{z_0}}$&$ { - 1}$&$0$&$0$&$0$&$2$&$ 1$ \\
\end{tabular}.
\end{equation}
Applying the prescription set out in \ref{eq:mon3.6}, \ref{eq:mon3.7} and \ref{eq:mon3.8}, we obtain the monopole equation for a $F_4$ instanton:
\begin{equation}
\label{eq:mon3.34}
\begin{aligned}
g_{instanton}^{{F_4}}  =\sum\limits_{{q_{j1}} \ge {q_{j,2}} \ge  - \infty \atop j = 1,3}^\infty  {\sum\limits_{{q_{2,1}} \ge {q_{2,2}} \ge {q_{2,3}} \ge  - \infty }^\infty  {\sum\limits_{{q_4} =  - \infty }^\infty  {{{{z}}_1}^{{q_{1,1}} + {q_{1,2}}}{{{z}}_2}^{{q_{2,1}} + {q_{2,2}} + {q_{2,3}}}{{{z}}_3}^{{q_{3,1}} + {q_{3,2}}}{{{z}}_4}^{{{{q}}_4}}P_{U\left( N \right)}^{{F_4}}{t^{\Delta \left( q \right)}}} } },
\end{aligned}
\end{equation}
where
\begin{equation}
\label{eq:mon3.35}
\begin{aligned}
P_{U(N)}^{{F_4}}\left( {q,t} \right)&=
\frac{{\prod\limits_{j = 1,3} {If\left[ {{q_{j,1}} \ne {q_{j,2}},1 + t} \right]} }}{{{{\left( {1 - t} \right)}^4}{{\left( {1 - {t^2}} \right)}^3}\left( {1 - {t^3}} \right)}}\\
&~\\
&~~ \times If\left[ {\exists i,j:{q_{2i}} \ne {q_{2j}},(1 + t + {t^2})} \right]\\
&~\\
&~~ \times If\left[ {!\exists i,j:{q_{2,i}} = {q_{2,j}},(1 + t)} \right]\\
\end{aligned}
\end{equation}
and
\begin{equation}
\label{eq:mon3.36}
\begin{aligned}
\Delta(q)&=
\frac{1}{2}\left( {\sum\limits_{i = 1}^2 {\left| {{q_{1,i}}} \right|}  + \sum\limits_{i = 1}^2 {\sum\limits_{j = 1}^3 {\left( {\left| {{q_{1,i}} - {q_{2,j}}} \right| + \left| {2{q_{2,j}} - {q_{3,i}}} \right|} \right) + \sum\limits_{i = 1}^2 {\left| {{q_{3,i}} - {{{q}}_4}} \right|} } } } \right)\\
&~~ - \sum\limits_{k = 1,3}^{} {\left| {{q_{k,1}} - {q_{k,2}}} \right| - \sum\limits_{i > j}^{} {\left| {{q_{2,i}} - {q_{2,j}}} \right|} }.
\end{aligned}
\end{equation}
The monopole construction for the $G_2$ instanton is based on the extended Cartan matrix, defined in accordance with the schema \ref{eq:mon3.2}, and its dual Coxeter labels, where we have labelled the affine simple root $z_0$:
\begin{equation}
\label{eq:mon3.37}
\begin{tabular}{c|c c c|c|}
${{z_1}}$&$ 2$&${ - 3}$&${ - 1}$&$ 2$ \\
${{z_2}}$&$ { - 1}$&$2$&$0$&$ 1$ \\
${{z_0}}$&$ { - 1}$&$0$&$2$&$ 1$ \\
\end{tabular}.
\end{equation}
Applying the prescription set out in \ref{eq:mon3.6}, \ref{eq:mon3.7} and \ref{eq:mon3.8}, we obtain the monopole equation for a $G_2$ instanton:
\begin{equation}
\label{eq:mon3.38}
\begin{aligned}
g_{instanton}^{{G_2}} =\sum\limits_{{q_{1,1}} \ge {q_{1,2}} \ge  - \infty }^\infty  {\sum\limits_{{{{q}}_2} =  - \infty }^\infty  {{{{z}}_1}^{{q_{1,1}} + {q_{1,2}}}{{{z}}_2}^{{{{q}}_2}}P_{U\left( N \right)}^{{G_2}}{t^{\Delta \left( q \right)}}} },
\end{aligned}
\end{equation}
where
\begin{equation}
\label{eq:mon3.39}
\begin{aligned}
P_{U(N)}^{{G_2}}\left( {q,t} \right)&=
\frac{{If\left[ {{q_{1,1}} \ne {q_{1,2}},\left( {1 + t} \right)} \right]}}{{{{\left( {1 - t} \right)}^2}\left( {1 - {t^2}} \right)}}
\end{aligned}
\end{equation}
and
\begin{equation}
\label{eq:mon3.40}
\begin{aligned}
\Delta(q)&=
\frac{1}{2}\sum\limits_{i = 1}^2 {\left( {\left| {{q_{1,i}}} \right| + \left| {3{q_{1,i}} - {{\rm{q}}_2}} \right|} \right)}  - \left| {{q_{1,1}} - {q_{1,2}}} \right|.
\end{aligned}
\end{equation}
We can use the root structures of $B_2$, $C_2$ and $G_2$ to illustrate how the monopole construction combines the Weyl group orbits of dominant weights into irreps that are symmetrisations of the adjoint representation. Recall we are labelling states in terms of their root space coordinates in Figures \ref{figB2roots}, \ref{figC2roots} and \ref{figG2roots}, rather than their weight space coordinates (Dynkin labels).

In all cases the RSIMS can be expressed as sums of orbits of dominant weights in the root lattice (weights in the interior of the positive root space). The conformal dimension remains constant around each orbit. More than one dominant weight can have the same conformal dimension. The orbits are combined, at multiplicities determined by the $P^G_{U(N)}$ factors, to give the adjoint representation and its symmetrisations. For all rank 2 groups, the adjoint is given by the orbits with conformal dimension 1 plus two orbits with conformal dimension 0. The isomorphism between $B_2$ and $C_2$ is evident upon interchange of simple roots (and Dynkin labels).
\subsection{Coulomb Branch Quiver Theories}
We have analysed these monopole constructions largely from a group theoretic perspective, however, in the case of the ADE series RSIMS, they correspond to the Coulomb branches of particular SUSY quiver gauge theories, being ${\cal N}=4$ superconformal gauge theories in 2+1 dimensions \cite{Cremonesi:2013lqa}. The Coulomb branches of these theories are HyperK\text{\" a}hler manifolds. The related brane configurations involve D2 branes against a background of D6 branes \cite{Cremonesi:2014xha}. As shown in \cite{Cremonesi:2014xha}, the BC series RSIMS correspond to quiver gauge theories for brane configurations which include orientifold planes. (The orientifold planes are required to ensure that the constructions can reproduce the root systems of the Lie algebras.)

In these theories, the quiver gauge theory is specified by the extended Dynkin diagram, with the dual Coxeter numbers ${{\mathord{\buildrel{\lower3pt\hbox{$\scriptscriptstyle\smile$}} \over a} }_i}$ associated to each node, determining the $U(N_i)$ gauge fields carried by the nodes. The zero central charge of the affine Lie algebra corresponds to an overall gauge invariance condition on the field combinations on the Coulomb branch. Since the affine root and its Dynkin label are redundant, by virtue of the degeneracy of the affine Cartan matrix, they can be gauged away, that is to say, we can describe the field combinations on the Coulomb branch purely by reference to the non-affine roots, combined into root monomials at some integer powers.

The delicate aspect of the monopole construction lies in the collection of root monomials into characters of representations of the Lie group that are precisely enumerated by the fugacity $t$, the exponents of which give the spin of the SU(2)-R global symmetry. This collection process depends crucially on the R-charges assigned to the BPS bare monopole operators carrying the GNO charges $q$. In schematic terms, these R-charges or {\it conformal dimensions} are given by application of the following general formula \cite{Cremonesi:2013lqa} to the quiver diagram:
\begin{equation}
\label{eq:mon3.41}
\Delta \left( q \right) = \underbrace {\frac{1}{2}\sum\limits_{i = 1}^r {\sum\limits_{{\rho _i} \in R}^{} {\left| {{\rho _i}(q)} \right|} } }_{\scriptstyle contribution~of~N = 4\atop
\scriptstyle hyper~multiplets} - \underbrace {\sum\limits_{\alpha  \in \Phi_+ }^{}{\left| {\alpha (q)} \right|}}_{\scriptstyle contribution~of~N = 4\atop
\scriptstyle{{vector~multiplets}}}.
\end{equation}
The first part of \ref{eq:mon3.41} is the R-charge of the ${\cal N}=4$ hyper multiplets. The second part is the R-charge of the ${\cal N}=4$ vector multiplets. It is instructive to compare \ref{eq:mon3.41} with \ref{eq:mon3.8}.

We can see that the first term in \ref{eq:mon3.8} shows precisely how the affine root is connected to other roots in the summation over matter fields. The contributions to the R-charge are from fields linking adjacent nodes in the quiver diagram and so correspond to bifundamental chiral operators within the ${\cal N}=4$ hypermultiplets.

The second term on the RHS of \ref{eq:mon3.41}, which is described as a sum over the positive root space, has been restated in \ref{eq:mon3.8} in terms of the $q$ charges. The charges $q$, which can be positive or negative, are assigned to the simple roots corresponding to the nodes of the Dynkin or quiver diagram. Each node in the diagram carries a $U(N)$ gauge field associated with the ${\cal N}=4$ vector multiplets. The symmetry breaking that arises internally to each $U(N)$ representation whenever its monopole flux $q$ contains a number of different charges serves to reduce the overall R-charge.

Importantly, the formula \ref{eq:mon3.8} clarifies the dimensional measures $|\rho_i(q)|$ and $|\alpha(q)|$ that are necessary for the RSIMS constructions to be faithful; these have to be implemented as the sum of absolute differences between the various $U(N_i)$ charges, described by their quantum numbers $q_{i,j}$, modulated by any differences in the root lengths encoded in the Cartan matrix.

Having observed that the R-charge collects sets of roots and their orbits, these still need to be assigned correctly to representations enumerated by t. This assignment is moderated or ``dressed" by the term \ref{eq:mon3.7}, which enumerates the degrees of the Casimirs of the $U(N)$ gauge groups that remain unbroken under each set of GNO charges $q$. (When a $U(N)$ symmetry is completely broken, a node has a $U(1)^N$ symmetry). The Casimirs in turn correspond to the set of symmetric invariant tensors of the adjoint representations of the surviving subgroup of the $U(N)$ symmetries.

\FloatBarrier
\section{RSIMS from Regular Semi-simple Subgroup Representations}
\label{sec:weylbranch}

We saw in Section \ref{sec:coulomb} how the RSIMS of a group can be constructed as Coulomb branch quiver theories on extended Dynkin diagrams. These extended Dynkin diagrams are by definition degenerate and this property can be used to establish mappings between the weight space of the Lie algebra of a parent (or {\it ambient}) group and the weight spaces of its subalgebras. As pointed out in \cite{Dynkin:1957um}, this mapping between algebras and subalgebras is equivalent to a mapping between the parent group and its subgroups. Rank is preserved through this procedure, which represents a form of symmetry breaking.

Such mappings are obtained by one or more {\it elementary transformations} \cite{Dynkin:1957um}. These are effected by removing a node from the extended Dynkin diagram of the parent group $G$, which corresponds to the elimination of a row and column from the extended Cartan matrix. The resulting matrix can be decomposed in block diagonal form as a direct sum of regular Cartan matrices of subgroups $\{G_1,\ldots, G_m \}$ where $m \geq 1$.  It then follows \cite{Fuchs:1997bb} that the character of any representation of $G$ maps to the character of some representation of the simple or semi-simple product group $G_1 \otimes  \ldots G_m$. Since only one row and column are removed from the extended Cartan matrix, rank is preserved by an elementary transformation.\footnote{These relationships differ from isomorphisms. While the map from the parent group to a subgroup is injective, it is generally not surjective, so that not all irreps of the subgroup can be mapped from representations of the parent.} A simple or semi-simple subgroup obtained by this method is described as {\it regular}.\footnote{This method does not yield {\it special} subalgebras, which all involve rank reduction.} A subgroup is further described as {\it maximal} if it is not possible to interpose another subgroup between it and the parent. Multiple elementary transformations can be chained to yield further regular, but non-maximal, semi-simple subgroups.

In the case of the A series, the resulting mappings are trivial, since the removal of a node from the extended Dynkin diagram invariably returns the original diagram (modulo some cyclic permutation of simple roots). In the case of other Classical and Exceptional group series, several non-trivial mappings may be possible, depending on the choice of the node removed. We list in Table \ref{table7} all the regular semi-simple proper subgroups of the Classical and Exceptional series arising from a single elementary transformation. While all the regular simple or semi-simple subgroups of Classical groups arising from a single elementary transformation are maximal, $F_4$, $E_7$ and $E_8$ have non-maximal regular subgroups that can also be reached via a maximal subgroup in a two step mapping.

\begin{table}
\caption{Regular Semi-simple Subgroups from Single Elementary Transformation}
\begin{center}
\begin{tabular}{|c|c|c|c|}
\hline
$ {Group}$&$ {Subgroups}$&$ {Type}$ \\
\hline
$ B_{r \ge 2}$&$B_{r-2} \otimes D_2, \ldots, B_1 \otimes D_{r-1}, D_r$&$Maximal$ \\
\hline
$ {{C_{r \ge 2}}}$&$C_{r-1} \otimes C_1, \ldots, C_{ \lceil r/2 \rceil} \otimes C_{ \lfloor r/2 \rfloor}$&$Maximal$ \\
\hline
$ {{D_{r \ge 4}}}$&$D_{r-2} \otimes D_2, \ldots, D_{ \lceil r/2 \rceil} \otimes D_{ \lfloor r/2 \rfloor}$&$Maximal$ \\
\hline
$ {{E_6}}$&$ {A_5} \otimes {A_1}, {A_2} \otimes {A_2} \otimes {A_2}$&$ Maximal$ \\
\hline
$ {{E_7}}$&$ {{D_6} \otimes {A_1}}, {{A_5} \otimes {A_2}}, {A_7} $&$Maximal$ \\
\hline
$ {{E_7}}$&$ {{A_3} \otimes {A_3} \otimes {A_1}}$&$Non-maximal$ \\
\hline
$ {{E_8}}$&$ {{E_7} \otimes {A_1}},{{E_6} \otimes {A_2}},{{A_4} \otimes {A_4}},{A_8} ,{D_8} $&$Maximal$ \\
\hline
$ {{E_8}}$&$ {A_7} \otimes {A_1},{A_5} \otimes {A_2}\otimes {A_1},{D_5} \otimes {A_3}$&$Non-maximal$ \\
\hline
$ {{F_4}}$&$ {{C_3} \otimes {A_1}}, {{A_2} \otimes {A_2}},B_4$&$Maximal$ \\
\hline
$ {{F_4}}$&$ {{A_3} \otimes {A_1}} $&$Non-maximal$ \\
\hline
$ {{G_2}}$&${A_1} \otimes {A_1}, A_2$&$Maximal$ \\
\hline
\end{tabular}
\end{center}
\label{table7}
\end{table}
Our focus here is on two particular types of mapping into regular subgroups. In this Section \ref{sec:weylbranch} we focus on mappings associated with maximal regular semi-simple subgroups. These include the decompositions of the Weyl group set out in Table \ref{table4} (for series other than type A). In Section \ref{sec:mHL} we shall focus on mappings to regular subgroups consisting of A series groups only, which are not generally maximal. In both cases we shall use HWGs to show how the irrep branching relationships resulting from these subgroup mappings permit elegant decompositions of the RSIMS of a group into the moduli spaces of its subgroups.

\subsection{RSIMS from Maximal Regular Semi-simple Subgroups}
\begin{figure}[htbp]
\begin{center}
\includegraphics[scale=0.55]{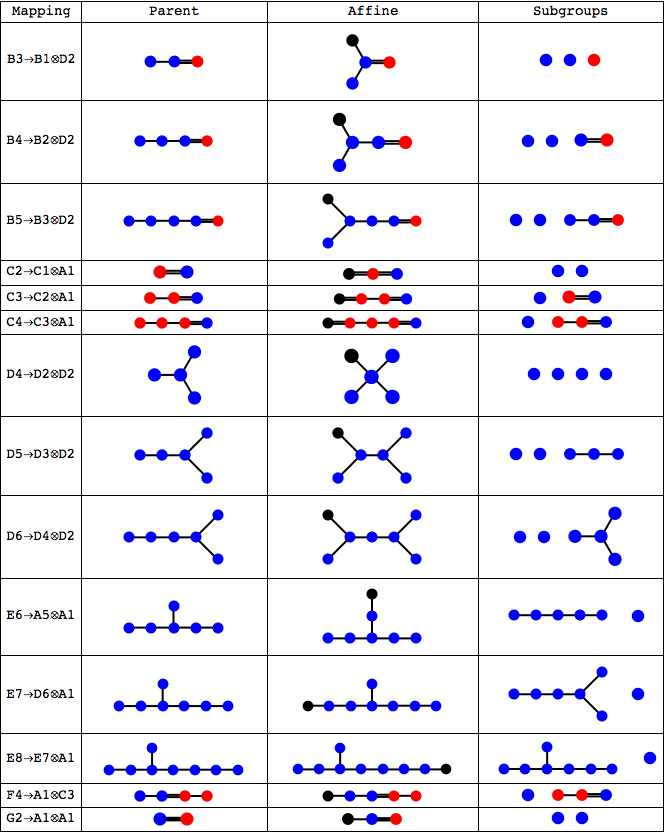}\\
\caption[Dynkin diagrams for subgroup mappings by adjoint node elimination.]{Dynkin diagrams for mapping of non-A series groups into maximal subgroups by adjoint node elimination. Blue nodes denote long roots with length 2. Red nodes denote short roots. A black node denotes the long root added in the affine construction, which is always linked to the adjoint node. The eliminated root is uniquely determined by the subgroups (up to graph automorphisms). Rank is preserved.}
\label{branchingWeyl}
\end{center}
\end{figure}

Regular subgroup mappings are determined by the choice of node for elimination from the extended Dynkin diagram. Elimination from the extended Dynkin diagram of a (non-A series) Classical or Exceptional group of the node corresponding to the Dynkin label of the adjoint representation gives a mapping into a subgroup that contains the group $G_0$ shown in Table \ref{table4}. These subgroups are all maximal and semi-simple. The Dynkin diagram manipulations are set out in Figure \ref{branchingWeyl}  and the resulting adjoint irrep branchings into irreps of the subgroups are set out in Table \ref{table8}.

\begin{sidewaystable}
\caption{Adjoint Representation Branching By Adjoint Node Elimination}
\begin{center}
\begin{tabular}{|c|c|c|c|}
\hline
$ {Group}$&$ {Adjoint}$&$ {Branching}$&$ {Decomposition}$ \\
\hline
$ {{B_3}}$&$ {[0,1,0]}$&$ {{B_1} \otimes {D_2}}$&$[ 2] +[2] + [2]+ [2][1][1]$\\
\hline
$ {{B_4}}$&$ {[0,1,0,0]}$&$ {{B_2} \otimes {D_2}}$&$ [0,2] +[2]+ [2]+[1,0][1][1]$ \\
\hline
$ {{B_5}}$&$ {[0,1,0,0,0]}$&$ {{B_3} \otimes {D_2}}$&$ [0,1,0] +[2]+ [2]+[1,0,0][1][1]$ \\
\hline
$ {{C_2} \cong {B_2}}$&$ {[2,0]}$&$ {{C_1} \otimes {A_1}}$&$ {[ 2] +[ 2] +[1][1]}$ \\
\hline
$ {{C_3}}$&$ {[2,0,0]}$&$ {{C_2} \otimes {A_1}}$&$ {[2,0] +[ 2] +[1,0][1]}$ \\
\hline
$ {{C_4}}$&$ {[2,0,0,0]}$&$ {{C_3} \otimes {A_1}}$&$ {[2,0,0] +[ 2] +[1,0,0][1]}$ \\
\hline
$ {{D_4}}$&$ {[0,1,0,0]}$&$ {{D_2} \otimes {D_2}}$&$ [2] +[2]+[2]+ [2]+[1][1][1][1]$ \\
\hline
$ {{D_5}}$&$ {[0,1,0,0,0]}$&$ {{D_3} \otimes {D_2}}$&$ [0,1,1] +[2]+[2]+[1,0,0][1][1]$ \\
\hline
$ {{D_6}}$&$ {[0,1,0,0,0,0]}$&$ {{D_4} \otimes {D_2}}$&$ [0,1,0,0] +[2]+[2]+[1,0,0,0][1][1]$ \\
\hline
$ {{E_6}}$&$ {[0,0,0,0,0,1]}$&$ {{A_5} \otimes {A_1}}$&$ [1,0,0,0,1] +[2]+[0,0,1,0,0][1]$ \\
\hline
$ {{E_7}}$&$ {[1,0,0,0,0,0,0]}$&$ {{D_6} \otimes {A_1}}$&$ [0,1,0,0,0,0] +[2]+[0,0,0,0,0,1][1]$ \\
\hline
$ {{E_8}}$&$ {[0,0,0,0,0,0,1,0]}$&$ {{E_7} \otimes {A_1}}$&$ [1,0,0,0,0,0,0] +[2]+[0,0,0,0,0,1,0][1]$ \\
\hline
$ {{F_4}}$&$ {[1,0,0,0]}$&$ {{C_3} \otimes {A_1}}$&$ {[2,0,0] +[ 2] +[0,0,1][1]}$ \\
\hline
$ {{G_2}}$&$ {[1,0]}$&$ {{A_1} \otimes {A_1}}$&$ {[ 2] +[ 2] +[3][1]}$ \\
\hline
\end{tabular}
\end{center}
\text{ Singlets omitted from descriptions of product group decompositions for brevity.}
\label{table8}
\end{sidewaystable}

These mappings based on elimination of the adjoint node do not exhaust the regular subgroups in Table \ref{table7} and mappings can be found to the other subgroups by eliminating other nodes.

Under all these subgroup mappings,  the adjoint representation of the parent group splits into the direct sum of the adjoint representations of the subgroups, plus a product group representation.

Importantly, each such mapping allows us to establish a bijection between the CSA coordinates of the weight space of the parent group and the CSA coordinates of the weight space of a maximal sub group.\footnote{An example of CSA coordinate map calculation is contained in Section \ref{sec:D4ex}.} However, while the mapping from irreps of the parent group to the representations of the product group is injective, it is not surjective, and one cannot generally map any representation of the product group back to a parent group representation; this is only possible for specific representations (such as those identified by the RSIMS deconstruction).
\FloatBarrier
\subsection{RSIMS deconstruction to Subgroup HWG}

Given such a coordinate mapping from a parent group $G$ of rank $r$ to a subgroup ${G_1}  \ldots \otimes{G_m}$ of equal rank, we can take $g_{instanton}^G (t, x_i)$, express it in terms of the CSA coordinates $\{ {y_1},\ldots, {y_r} \}$ of its subgroup, and use a character generating function $g_{\cal X}^{{G_1} \otimes \ldots {G_m}}( {{m_i}, {y_j}})$ for the irreps of the subgroup to project $g_{instanton}^G (t,y_i)$ onto the irreps of the subgroup. The subgroup irreps are tracked using the Dynkin label fugacities $\{ {m_1},\ldots, {m_r} \}$ and the projection coefficients obtained are polynomials in the fugacity $t$.

The analysis depends on the completeness of the characters $[n_1, \ldots ,n_r]$ of the subgroup, which permits the decomposition of the instanton moduli space in terms of the coefficients $C_{{n_1}, \ldots, {n_r}}$ defined from:
\begin{equation}
\label{eq:weyl1}
\begin{aligned}
{g^G_{instanton}}(t,{x_i}) & \equiv \sum\limits_{k = 0}^\infty  [k~adj](x_i)~{t^k},\\
{g^G_{instanton}}(t,{y_i}) &  \equiv \sum\limits_{{n_i}}^{} {C_{{n_1}, \ldots, {n_r}}}(t) [{{n_1}, \ldots ,{n_r}}] (y_i),
\end{aligned}
\end{equation}
and upon Weyl integration, which allows us to use character generating functions to project out an HWG function in terms of $\{t,m_i\}$:
\begin{equation}
\label{eq:weyl2}
\begin{aligned}
g_{instanton}^G\left( {t,{m_j}} \right) &\equiv \sum\limits_{{n_i}}^{} {{C_{{n_1}, \ldots ,{n_r}}}} (t)~m_1^{{n_1}} \ldots m_r^{{n_r}}\\
&  = \oint\limits_{{G_1} \otimes  \ldots {G_m}} {d\mu (y_i)}~{g^{{G_1} \otimes  \ldots {G_m}}_{\cal X}}( {{m_j},y^*_i})~g_{instanton}^G (t,y_i).
\end{aligned}
\end{equation}
For further detail on the use of character generating functions to project out HWGs the reader is referred to \cite{Hanany:2014dia}. 

The HWGs for all the maximal regular simple and semi-simple subgroup mappings of Classical group RSIMS are set out in Table \ref{table9} and those for a number of Exceptional group RSIMS are set out in Table \ref{table10}. These include all the maximal regular subgroup mappings identified in Table \ref{table7} (of which those in Table \ref{table8} are a subset). For convenience, the HWGs are presented using their PLs. There are many observations that can be made about the structure of these highest weight moduli spaces. 
\begin{sidewaystable}
\caption{HWGs for Classical RSIMS Deconstructions into Maximal Regular Subgroups}
\begin{center}
\begin{tabular}{|c|c|c|c|}
\hline
$ {RSIMS}$&${ \begin{array}{c}Subgroup\\ \left[ m \right]\left[ n \right]\end{array}}$&$PL[HWG]$&$\begin{array}{c}HWG\\Dimension\end{array}$ \\
\hline
$ {{B_2}}$&$D_2$&${(n_1^2 + n_2^2 + {n_1}{n_2})t - n_1^2 n_2^2{t^2}}$&$2$\\
$ {{B_3}}$&$D_3$&${({n_1} + {n_2}{n_3})t}$&$2$\\
$ {{B_{r \ge 4}}}$&$D_r$&${({n_1} + {n_2})t}$&$2$\\
\hline
$ {{B_3}}$&${{B_1} \otimes {D_2}}$&${(m_1^2 + n_1^2 + n_2^2 + m_1^2{n_1}{n_2})t + (1 + m_1^2{n_1}{n_2}){t^2} - m_1^4n_1^2n_2^2{t^4}}$&$5$\\
$ {{B_4}}$&${{B_1} \otimes {D_3}}$&${(m_1^2 + {n_2}{n_3} + m_1^2{n_1})t + (1 + n_1^2 + m_1^2{n_1}){t^2} - m_1^4n_1^2{t^4}}$&$5$\\
$ {{B_{r \ge5}}}$&${{B_1} \otimes {D_{r - 1}}}$&${(m_1^2 + {n_2} + m_1^2{n_1})t + (1 + n_1^2 + m_1^2{n_1}){t^2} - m_1^4n_1^2{t^4}}$&$5$\\
\hline
$ {{B_4}}$&${{B_2} \otimes {D_2}}$&${(m_2^2 + n_1^2 + n_2^2 + {m_1}{n_1}{n_2})t + (1 + m_1^2 + {m_1}{n_1}{n_2}){t^2} - m_1^2n_1^2n_2^2{t^4}}$&$6$\\
$ {{B_5}}$&${{B_2} \otimes {D_3}}$&${(m_2^2 + {n_2}{n_3} + {m_1}{n_1})t + (1 + m_1^2 + n_1^2 + {m_1}{n_1}){t^2} - m_1^2n_1^2{t^4}}$&$6$\\
${{B_{r \ge 6}}}$&${{B_2} \otimes {D_{r - 2}}}$&${(m_2^2 + {n_2} + {m_1}{n_1})t + (1 + m_1^2 + n_1^2 + {m_1}{n_1}){t^2} - m_1^2n_1^2{t^4}}$&$6$\\
\hline
${{B_{r \ge 5}}}$&${{B_{r - 2}} \otimes {D_2}}$&${({m_2} + n_1^2 + n_2^2 + {m_1}{n_1}{n_2})t + (1 + m_1^2 + {m_1}{n_1}{n_2}){t^2} - m_1^2n_1^2n_2^2{t^4}}$&$6$\\
${{B_{r \ge 6}}}$&${{B_{r - 3}} \otimes {D_3}}$&${({m_2} + {n_2}{n_3} + {m_1}{n_1})t + (1 + m_1^2 + n_1^2 + {m_1}{n_1}){t^2} - m_1^2n_1^2{t^4}}$&$6$\\
${{B_{r \ge 7}}}$&${{B_{r - s \ge 3}} \otimes {D_{s \ge 4}}}$&${({m_2} + {n_2} + {m_1}{n_1})t + (1 + m_1^2 + n_1^2 + {m_1}{n_1}){t^2} - m_1^2n_1^2{t^4}}$&$6$\\
\hline
${{C_{r \ge 2}}}$&${{C_{r - s}} \otimes {C_s}}$&${(m_1^2 + n_1^2 + {m_1}{n_1})t - m_1^2n_1^2{t^2}}$&$2$\\
\hline
${{D_4}}$&${{D_2} \otimes {D_2}}$&${(m_1^2 + m_2^2 + n_1^2 + n_2^2 + {m_1}{m_2}{n_1}{n_2})t + (1 + {m_1}{m_2}{n_1}{n_2}){t^2} - m_1^2m_2^2n_1^2n_2^2{t^4}}$&$6$\\
${{D_5}}$&${{D_3} \otimes {D_2}}$&${({m_2}{m_3} + n_1^2 + n_2^2 + {m_1}{n_1}{n_2})t + (1 + m_1^2 + {m_1}{n_1}{n_2}){t^2} - m_1^2n_1^2n_2^2{t^4}}$&$6$\\
${{D_{r \ge 6}}}$&${{D_{r - 2}} \otimes {D_2}}$&${({m_2} + n_1^2 + n_2^2 + {m_1}{n_1}{n_2})t + (1 + m_1^2 + {m_1}{n_1}{n_2}){t^2} - m_1^2n_1^2n_2^2{t^4}} $&$6$\\
${{D_{r \ge 7}}}$&${{D_{r - 3}} \otimes {D_3}}$&$ {({m_2} + {n_2}{n_3} + {m_1}{n_1})t + (1 + m_1^2 + n_1^2 + {m_1}{n_1}){t^2} - m_1^2n_1^2{t^4}} $&$6$\\
${{D_{r \ge 8}}}$&${{D_{r -s \ge 4}} \otimes {D_{s \ge 4}}}$&$ {({m_2} + {n_2} + {m_1}{n_1})t + (1 + m_1^2 + n_1^2 + {m_1}{n_1}){t^2} - m_1^2n_1^2{t^4}} $&$6$\\
\hline
\end{tabular}
\end{center}
\text{ Subscripts are used to distinguish Dynkin label fugacities within each subgroup.}
\label{table9}
\end{sidewaystable}

\begin{sidewaystable}
\caption{HWGs for Exceptional RSIMS Deconstructions into Maximal Regular Subgroups}
\begin{center}
\begin{tabular}{|c|c|c|c|}
\hline
$ {RSIMS}$&$\begin{array}{c}Subgroup\\\left[ m \right]\left[ n \right]\end{array}$&$PL[HWG]$&$\begin{array}{c}HWG\\Dimension\end{array}$ \\
\hline
$ {{E_6}}$  &$ {A_5} \otimes {A_1}$ &$    {({m_1}{m_5} + n_1^2 + {m_3}{n_1})t + (1 + {m_2}{m_4} + {m_3}{n_1}){t^2} + {m_3}^2{t^3} - m_3^2n_1^2{t^4}}       $&$6$ \dag\\
$ {{E_6}}$&$ {A_2} \otimes {A_2} \otimes {A_2}$&$     not~complete~intersection       $&$13$\\
\hline
$ {{E_7}}$&$ {A_7}$&${({m_1}{m_7} + {m_4})t + (1 + {m_4} + {m_2}{m_6}){t^2} + {m_3}{m_5}{t^3}}$&$6$\\
$ {{E_7}}$ &${{D_6} \otimes {A_1}}$  &${({m_2} + n_1^2 + {m_5}{n_1})t + (1 + {m_4} + {m_5}{n_1}){t^2} + {m_5}^2{t^3} - m_5^2n_1^2{t^4}}$&$6$ \dag \\
$ {{E_7}}$&${{A_5} \otimes {A_2}}$&$    \ldots       $&$\ge 16$\\
\hline
$ {{E_8}}$&$ {D_8}$&${({m_2} + {m_7})t + (1 + {m_4} + {m_7}){t^2} + {m_6}{t^3}}$ &$6$\\
$ {{E_8}}$  &${{E_7} \otimes {A_1}} $  &${({m_1} + n_1^2 + {m_6}{n_1})t + (1 + {m_5} + {m_6}{n_1}){t^2} + {m_6}^2{t^3} - m_6^2n_1^2{t^4}}$&$6$ \dag \\
$ {{E_8}}$&$ {E_6} \otimes {A_2}$&$   \ldots        $&$\ge 19  $\\
$ {{E_8}}$&$ {A_8} $&$   \ldots        $&$\ge 22  $\\
$ {{E_8}}$&$ {A_4} \otimes {A_4}$&$   \ldots        $&$\ge 38  $\\
\hline
$ {{F_4}}$&$ {B_4}$&${\left( {{m_2} + {m_4}} \right)t}$&$2$\\
$ {{F_4}}$ &$ {C_3} \otimes {A_1}$  &${(m_1^2 + n_1^2 + {m_3}{n_1})t + (1 + m_2^2 + {m_3}{n_1}){t^2} + {m_3}^2{t^3} - m_3^2n_1^2{t^4}}$&$6$ \dag \\
$ {{F_4}}$&$ {A_2}\otimes {A_2}$&$   not~complete~intersection $&$10$\\
\hline
$ {{G_2}}$&$ {A_2}$&${\left( {{m_1} + {m_2} + {m_1}{m_2}} \right)t}$&$3$\\
$ {{G_2}}$  &$ {{A_1} \otimes {A_1}}$  &${(m_1^2 + n_1^2 + m_1^3{n_1})t + (1 + m_1^3{n_1}){t^2} - m_1^6n_1^2{t^4}}$&$4$ \dag \\
\hline

\end{tabular}
\end{center}
\text{ Subscripts are used to distinguish Dynkin label fugacities within each subgroup.}\\
\text{\dag: Branchings obtained by adjoint node elimination.}
\label{table10}
\end{sidewaystable}
Firstly, these moduli spaces are all generated by a small number of representations of the product group. They include, in all cases, the adjoint representations of each of the constituents of the product group at order $t$.

Next, taking a geometric perspective, all the moduli spaces of Classical RSIMS deconstructions are either freely generated, being products of geometric series, or complete intersections, being quotients of products of geometric series. In all cases there is a further generator in addition to the adjoints at order $t$ involving the vector representation(s).  In the case of symplectic groups, there is a relation at order $t^2$. For orthogonal groups, there may also be additional generators at order $t^2$ and a relation at order $t^4$ at most. 

The dimensions of the HWG moduli spaces, which are given by the number of generators less relations, vary from two in the case of the symplectic groups up to at most six for orthogonal groups. The apparent complexity of many of the decompositions can be simplified further. Assuming minimum ranks of 2 and 3 respectively for any B and D series subgroups, we can write the HWG for a Classical RSIMS deconstruction into a maximal \emph{pair} of subgroups in the form:
\begin{equation}
\label{eq:weyl3}
\begin{aligned}
g_{instanton}^{B~or~D} & \to PE\left[ {\left( {{{\theta  + \theta'  + v}} \otimes {{v'}}} \right)t + \left( {1 + {{g}} + {{g'}} + {{v}} \otimes {{v'}}} \right){t^2} - g \otimes g' {t^4}} \right],\\
g_{instanton}^{C} & \to PE\left[ {\left( \theta  + \theta'  + v \otimes v' \right)t - g \otimes g' {t^2}} \right],
\end{aligned}
\end{equation}
where the adjoint, vector and graviton (symmetrised vector) representations of the two (primed and unprimed) subgroups are represented by $\{ \theta, v, g \}$ respectively. Importantly, since the form of the HWG does not change for higher rank BCD series groups, we conjecture that these expressions give us complete descriptions of RSIMS decompositions into regular semi-simple subgroups for all Classical Lie algebras. 

Some of the HWGs for  deconstructions of Exceptional RSIMS follow the same pattern as the HWGs for Classical RSIMS, being freely generated or complete intersections, and having dimensions between two and six. Notably, these simple HWGs include those obtained by adjoint node elimination, as in Table \ref{table8}. They also include the HWGs of branchings into a simple regular subgroup (other than for $E_8$ to $A_8$). However, for the E and F series, some maximal regular subgroups lead to complicated HWGs; these are not complete intersections, have generators at higher orders of $t$ and have dimensions that vary up to at least $38$ (as explained below). The HWGs for the non-maximal regular subgroups are also complicated. So far, we have not been able to calculate all these HWGs.
\FloatBarrier
\subsection{Dimensions of HWGs and Hilbert Series for RSIMS}
It is interesting to relate the dimensions of an HWG to the dimensions of the Hilbert series for the same RSIMS. Recall that the dimension of an (unrefined) Hilbert series for an RSIMS is always equal to twice the sum of the dual Coxeter labels for the group \cite{Hanany:2014dia}. The difference in dimension of the two moduli spaces is accounted for by the degree of the dimensional polynomial for the weight space spanned by the subgroup irreps. As an example, for irreps of $A_2$, with Dynkin labels $[n_1,n_2]$, we have the dimensional formula:
\begin{equation}
\label{eq:hl5.1}
Dim[n_1,n_2]= (1 + n_1) (1 + n_2) (2 + n_1 + n_2)/2,
\end{equation}
and so the degrees of the dimensional polynomial are 2 for irreps of the type $[n_1>0,0]$ or $[0,n_2>0]$ and 3 for irreps of the type $[n_1>0,n_2>0]$, which include the adjoint representation. This degree of 3 equals the difference between the dimension of the Hilbert series for the $A_2$ RSIMS $(1+4t +t^2)/(1-t)^4$, which is 4, and the unit dimension of the corresponding HWG $1/(1-m_1 m_2 t)$.

Assuming that a weight lattice is saturated (i.e. that all Dynkin labels are non-zero), the degree of the dimensional polynomial is always equal to the number of positive roots. By using the standard dimensional polynomials we can reconcile the dimensions of the various moduli spaces as set out in Table \ref{table11}. It is important to note that if the HWG irreps do not saturate the subgroup weight lattice, this reduces the degree of the relevant dimensional polynomial.

\begin{sidewaystable}
\caption{Dimensions of Hilbert Series and Subgroup HWGs for RSIMS}
\begin{center}
\begin{tabular}{|c|c|c|c|c|c|}
\hline
$ {Group}$&$\begin{array}{c}HS\\Dimension\\(a)+(b)\end{array}$&$Subgroup$&$\begin{array}{c}HWG\\Dimension\\(a)\end{array}$&$\begin{array}{c}HWG\\Irrep\\Structure \end{array}$&$\begin{array}{c}Degree~of\\Dimensional\\Polynomial\\(b)\end{array}$ \\
\hline
$ {{B_{r \ge 2}}}$&$4r-4$&$ D_r$&$2$&$[n, n,\ldots,0] ^\dag $&$4r-6$\\
$B_{r \ge 3}$&$4r-4$&$ {{B_1} \otimes {D_{r-1}}}$&$5$&$[n][n,n,\ldots,0] ^\dag$&$1 \oplus 4(r-1)-6$\\
$ B_{r \ge 4}$&$4r-4$&$B_{r-s \ge 2} \otimes D_{s \ge2}$&$6$&$[n,n,\ldots,0][n,n,\ldots,0] ^\dag$&$ 4(r-s)-4 \oplus 4s-6$ \\
\hline
$ {{C_{r \ge 2}}}$&$2r$&$C_{r-s} \otimes C_s$&$2$&$[n,\ldots,0][n,\ldots,0]$&$2(r-s)-1 \oplus 2s-1 $ \\
\hline
$ {{D_{r \ge 4}}}$&$4r-6$&$D_{r-s \ge 2} \otimes D_{s \ge 2}$&$6$&$ [n,n,\ldots,0] [n,n,\ldots,0] ^\dag$&$4(r-s)-6 \oplus 4s-6$\\
\hline
$ {{E_6}}$&$22$&$ {A_5} \otimes {A_1}$&$ 6$&$[n,n,n,n,n][n] $&$15 \oplus 1$ \\
$ {{E_6}}$&$22$&$ {A_2} \otimes {A_2} \otimes {A_2}$&$13$&$[n,n][n,n][n,n] $&$3 \oplus 3\oplus 3$ \\
\hline
$ {{E_7}}$&$34$&$ {{A_7} } $&$6$&$[n,n,n,n,n,n] $ &$28$\\
$ {{E_7}}$&$34$&$ {{D_6} \otimes {A_1}} $&$6$&$[0,n,0,n,n,0][n] $ &$27 \oplus 1$\\
\hline
$ {{E_8}}$&$58$&$ {{D_8}} $&$6$&$[0,0,0,n,0,n,n,0] $ &$52$\\
$ {{E_8}}$&$58$&$ {{E_7} \otimes {A_1}} $&$6$&$[n,0,0,0,n,n,0][n] $ &$51 \oplus 1$\\
\hline
$ {{F_4}}$&$16$&$ {{B_4}}$&$2$&$ [0,n,0,n]$&$14$ \\
$ {{F_4}}$&$16$&$ {{C_3} \otimes {A_1}}$&$6$&$ [n,n,n][n]$&$9 \oplus 1$ \\
$ {{F_4}}$&$16$&$ {{A_2} \otimes {A_2}}$&$10$&$ [n,n][n,n]$&$3 \oplus 3$ \\
\hline
$ {{G_2}}$&$6$&${A_2}$&$3$&$[n][n] $&$3$ \\
$ {{G_2}}$&$6$&${A_1} \otimes {A_1}$&$4$&$[n][n] $&$1 \oplus 1$ \\
\hline
\end{tabular}
\end{center}
\text{\dag: HWG irrep structure for $D_3$ is $[n,n,n]$.}\\
\text{$n$ denotes a non-zero Dynkin label.}\\
\text{Not all maximal regular subgroups are included for Exceptional series.}
\label{table11}
\end{sidewaystable}
Thus, we can explicate the relationship between a given mapping, the weight lattice of the subgroup and the difference in dimensions between the Hilbert series for the RSIMS and the subgroup HWG. When a subgroup has a weight lattice with a dimensional polynomial of low degree, this is balanced by an increase in the dimension of the HWG. Given some mapping, the degree of the dimensional polynomial of the saturated weight lattices of the subgroup places a lower bound on the dimension of the HWG, as indicated in Table \ref{table10} for the unknown HWGs. 

For A series groups, the number of positive roots is only $(r^2+r)/2$, compared with  $r^2$ or $r^2-r/2$ for B/C or D series groups, and so A series dimensional polynomials tend to be of lower degree. Thus, many Exceptional group mappings to A series subgroups lead to HWGs with a high dimension. While these are all calculable in principle, using \ref{eq:weyl1} and \ref{eq:weyl2} , this can be difficult in practice due to computing constraints. This raises the question as to whether it is possible to deconstruct RSIMS using moduli spaces that have a higher dimensional degree than Lie group representations and so lead to low dimensional HWGs. We find that such moduli spaces can be provided by modified Hall-Littlewood polynomials.
\FloatBarrier
\section{RSIMS from A Series Hall-Littlewood Polynomials}
\label{sec:mHL}
Constructions for the RSIMS of $E_6$, $E_7$ and $E_8$ instantons based on Hall-Littlewood polynomials have been given in \cite{Gadde:2011uv}. These draw upon branching relationships between the characters of irreps of these groups and those of A series subgroups. The constructions in \cite{Gadde:2011uv} are guided by a conjectured characterisation of punctures on spheres, which helps to identify combinations of  A series modified Hall-Littlewood polynomials that yield the desired moduli spaces. We take a different approach and carry out the direct decompositions of RSIMS, all of which have known group theoretic constructions, as discussed earlier, in terms of the modified Hall-Littlewood polynomials of A series groups. Hall-Littlewood polynomials can also be constructed for other Classical or Exceptional groups, but the analysis herein is limited to those of unitary groups. Our strategy exploits the fact that Hall-Littlewood polynomials provide a basis for single parameter class functions \cite{Macdonald:1995fk}, such as RSIMS. In order to find the coefficients defining these decompositions, we construct a set of generating functions for Hall-Littlewood polynomials and exploit their orthogonality properties under Weyl integration, using an appropriate measure. Our decompositions then follow group mappings into regular semi-simple subgroups, in a similar manner to the previous Section.

\subsection{Hall-Littlewood Polynomials and their Generating Functions}
Hall-Littlewood polynomials are symmetric polynomials in a set of coordinates that are parameterised by an additional variable \cite{Macdonald:1995fk}, and so correspond in a natural way to plethystic class functions built from the CSA coordinates for characters of unitary groups combined with a counting fugacity $t$. Hall-Littlewood polynomials can be labelled by the Dynkin labels of irreps of $U(N)$, or, equivalently, by partitions of $N$ objects, or by Young tableaux. They are most helpfully defined in terms of their orthogonality properties under Weyl integration using an explicit measure, as presented in \cite{Gadde:2011uv}, for example. There are various choices of normalisation possible: \cite{Gadde:2011uv} chooses a normalisation under which the Hall-Littlewood polynomials are strictly orthonormal; \cite{Macdonald:1995fk} chooses a normalisation under which they become symmetric monomial functions for $t=1$. We shall use a third normalisation scheme, also used in \cite{Cremonesi:2014kwa}, that follows naturally from their generating functions. Under all these normalisation schemes, the Hall-Littlewood polynomials revert to Schur polynomials, i.e. the characters of irreps of $U(N)$, for $t=0$.

Hall-Littlewood polynomials incorporating the characters of irreps of $SU(N)$ are closely related to those for $U(N)$, however, care needs to be taken over the choice of coordinates, labelling of partitions and normalisation. We shall ultimately work with the Hall-Littlewood polynomials and related functions for $SU(N)$, however, we derive their properties from those of the polynomials for $U(N)$.

We set out in Table \ref{table12} the structure of the Hall-Littlewood measure. This is the product of the usual Haar measure for $U(N)$ (given by the first two factors) with an additional plethystic function parameterised by $t$. Clearly the parameter $t$ is key in determining the basis functions on a space with the Hall-Littlewood measure. Thus, it can be seen that the measure reverts to the $U(N)$ Haar measure for $t=0$, or to the $U(1)^N$ Haar measure for $t=1$. The corresponding basis functions $ {HL_\lambda }(t)$ become either Schur polynomials ${s_\lambda }$, or monomials ${m_\lambda }$, respectively, in these limits \cite{Macdonald:1995fk}.

\begin{table}
\caption{Components of Hall-Littlewood Measure for $U(N)$}
\begin{center}
\begin{tabular}{|c|ccccc|}
\hline
$ {}$&$ {U{{\left( 1 \right)}^N}}$&${}$&${U\left( N \right)}$&${}$&${Hall - Littlewood}$\\
\hline
$ {}$&$ {}$&${}$&${}$&${}$&${}$\\
$ {Measure}$&$ {\prod\limits_i {\oint\limits_{} {\frac{{d{x_i}}}{{{x_i}}}} } }$&$ \times $&${\frac{1}{{N!}}\prod\limits_{j \ne k} {\left( {1 - {x_j}/{x_k}} \right)} }$&$ \times $&${\prod\limits_{i \ne j} {\frac{1}{{1 - t{x_i}/{x_j}}}} }$ \\
$ {(Plethystic)}$&$ {}$&${}$&${\frac{1}{{N!}}PE\left[ {-\left(adjoint - rank \right)} \right]}$&$ {} $&${PE\left[ {\left( {adjoint - rank} \right) t} \right]}$ \\
\hline
$ {Basis}$&$ {{m_\lambda } = {HL_\lambda }\left( 1 \right)}$&${}$&${{s_\lambda } = {HL_\lambda }\left( 0 \right)}$&${}$&${{HL_\lambda }\left( t \right)}$\\
\hline

\end{tabular}
\end{center}
\label{table12}
\end{table}

Hall-Littlewood polynomials which are orthogonal with respect to this defined measure are given by  \cite{Gadde:2011uv}:

\begin{equation}
\label{eq:hl5.1}
{HL_\lambda }\left( {{x_i}, t} \right) = \sum\limits_{w \in {S_N}} {w\left( {x_1^{{\lambda _1}} \ldots x_N^{{\lambda _N}}\prod\limits_{i < j} {\frac{{{x_i} - t{x_j}}}{{{x_i} - {x_j}}}} } \right)}, 
\end{equation}
where the $x_i$ are CSA coordinates for $U(N)$ and $\lambda  = \left( {{\lambda _1}, \ldots ,{\lambda _N}} \right)$ is the partition corresponding to Dynkin labels $[n_1,\ldots,n_N]$ for $U(N)$ through the relationship 

\begin{equation}
\label{eq:hl5.1a}
\left( {{\lambda _1}, \ldots ,{\lambda _j}, \ldots ,{\lambda _N}} \right) = \left( {\sum\limits_{i = 1}^N {{n_i}, \ldots ,\sum\limits_{i = j}^N {{n_i}, \ldots ,{n_N}} } } \right).
\end{equation}
(This bijection allows us to refer to a Hall-Littlewood polynomial by either $HL_\lambda$ or $HL_{[n]}$.) The sum in \ref{eq:hl5.1} is taken over the Weyl group of $U(N)$, which is the symmetric group $S_N$. The orthogonality of the ${HL_\lambda }$ and their complex conjugates, under an inner product incorporating the Hall-Littlewood measure, is given by:
\begin{equation}
\label{eq:hl5.2}
\oint\limits_{U\left( N \right)} {d{\mu _{HL}}}~{HL_\lambda }\left( {{x_i}, t} \right){HL_\mu }\left( {x^*{_i}, t} \right) = {\delta _{\lambda \mu }}{v_\lambda }\left( t \right),
\end{equation}
where we are using abbreviated notation ${d{\mu _{HL}}}$ for the Hall-Littlewood measure,
\begin{equation}
\label{eq:hl5.3}
\oint\limits_{U\left( N \right)} {d{\mu _{HL}}}  \equiv \prod\limits_i {\oint\limits_{} {\frac{{d{x_i}}}{{{x_i}}}} } \frac{1}{{N!}}\left( {\prod\limits_{j \ne k} {\left( {1 - {x_j}/{x_k}} \right)} } \right)\left( {\prod\limits_{j \ne k} {\frac{1}{{1 - t{x_j}/{x_k}}}} } \right)
\end{equation}
and we have introduced the normalisation function ${v_\lambda}(t)$:
\begin{equation}
\label{eq:hl5.4}
{v_\lambda }\left( t \right) = \prod\limits_{i \ge 0} {\prod\limits_{j = 1}^{m_i(\lambda)} {\frac{{1 - {t^j}}}{{1 - t}}} }.
\end{equation}
In the ${v_\lambda}(t)$ function, the product is taken over each distinct integer $i$, including zero, appearing in the partition $\lambda$ according to its multiplicity $m_i$ \cite{Macdonald:1995fk}.\footnote{In \cite{Macdonald:1995fk} the ${HL_\lambda }$ are normalised by dividing by ${v_\lambda }(t)$ and in  \cite{Gadde:2011uv} they are normalised by dividing by ${\sqrt {{v_\lambda }(t)}}$.} In effect, ${v_\lambda }(t)$ is determined by the number and location of zeros amongst the Dynkin labels corresponding to a given partition. It is important to distinguish Dynkin labels for Hall-Littlewood polynomials from those for $U(N)$ characters; we shall ultimately wish to work with both types of label to describe the relationships between the two types of class function.

The Hall-Littlewood polynomials \ref{eq:hl5.1} provide a complete basis for class functions that combine the characters of a unitary group with coefficients given by polynomials in the parameter $t$ \cite{Macdonald:1995fk}.

We now follow the HWG methodology introduced in \cite{Hanany:2014dia} and define the fugacities $\{h_1,\ldots,h_N\}$ for the Dynkin labels $[n_1,\ldots,n_N]_{HL}$. Note that we prefer to use Dynkin label fugacities $h_i$ for Hall-Littlewood polynomials and $m_i$ for characters. We then convert \ref{eq:hl5.1} from partition to Dynkin label notation and rearrange to obtain a highest weight generating function for the ${HL_\lambda }$ or ${HL_{[n]} }$ :
\begin{equation}
\label{eq:hl5.5}
\begin{aligned}
{g_{HL}}\left( {{x_i},t,{h_i}} \right) &\equiv 
\sum\limits_{{n}} {{{HL}_{\left[ {{n_1}, \ldots ,{n_N}} \right]}}\left( {{x_i},t} \right)h_1^{{n_1}} \ldots h_N^{{n_N}}}\\
&= \sum\limits_{w \in {S_N}} {w\left( {\sum\limits_{{n}} {h_1^{{n_1}} \ldots h_N^{{n_N}}} x_1^{{n_1} +  \ldots {n_N}} \ldots x_N^{{n_N}}\prod\limits_{i < j} {\frac{{1 - t{x_j}/{x_i}}}{{1 - {x_j}/{x_i}}}} } \right)} \\
&  = \sum\limits_{w \in {S_N}} {w\left( {\prod\limits_{k = 1}^N {\frac{1}{{1 - {h_k}\prod\limits_{l = 1}^k {{x_l}} }}} \prod\limits_{i < j} {\frac{{1 - t{x_j}/{x_i}}}{{1 - {x_j}/{x_i}}}} } \right)}.
\end{aligned}
\end{equation}
From \ref{eq:hl5.2}, it follows that the complex conjugates of the generating functions $g_{HL}({x_i},t,{h_i})$ have the orthogonality property:
\begin{equation}
\label{eq:hl5.6}
\begin{aligned}
\oint\limits_{U\left( N \right)} {d{\mu _{HL}}}~g_{HL}\left( {{x^*}_i,t,{h_i}} \right){HL_\lambda }\left( {{x_i},t} \right) ={v_\lambda }\left( t \right) {h^\lambda },
\end{aligned}
\end{equation}
where we have defined ${h^\lambda } \equiv \prod\limits_i {h_i^{{n_i}(\lambda)}}$.
We can obtain more a useful contragredient generating function, which generates polynomials that are orthonormal (rather than just orthogonal) to the ${HL_\lambda}({x}_i,t)$, by gluing together the $g_{HL}({x^*}_i,t,{h_i})$ with a generating function for the inverse of the ${v_\lambda }$.

Let us briefly describe this gluing procedure. Suppose we have two power series in $t$ given by $A(t) = \sum\limits_{n = 0}^\infty  {{a_n}{t^n}}$ and $B(t) = \sum\limits_{n = 0}^\infty  {{b_n}{t^n}}$. We can glue the coefficients together into a single series by introducing conjugate $U(1)$ fugacities into the counting variables for the two series and then using Weyl integration to project out the $U(1)$ singlets of their product. Thus:
\begin{equation}
\label{eq:hl5.7}
\begin{aligned}
 \sum\limits_n {{a_n}} \left( x \right){b_n}\left( x \right){t^n}\\
 &=\oint\limits_{U\left( 1 \right)} {d\mu(q) \sum\limits_n {{a_n}} \left( x \right){{\left( {qt^{1/2}} \right)}^n}} \sum\limits_m {{b_m}} \left( x \right){\left( {{q^{ - 1}}t^{1/2}} \right)^m}. \\
\end{aligned}
\end{equation}
Applying such a transformation to the problem at hand, we define:
\begin{equation}
\label{eq:hl5.8}
\begin{aligned}
\overline {g_{HL}}  \left( {{x^*_i},t,{h_i}} \right) &\equiv \sum\limits_\lambda  {{HL_\lambda }\left( {{x^*_i},t} \right)} {h^\lambda }/{v_\lambda}(t),\\
\end{aligned}
\end{equation}
and
\begin{equation}
\label{eq:hl5.9}
\begin{aligned}
{v^{ - 1}}\left( {t,{h_i}} \right) &\equiv \sum\limits_\lambda  {{h^\lambda }} /{v_\lambda }\left( t \right).
\end{aligned}
\end{equation}
It then follows that we have the desired orthonormality relations:
\begin{equation}
\label{eq:hl5.10}
\begin{aligned}
\oint\limits_{U\left( N \right)} {d{\mu _{HL}}~\overline {g_{HL}} \left( {x_i^*,t,{h_i}} \right){HL_\lambda }\left( {{x_i},t} \right)}  = {h^\lambda },
\end{aligned}
\end{equation}
where the $\overline{g_{HL}} ({x^*_i},t,{h_i})$ can be calculated by the gluing procedure \ref{eq:hl5.7}:
\begin{equation}
\label{eq:hl5.11}
\begin{aligned}
\overline {g_{HL}} \left( {{x^*_i},t,{h_i}} \right) = \oint\limits_{U{{\left( 1 \right)}^N}} {d\mu ({{q_i}})}~ {v^{ - 1}}\left( {t,{h_i}^{1/2}{q_i^{-1}}} \right)g_{HL}\left( {{x^*_i},t,{h_i}^{1/2} q_i} \right).
\end{aligned}
\end{equation}
In this procedure, we introduce a dummy set of $U(1)^N$ coordinates $\{q_1,\ldots,q_N\}$ and map these to the Dynkin label fugacities $\left\{ {{h_1}, \ldots ,{h_N}} \right\} \to \left\{ {{q_1}{h_1}^{1/2}, \ldots ,{q_N}{h_N}^{1/2}} \right\}$. We map a conjugate set of $U(1)^N$ coordinates to the fugacities in the $v^{ - 1}(t,{h_i})$ generating function. Weyl integration using the $U{\left( 1 \right)^N}$ measure then selects singlets, for which the ${v_\lambda}$ factors exactly cancel.

The final input required for calculations is provided by the generating functions ${v^{ - 1}}\left( {t,{h_i}} \right)$. These are shown in Table \ref{table13} for some low rank unitary groups.
%
\begin{table}[htp]
\caption{Generating Functions for $1/{v_\lambda}(t)$}
\begin{center}
\begin{tabular}{|c|c|}
\hline
$ {Group}$&$ {{v^{ - 1}}\left( {t,{h_i}} \right)}$ \\
\hline
$ {U\left( 1 \right)}$&$ 1$ \\
\hline
$ {U\left( 2 \right)}$&$ {\frac{{1 + {h_1}t}}{{(1 - {h_1})(1 - {h_2})(1 + t)}}}$ \\
\hline
$ {U\left( 3 \right)}$&$ {\frac{{1 + {h_1}t + {h_2}t + {h_1}{t^2} + {h_2}{t^2} + {h_1}{h_2}{t^3}}}{{(1 - {h_1})(1 - {h_2})(1 - {h_3})(1 + t)\left( {1 + t + {t^2}} \right)}}}$ \\
\hline
$ {U\left( 4 \right)}$&$ {\frac{{\left( \begin{array}{c}
1 + {h_1}t + {h_2}t + {h_3}t + {h_1}{t^2} + 2{h_2}{t^2} + {h_3}{t^2} + {h_1}{h_3}{t^2}\\
 + {h_1}{t^3} + {h_2}{t^3} + {h_1}{h_2}{t^3} + {h_3}{t^3} + {h_1}{h_3}{t^3} + {h_2}{h_3}{t^3}\\
 + {h_2}{t^4} + {h_1}{h_2}{t^4} + 2{h_1}{h_3}{t^4} + {h_2}{h_3}{t^4}\\
 + {h_1}{h_2}{t^5} + {h_1}{h_3}{t^5} + {h_2}{h_3}{t^5} + {h_1}{h_2}{h_3}{t^6}
\end{array} \right)}}{{(1 - {h_1})(1 - {h_2})(1 - {h_3})(1 - {h_4}){{(1 + t)}^2}(1 + {t^2})(1 + t + {t^2})}}}$ \\
\hline
\end{tabular}
\end{center}
\label{table13}
\end{table}%
Generating functions for higher rank $U(N)$ groups can be obtained as required from the formula:
\begin{equation}
\label{eq:hl5.12}
\begin{aligned}
{{ v} ^{ - 1}}\left( {t,{h_i}} \right) = \prod\limits_{i = 1}^N {\frac{1}{{1 - {h_i}}} + \sum\limits_{\left[ n \right] = \left[ {0, \ldots ,0} \right]}^{\left[ {1, \ldots ,1} \right]} {\frac{1}{{{v_{\left[ n \right]}}\left( t \right)}}\prod\limits_{i = 1}^N {\frac{{h_i^{{n_i}}}}{{\left( {1 - {n_i}{h_i}} \right)}}} } },
\end{aligned}
\end{equation}
where the summation is carried out over all possible combinations of zero and unit Dynkin labels: $[ n] = \left\{ {\left[ {{n_1}, \ldots ,{n_N}} \right]:{n_i} = 0~or~1} \right\}$ and the ${{v_{\left[ n \right]}}(t)}$ follow from \ref{eq:hl5.4}.

The orthonormal generating functions $\overline{g_{HL}} ({x^*_i},t,{h_i})$ allow us to decompose any class function $F\left( {{x_i},t} \right)$ into a weighted sum of Hall-Littlewood polynomials. We first define the decomposition coefficients ${C_\lambda }(t)$ from:
\begin{equation}
\label{eq:hl5.13}
\begin{aligned}
F\left( {{x_i},t} \right) \equiv \sum\limits_\lambda ^{} {{C_\lambda }\left( t \right){HL_\lambda }\left( {{x_i},t} \right)}.
\end{aligned}
\end{equation}
We can then obtain a highest weight generating function ${C(t,h_i)}$ for the ${C_\lambda }(t)$ using the $\overline{g_{HL}}({x^*_i},t,{h_i})$ generating functions and the property \ref{eq:hl5.10}: 
\begin{equation}
\label{eq:hl5.14}
\begin{aligned}
{C(t,h_i)} &\equiv \sum\limits_\lambda ^{} {{C_\lambda }\left( t \right)} {h^\lambda } = \oint\limits_{U\left( N \right)} {d{\mu _{HL}}}~ {\overline{g_{HL}}}\left( {{x^*}_i,t,{h_i}} \right)F\left( {{x_i},t} \right).
\end{aligned}
\end{equation} 
Individual ${C_\lambda }(t)$ can be extracted from ${C(t,h_i)}$ by Taylor expansion, followed by matching the coefficients of the monomials ${h^\lambda}$. Furthermore, to establish consistency, we can also implement a second gluing procedure to recover the initial generating function $F\left( {{x_i},t} \right)$ from the generating functions ${C(t,h_i)}$ and $g_{HL}( {{x_i},t,{h_i}})$:
\begin{equation}
\label{eq:hl5.15}
\begin{aligned}
F\left( {{x_i},t} \right) = {\left. {\oint\limits_{U{{\left( 1 \right)}^N}} {d\mu }~C\left( {t,h/q_i} \right)g_{HL}\left( {{x_i},t,h {q_i}} \right)} \right|_{h = 1}}.
\end{aligned}
\end{equation} 

Having introduced the Hall-Littlewood polynomials, and shown how to construct their generating functions so that we can work with them, it is convenient, for the purpose of the construction of RSIMS, to follow the approach in \cite{Gadde:2011uv} and to define a modified set of symmetric functions that are closely related to the $HL_\lambda$, but which incorporate the fugacity $t$ in their denominators and are orthonormal under a different measure. Specifically, we rearrange the orthonormality relations:
\begin{equation}
\label{eq:hl5.16}
\begin{aligned}
\oint\limits_{U\left( N \right)} {d{\mu _{HL}}}~{HL_\lambda }\left( {{x_i},t} \right){\overline {HL _\mu} }({x^*}_i,t) = {\delta _{\lambda \mu }},
\end{aligned}
\end{equation}
as:
\begin{equation}
\label{eq:hl5.17}
\begin{aligned}
\oint\limits_{U\left( N \right)} {d{\mu _{mHL}}}~{mHL_\lambda }\left( {{x_i},t} \right){\overline {mHL_\mu}}({x^*}_i,t) = {\delta _{\lambda \mu }},
\end{aligned}
\end{equation}
where
\begin{equation}
\label{eq:hl5.18}
\begin{aligned}
\oint\limits_{U\left( N \right)} {d{\mu _{mHL}}}  \equiv \prod\limits_i {\oint\limits_{} {\frac{{d{x_i}}}{{{x_i}}}} } \frac{1}{{N!}}\left( {\prod\limits_{j \ne k} {\left( {1 - {x_j}/{x_k}} \right)} } \right)\left( {\prod\limits_{j \ne k} {\left( {1 - t{x_j}/{x_k}} \right)} } \right),
\end{aligned}
\end{equation}
and
\begin{equation}
\label{eq:hl5.19}
\begin{aligned}
{mHL_\lambda }\left( {{x_i},t} \right) \equiv {\left({\prod\limits_{j \ne k} {\frac{1}{{1 - t{x_j}/{x_k}}}} }\right)}~{HL_\lambda }\left( {{x_i},t} \right), \text {etc.}.
\end{aligned}
\end{equation}
The $mHL_\lambda$ functions have the same dependence on the partition $\lambda$ as the ${HL_\lambda }$ polynomials, but incorporate the plethystic function $PE[(adjoint-rank)~t]$ as a pre-factor. This has the effect of multiplying all the ${HL_\lambda }$ by symmetrisations of the adjoint. This feature can make the $mHL_\lambda$ functions extremely useful in the subgroup deconstruction of RSIMS, since the necessary symmetrisations of the adjoint irreps of the subgroup are automatically incorporated in the $mHL_\lambda$. This can, in certain cases, permit a dramatic reduction in the dimensions of the HWG describing an RSIMS deconstruction, as will be shown.

The generating functions ${g_{mHL}}({x_i},t,h_i)$ follow in a straightforward manner:
\begin{equation}
\label{eq:hl5.20}
\begin{aligned}
g_{mHL}\left( {{x_i},t,{h_i}} \right) \equiv \left( {\prod\limits_{j \ne k} {\frac{1}{{1 - t{x_j}/{x_k}}}} }\right)  g_{HL}\left( {{x_i},t,{h_i}} \right).
\end{aligned}
\end{equation}
In order to obtain the Hall-Littlewood polynomials of $SU(N)$, rather than $U(N)$, we need to make certain changes to the expressions \ref{eq:hl5.1} to \ref{eq:hl5.20}. First, we replace the coordinates $x_i$ of $U(N)$ by the monomials of the character of the $SU(N)$ fundamental. This substitution forces the last Dynkin label $n_N$ to zero; this label is conventionally dropped when describing irreps of $SU(N)$, although it needs to be reinstated when calculating the normalisation factors ${v_\lambda }(t)$. Finally, we replace the Haar measure of $U(N)$ by that of $SU(N)$.

\subsection {Modified Hall-Littlewood Polynomials and Characters of $SU(N)$}
All the three types of symmetric function studied herein (characters, HL and mHL) provide complete bases for the class functions of a group. It is useful to be able to express these functions in terms of each other. If we have knowledge of the coefficients (which are generally quotients of polynomials in $t$) for such decompositions, we can describe a moduli space in the most convenient basis, while retaining the ability to translate to the other bases. HWGs provide an efficient method both for encoding these relationships and for working with them.

The general prescription for the decomposition of an mHL polynomial into characters follows similar principles to \ref{eq:hl5.13} and \ref{eq:hl5.14}. Thus, suppose we wish to find the coefficients $C_{[ n],[ n']}(t)$ for the decomposition of $mHL_{[n]}(x, t)$ in terms of characters:
\begin{equation}
\label{eq:hl5.13a}
\begin{aligned}
{mHL_{\left[ n \right]}}\left( {x,t} \right) = \sum\limits_{\left[ {n'} \right]}^{} {{C_{\left[ n \right],\left[ {n'} \right]}}} \left( t \right){{{\cal X}}_{\left[ {n'} \right]}}\left( x \right).
\end{aligned}
\end{equation} 
We have already constructed generating functions, both for mHL polynomials and for characters:
\begin{equation}
\label{eq:hl5.14a}
\begin{aligned}
g_{mHL}^A\left( {x,t,h} \right) & \equiv \sum\limits_{\left[ n \right]}^{} {{mHL_{\left[ n \right]}}\left( {x,t} \right)}{h^n},\\
g_{{\cal X}}^A\left( {x,m} \right) & \equiv \sum\limits_{\left[ n' \right]}^{} {{{{\cal X}}_{\left[ n' \right]}}\left( x \right)}{m^{n'}}.
\end{aligned}
\end{equation}
So, we can use Weyl integration to combine these to yield a generating function for the $C_{[ n],[ n']}(t)$ coefficients:
\begin{equation}
\label{eq:hl5.14aa}
\begin{aligned}
g_{mHL \to \cal X}^A\left( {t,h,m} \right) & \equiv \sum\limits_{\left[ n \right]}^{} {\sum\limits_{\left[ {n'} \right]}^{} {{C_{\left[ n \right],\left[ {n'} \right]}}} \left( t \right)} {h^n}{m^{n'}}\\
&  = \oint\limits_A {d\mu } \,g_{mHL}^A\left( {x,t,h} \right)g_{{\cal X}}^A\left( {x^*,m} \right).
\end{aligned}
\end{equation}

To illustrate, we set out in Table \ref{table14} the HWGs $g_{mHL \to {\cal X}}^A$ for the decomposition of mHL polynomials for $SU(2)$ and $SU(3)$ into characters. Thus the HWG for the mHL of $SU(2)$ is the product of two factors  $1/(1-m^2 t)$ and $ 1/(1- h m)$, where $h$ is a fugacity for the Dynkin labels of $SU(2)$ mHL and $m$ is a fugacity for the Dynkin labels of $SU(2)$ characters. The first factor matches the HWG for the $SU(2)$ RSIMS. The second factor gives the dependence of the $SU(2)$ mHL on the characters of  $SU(2)$. So, for example, $mHL_{[1]}=([1]+[3]t+[5]t^2+[7]t^3+\ldots)$.

\begin{table}[htp]
\caption{HWGs for Decomposition of mHL into Characters}
\begin{center}
\begin{tabular}{|c|c|}
\hline
$ {Group}$& $g_{mHL(h) \to {\cal X}(m)}^A$ \\
\hline
$ {SU\left( 2 \right)}$&$ \frac{1}{(1- h m) \left(1-m^2 t\right)}$ \\
\hline
$ {SU\left( 3 \right)}$&$\frac{

\left( {\begin{array}{{c}}
1 + {m_1}{m_2}{t^2} - {h_1}{m_1}m_2^3{t^3} - {h_2}m_1^3{m_2}{t^3} - {h_1}{h_2}m_1^2m_2^2{t^2}\\
 + m_1^2m_2^2{t^4} - {h_1}m_1^3m_2^2{t^4} - {h_2}m_1^2m_2^3{t^4} + {h_1}{h_2}m_1^3m_2^3{t^4}\\
 - {h_1}m_1^2m_2^4{t^5} - {h_2}m_1^4m_2^2{t^5} + {h_1}{h_2}m_1^2m_2^5{t^5} + {h_1}{h_2}m_1^5m_2^2{t^5}\\
 + {h_1}{h_2}m_1^4m_2^4{t^6}
\end{array}} \right)

}{ (1-m_1 m_2 t) (1-m_1^3 t^3) (1-m_2^3 t^3) (1-h_1 m_1) (1-h_2 m_2) (1-h_1 m_2^2 t) (1-h_2 m_1^2 t)}$ \\
\hline
\end{tabular}
\end{center}
\label{table14}
\end{table}%

It is important to note that the HWGs which provide the inverse maps from characters to mHL polynomials are different, since the orthonormal mHL polynomials are not simply given by complex conjugation and the measure also differs. For example, the inverse HWG from characters of $SU(2)$ to mHL polynomials of $SU(2)$ is given by ${g}_{ {\cal X}(m) \to mHL(h) }^A=(1-h^2 t)/(1- h m)$, as can be verified.

These HWGs show that the mHL polynomials include the RSIMS factor ${1/(1 - {m_1}{m_r} t)}$. This can help to reduce the dimension of the HWG for the decomposition of an RSIMS into mHL polynomials, as discussed earlier. We can find other decompositions, as desired, by working in a similar manner with different combinations of HL, mHL and characters.

\subsection{Modified Hall-Littlewood Polynomials and $T(SU(N))$}
\begin{figure}
\begin{center}
\includegraphics[scale=0.6]{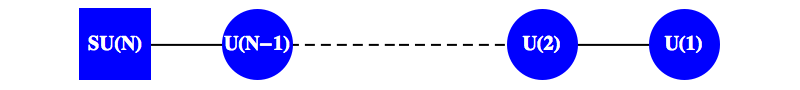}\\
\caption[Quiver diagram for $T(SU(N))$.]{The $T(SU(N))$ quiver consists of a $SU(N)$ flavour node connected to gauge nodes $U(N-1)$ through $U(1)$.}
\label{fig:tsun}
\end{center}
\end{figure}

One of the remarkable aspects of modified Hall-Littlewood polynomials is that they correspond to the Coulomb branches of SUSY ${\cal N}=4$ quiver gauge theories in 2+1 dimensions known as $T(SU(N))$ \cite{Cremonesi:2014kwa}. These Coulomb branch constructions have similarities with the RSIMS constructions described in Section \ref{sec:coulomb}. However, the leading node carries $SU(N)$ flavour charges and connects to a linear chain of gauge nodes carrying {\it monopole} charges from $U(N-1)$ down to $U(1)$ as in Figure \ref{fig:tsun}. These quivers are balanced (as described earlier).

Following \cite{Cremonesi:2014kwa}, we obtain the $T(SU(N))$ series of functions from such a quiver by adapting the Coulomb branch prescription, as set out in \ref{eq:mon3.6} to \ref{eq:mon3.8}, to include external charges described by a partition $\lambda \equiv ({\lambda_1,\ldots,\lambda_N})$. With a little further work, the construction can be rearranged into a recursive set of relations for $T(SU(N))$:
\begin{equation}
\label{eq:hl5.20b}
\begin{aligned}
&T\left( {SU\left( N \right)} \right)\left( {{\lambda},{z},t} \right)=\\
&{x^{\sum\limits_{j = 1}^N {{\lambda _j}} }}\sum\limits_{{q_1} \ge  \ldots  \ge {q_{N - 1}} \ge  - \infty }^\infty  {\left( \begin{array}{l}
{P_{U\left( {N - 1} \right)}}\left( {{q_1}, \ldots ,{q_{N - 1}}} \right){x^{ - \frac{N}{{N - 1}}\sum\limits_{i = 1}^{N - 1} {{q_i}} }}{t^{\sum\limits_{i = 1}^N {\sum\limits_{j = 1}^{N - 1} {\left| {{\lambda _i} - {q_j}} \right|/2} }  - \sum\limits_{i = 1}^{N - 1} {\sum\limits_{j = 1}^{i - 1} {\left| {{q_i} - {q_j}} \right|} } }}\\
\\
 \times ~~T\left( {SU\left( {N - 1} \right)} \right)\left( {{q_1}, \ldots ,{q_{N - 1}},{z_2}, \ldots ,{z_{N - 1}},t} \right)
\end{array} \right)}.
\end{aligned}
\end{equation}
In this formula, $z \equiv ({z_1,\ldots,z_{N-1}})$ is a system of $SU(N)$ simple roots, the CSA coordinate for the highest weight of the $SU(N)$ fundamental is:
\begin{equation}
\label{eq:hl5.20c}
\begin{aligned}
x& = {\left( {\prod\limits_{i = 1}^{N - 1} {z_i^{N - i}} } \right)^{\frac{1}{N}}},
\end{aligned}
\end{equation}
and the symmetry factors, which depend on each partition of gauge field charges $({q_1,\ldots,q_{N}})$, are given by:
\begin{equation}
\label{eq:hl5.20c}
\begin{aligned}
{P_{U\left( N \right)}}\left( {{q_1}, \ldots ,{q_{N }}} \right) = \prod\limits_{i = 1}^{N } {\frac{1}{{1 - {t^{{d_i}\left( {{q_1}, \ldots ,{q_{N}}} \right)}}}}}.
\end{aligned}
\end{equation}
The recursion relations assume the $({q_1,\ldots,q_N})$ form an ordered partition, but range over both positive and negative integers. In each case the summation corresponds to one of the gauge nodes. We set $T(SU(1))=1$ and it then follows that the first non-trivial member of the series is: 
\begin{equation}
\label{eq:hl5.20a}
\begin{aligned}
T\left( {SU\left( 2 \right)} \right)\left( {{\lambda _1},{\lambda _2},{z_1},t} \right) & = x^{\left( {{\lambda _1} + {\lambda _2}} \right)}\frac{1}{{\left( {1 - t} \right)}}\sum\limits_{q =  - \infty }^\infty  {x^{-2q}{t^{\left| {{\lambda _1} - q} \right|/2 + \left| {{\lambda _2} - q} \right|/2}}},
\end{aligned}
\end{equation}
where $z_1=x^2$.

As shown in \cite{Cremonesi:2014kwa}, the $T(SU(N))$ quivers correspond to modified Hall-Littlewood polynomials of $SU(N)$ if the flavour partition $({\lambda_1,\ldots,\lambda_N})$ is chosen such that $\lambda_N=0$, whereupon all the other partition labels are non-negative and $({\lambda_1,\ldots,\lambda_{r=N-1}})$ map to highest weight Dynkin labels $[n_1,\ldots,n_{r}]$ for $SU(N)$, through the relationship $\left( {{\lambda _1}, \ldots ,{\lambda _j}, \ldots ,{\lambda _{r}}} \right) = \left( {\sum\limits_{i = 1}^{r} {{n_i}, \ldots ,\sum\limits_{i = j}^{r} {{n_i}, \ldots ,{n_{r}}} } } \right)$. The correspondence is modulated by a pre-factor, so that the precise relationship is:

\begin{equation}
\label{eq:hl5.20d}
\begin{aligned}
mH{L_{\left[ {{n_1}, \ldots ,{n_r}} \right]}}\left( {{z},t} \right) = {t^{ -\rho. {G_{ij}} . \left[ {{n_1}, \ldots ,{n_r}} \right]}}T\left( {SU\left( r+1 \right)} \right)\left( {\lambda \left( {{n_1}, \ldots ,{n_r}} \right),{z},t} \right).
\end{aligned}
\end{equation}
The exponent of the pre-factor is given by the contraction of the Weyl vector $\rho$, which is $(1,\ldots ,1)$ in a canonical basis of CSA coordinates, with the Dynkin labels of the mHL polynomial, using the group metric tensor $G_{ij}$.\footnote{While mHL polynomials with similar properties can be defined for other groups, the $T(G)$ quiver theories that have been proposed for these polynomials, other than for isomorphisms with the A series, face some critical issues.} We return to the subject of $T(SU(N))$ in the concluding Section.

\subsection{Extended Dynkin Diagrams and A Series Subgroups}
\begin{figure}[htbp]
\begin{center}
\includegraphics[scale=0.75]{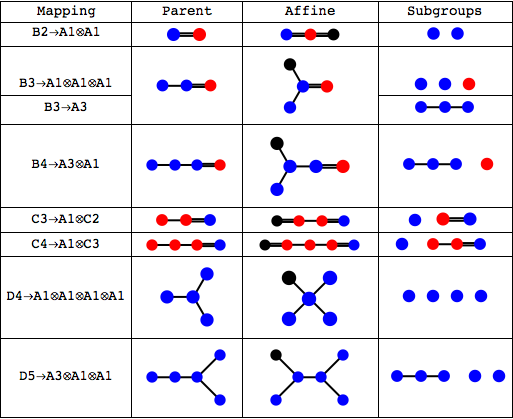}\\
\caption[Dynkin diagrams for BCD group mappings into A series subgroups.]{Mappings of selected BCD groups via their extended or affine Dynkin diagrams into A series subgroups by single elementary transformations. Blue nodes denote long roots with length 2. Red nodes denote short roots. A black node denotes the long root added in the extended Dynkin diagram. The eliminated root is uniquely determined (up to graph automorphisms) by the subgroups. Rank is preserved. (Multiple elementary transformations can be used to map C series groups fully.)}
\label{branchingBCD}
\end{center}
\end{figure}

\begin{figure}[htbp]
\begin{center}
\includegraphics[scale=0.60]{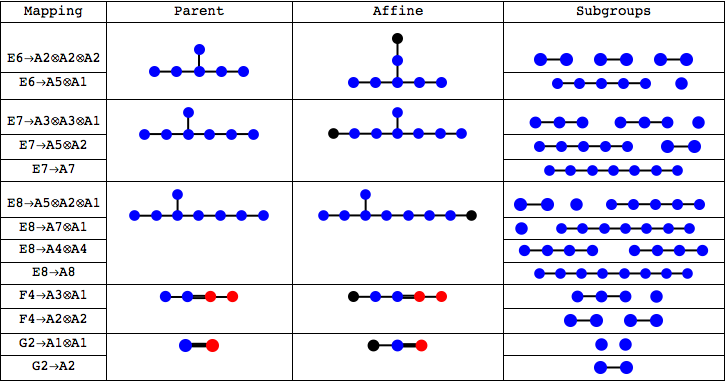}\\
\caption[Dynkin diagrams for EFG group mappings into A series subgroups.]{Mappings of Exceptional groups via their extended or affine Dynkin diagrams into A series subgroups by single elementary transformations. Blue nodes denote long roots with length 2. Red nodes denote short roots. A black node denotes the long root added in the extended Dynkin diagram. The eliminated root is uniquely determined (up to graph automorphisms) by the subgroups. Rank is preserved.}
\label{branchingEFG}
\end{center}
\end{figure}

We have seen in Section \ref{sec:weylbranch} how the RSIMS of a Classical or Exceptional group can be decomposed in terms of  the irreps of a subgroup. In order to explore RSIMS decompositions in terms of the mHL of $SU(N)$ we must work with regular A series subgroups. These are not generally maximal. Proceding as before, we describe a selection of the relevant elementary transformations in Figures \ref{branchingBCD} and \ref{branchingEFG}, which give the Dynkin diagrams, and in Table \ref{table15}, which shows the resulting branching of the adjoint representation of the parent into subgroup irreps. In the case of C series groups of rank greater than two, each elementary transformation splits off a single $A_1$ subgroup, and therefore multiple such elementary transformations are in general required to map a C series group to its A series subgroups.

We have not included in Table \ref{table15} the elementary transformation of A series groups into themselves. We have however included non-maximal subgroups that can only be reached via an intermediary subgroup, such as ${C_3} \to {C_2} \otimes {A_1} \to {A_1} \otimes {A_1} \otimes {A_1}$. It follows that, by using multiple elementary transformations, any group can be mapped into one or more regular A series simple or semi-simple subgroups.

Importantly, each such mapping establishes a diffeomorphism between the CSA coordinates of the parent group and those of its subgroups. However, while the coordinate map is bijective, the mapping of irreps from the parent group into the irreps of the A series product group is only injective; one cannot generally map all the representations of the subgroup back to those of the parent; this is possible only for specific representations (such as those arising in the RSIMS construction).

\begin{sidewaystable}
\caption{Branching of Adjoint Representation to A Series Subgroups}
\begin{center}
\begin{tabular}{|c|c|c|c|}
\hline
$ {Group}$&$ {Adjoint }$&$ {Subgroup}$&$ {Branching}$ \\
\hline
$ {{B_2} \cong {C_2}}$&$ {\left[ {0,2} \right]}$&$ {{A_1} \otimes {A_1}}$&$ {\left[ 2 \right] + \left[ 2 \right] + \left[ 1 \right]\left[ 1 \right]}$ \\
\hline
$ {{B_3}}$&$ {\left[ {0,1,0} \right]}$&$ {{A_1} \otimes {A_1} \otimes {A_1}}$&$
\left[ 2 \right] + \left[ 2 \right] + \left[ 2 \right]+ \left[ 1 \right]\left[ 2 \right]\left[ 1 \right]$\\
\hline
$ {{B_3}}$&$ {\left[ {0,1,0} \right]}$&$A_3$&$[1,0,1] + [0,1,0]$\\
\hline
$ {{B_4}}$&$ {\left[ {0,1,0,0} \right]}$&$ {{A_3} \otimes {A_1}}$&$ 
\left[ {1,0,1} \right] + \left[ 2 \right]+ \left[ {0,1,0} \right]\left[ 2 \right]$ \\
\hline
 $ {{C_3}}$&$ {\left[ {2,0,0} \right]}$&$  {A_1} \otimes {A_1} \otimes {A_1}$&$ 
{[2] + [2] + [2] + [1][1] + [1][1] + [1][1]}$ \\
\hline
$ {{C_4}}$&$ {\left[ {2,0,0,0} \right]}$&$ 
 {A_1} \otimes {A_1} \otimes {A_1} \otimes {A_1}
 $&$
{ [2]+[2]+[2]+[2]+[1][1]+[1][1]+[1][1]+[1][1]+[1][1]+[1][1]}
$ \\
\hline
$ {{D_4}}$&$ {\left[ {0,1,0,0} \right]}$&$ {{A_1} \otimes {A_1} \otimes {A_1} \otimes {A_1}}$&$ \begin{array}{c}
\left[ 2 \right] + \left[ 2 \right] + \left[ 2 \right] + \left[ 2 \right]\\
 + \left[ 1 \right]\left[ 1 \right]\left[ 1 \right]\left[ 1 \right]
\end{array}$ \\
\hline
$ {{D_5}}$&$ {\left[ {0,1,0,0,0} \right]}$&$ {{A_3} \otimes {A_1} \otimes {A_1}}$&$ {\left[ {1,0,1} \right] + \left[ 2 \right] + \left[ 2 \right] + \left[ {0,1,0} \right]\left[ 1 \right]\left[ 1 \right]}$ \\
\hline
$ {{E_6}}$&$ {\left[ {0,0,0,0,0,1} \right]}$&$ {{A_2} \otimes {A_2} \otimes {A_2}}$&$ \begin{array}{c}
\left[ {1,1} \right] + \left[ {1,1} \right] + \left[ {1,1} \right]+ \left[ {1,0} \right]\left[ {1,0} \right]\left[ {1,0} \right] + \left[ {0,1} \right]\left[ {0,1} \right]\left[ {0,1} \right]
\end{array}$ \\
\hline
$ {{E_7}}$&$ {\left[ {1,0,0,0,0,0,0} \right]}$&$ {{A_3} \otimes {A_3} \otimes {A_1}}$&$ \begin{array}{c}
\left[ {1,0,1} \right] + \left[ {1,0,1} \right] + \left[ 2 \right]\\
 + \left[ {1,0,0} \right]\left[ {1,0,0} \right]\left[ 1 \right] + \left[ {0,0,1} \right]\left[ {0,0,1} \right]\left[ 1 \right] + \left[ {0,1,0} \right]\left[ {0,1,0} \right]
\end{array}$ \\
\hline
$ {{E_8}}$&$ {\left[ {0,0,0,0,0,0,1,0} \right]}$&$ {{A_5} \otimes {A_2} \otimes {A_1}}$&$ \begin{array}{c}
\left[ {1,0,0,0,1} \right] + \left[ {1,1} \right] + \left[ 2 \right]\\
 + \left[ {1,0,0,0,0} \right]\left[ {1,0} \right]\left[ 1 \right] + \left[ {0,0,0,0,1} \right]\left[ {0,1} \right]\left[ 1 \right]\\
 + \left[ {0,1,0,0,0} \right]\left[ {0,1} \right] + \left[ {0,0,0,1,0} \right]\left[ {1,0} \right] + \left[ {0,0,1,0,0} \right]\left[ 1 \right]
\end{array}$ \\
\hline
$ {{F_4}}$&$ {\left[ {1,0,0,0} \right]}$&$ {{A_3} \otimes {A_1}}$&$ {\left[ {1,0,1} \right] + \left[ 2 \right] + \left[ {0,1,0} \right]\left[ 2 \right] + \left[ {1,0,0} \right]\left[ 1 \right] + \left[ {0,0,1} \right]\left[ 1 \right]}$ \\
$ {}$&$ {}$&$ {{A_2} \otimes {A_2}}$&${\left[ {1,1} \right] + \left[ {1,1} \right] + \left[ {1,0} \right]\left[ {2,0} \right] + \left[ {0,1} \right]\left[ {0,2} \right]}$ \\
\hline
$ {{G_2}}$&$ {\left[ {1,0} \right]}$&$ {{A_1} \otimes {A_1}}$&$ {\left[ 2 \right] + \left[ 2 \right] + \left[ 3 \right]\left[ 1 \right]}$ \\
$ {{}}$&$ $&$ {{A_2}}$&$ {\left[1,1 \right] + \left[ 1,0 \right] + \left[0,1 \right]}$ \\
\hline
\end{tabular}
\end{center}
\text{Singlets omitted from descriptions of product group decompositions.}\\
\text{This list contains examples and is not exhaustive}\\
\label{table15}
\end{sidewaystable}
\FloatBarrier
\subsection{RSIMS Decomposition to modified Hall-Littlewood polynomials}
We are now ready to show how modified Hall-Littlewood polynomials can be deployed, together with the branching relations described above, to construct the RSIMS for any group out of mHL polynomials. We start by generalising the three way schema given in \cite{Gadde:2011uv} and define the decomposition:
\begin{equation}
\label{eq:hl5.21}
\begin{aligned}
g_{instanton}^{G} \left( {t,{x_i}} \right) &\equiv \sum\limits_{k = 0}^\infty  {\left[ {k~adj} \right]} ~{t^k},\\
g_{instanton}^{G \to A \otimes B \otimes  \cdots D}\left( {t,{y_i}} \right)& = \sum\limits_{\left[ {{n_\lambda }} \right]\left[ {{n_\mu }} \right] \ldots \left[ {{n_\sigma }} \right]}^{} C_{\lambda ,\mu , \ldots ,\sigma }(t) ~{mHL_{\left[ {{n_\lambda }} \right]}^A}\left( {y_i,t} \right)~{mHL_{\left[ {{n_\mu }} \right]}^B}\left( {y_j,t} \right) \cdots {mHL_{\left[ {{n_\sigma }} \right]}^D}\left( {y_k,t} \right).
\end{aligned}
\end{equation}
The characters of representations of the parent group $G$, of rank $r$, correspond to characters of the semi-simple group ${{A } \otimes {B} \otimes  \cdots {D }}$, and so $g_{instanton}^{G }$ can be expressed using the subgroup CSA coordinates $\{y_i:i=1,\ldots,r\}$. The $C_{\lambda ,\mu , \ldots ,\sigma }(t)$ coefficients range over all irreps $\left\{ {\left[ {{n_\lambda }} \right], \left[ {{n_\mu }} \right] ,\ldots ,\left[ {{n_\sigma }} \right]}\right\}$ of the respective subgroups, identified by their Dynkin labels or partitions.

We derive the $C_{\lambda ,\mu , \ldots ,\sigma }(t)$ coefficients using a general procedure that gives the decomposition of the RSIMS of any group into modified Hall-Littlewood polynomials of subgroups. To do this, we exploit the fact that $g_{instanton}^G(t,{x_i})$ has a known generating function, and so we can use the generating functions $g_{mHL}$ and their orthonormal conjugates $\overline {g_{mHL}}$, described earlier, to obtain a generating function for the $C_{\lambda ,\mu , \ldots,\sigma }(t)$. This follows from \ref{eq:hl5.21} as:
\begin{equation}
\label{eq:hl5.22}
\begin{aligned}
C\left( {{h_{{A}}},{h_{{B}}}, \ldots ,{h_{{D}}},t} \right) &\equiv \sum\limits_{\left[ {{n_\lambda }} \right]\left[ {{n_\mu }} \right] \cdots \left[ {{n_\sigma }} \right]}^{} {{C_{\lambda ,\mu , \ldots ,\sigma }}\left( t \right){h_A}^{n_\lambda} {h_B}^{n_\mu}  \cdots {h_D}^{n_\sigma} }\\
&  = \oint\limits_{A \otimes B \otimes  \cdots D} {d{\mu _{mHL}}}~\overline{ g_{mHL}^A} \left( {{y_i}^*,t,{h_{{A}}}} \right)\overline {g_{mHL}^B} \left( {{y_j}^*,t,{h_{{B}}}} \right) \cdots \\
&~~~~~\times ~~~~~~~\overline{ g_{mHL}^D} \left( {{y_k}^*,t,{h_{{D}}}} \right)g_{instanton}^{G \to A \otimes B \otimes  \cdots D}\left( {t,{y_i}} \right).\\
\end{aligned}
\end{equation}
The expression \ref{eq:hl5.22} can be evaluated to obtain a rational function in terms of the fugacities $\{ {h_{{A}}},{h_{{B}}}, \ldots ,{h_{{D}}},t\}$. Individual $C_{\lambda ,\mu , \ldots ,\sigma }(t)$ coefficients can be extracted by equating powers in $\{ {h_{{A}}},{h_{{B}}}, \ldots ,{h_{{D}}}\}$ following Taylor expansion. A key advantage of this approach is that the generating function gives the $C_{\lambda ,\mu , \ldots ,\sigma }(t)$ to all orders in $t$.  We focus on constructions that map RSIMS to semi-simple A series subgroups and their $g_{mHL}$ functions.
\subsubsection{$D_4$ Example}
\label{sec:D4ex}
We outline below the construction of the RSIMS of $D_4$ from the $mHL_\lambda$ functions of four $A_1$ subgroups. Specifically, we wish to calculate the coefficients $C_{\lambda, \mu, \nu, \rho}(t)$ such that:
\begin{equation}
\label{eq:hl5.23}
\begin{aligned}
g_{instanton}^{{D_4}}\left( {a,b,c,d,t} \right) = \sum\limits_{\lambda ,\mu ,\nu ,\rho } {{C_{\lambda, \mu, \nu ,\rho }}\left( t \right)} ~{{mHL_\lambda}^{A_1} }\left( {a,t} \right)~{{mHL_\mu}^{A_1} }(b,t)~{{mHL_\nu}^{A_1} }(c,t)~{{mHL_\rho}^{A_1} }(d,t).
\end{aligned}
\end{equation}
We start with the expansion for $g_{instanton}^{D_4}\left( {w,x,y,z,t} \right)$ obtained by the methods in Section \ref{sec:plethystics}, where $\{w,x,y,z,t \}$ are CSA coordinates for $D_4$ (we do not show this here since it is rather lengthy). By eliminating the second node, we obtain the root and CSA coordinate mappings in Table \ref{table16} from the extended Cartan matrix for $D_4$ and the Cartan matrices for the four $A_1$ subgroups.

\begin{table}[htp]
\caption{$D_4$ to ${A_1}^{\otimes 4}$ root and CSA coordinate mappings}
\begin{center}
\begin{tabular}{|cc|cc|}
\hline
$ \begin{array}{c}{D_4}\\roots\end{array}
$&$\begin{array}{c}{D_4}\\coords\end{array}
$&$ \begin{array}{c}{A_1} \otimes {A_1} \otimes {A_1} \otimes {A_1}\\roots\end{array}
$&$\begin{array}{c}{A_1} \otimes {A_1} \otimes {A_1} \otimes {A_1}\\coords\end{array}$ \\
\hline
$ {{z_1}}$&${{w^2}x}$&$ {{z_a}}$&${{a^2}}$ \\
$ {{z_2}}$&${{x^2}/wyz}$&$  - $&$ - $ \\
$ {{z_3}}$&${{y^2}/x}$&$ {{z_b}}$&${{b^2}}$ \\
$ {{z_4}}$&${{z^2}/x}$&$ {{z_c}}$&${{c^2}}$ \\
$ {{z_0}}$&${1/x}$&$ {{z_d}}$&${{d^2}}$ \\
\hline
\end{tabular}
\end{center}
\label{table16}
\end{table}
We solve the root mapping to obtain the coordinate mapping $\left\{w\to \frac{a}{d},x\to \frac{1}{d^2},y\to \frac{b}{d},z\to \frac{c}{d}\right\}$ and use this to transform $g_{instanton}^{D_4}\left( {w,x,y,z,t} \right)$ to $g_{instanton}^{D_4}\left( {a,b,c,d,t} \right)$. We then introduce generating functions for the $\overline {g_{mHL}}$ using the Dynkin label fugacities ${h_A,h_B,h_C,h_D}$ and specialise \ref{eq:hl5.22} as:
\begin{equation}
\label{eq:hl5.24}
\begin{aligned}
C({h_A},{h_B},{h_C},{h_D},t) &\equiv \sum\limits_{\lambda ,\mu ,\nu ,\rho  = 0}^\infty  {{C_{\lambda ,\mu ,\nu ,\rho }}\left( t \right)~{h_A}^\lambda {h_B}^\mu {h_C}^\nu {h_D}^\rho } \\
& = \oint\limits_{{A_1} \otimes {A_1} \otimes {A_1} \otimes {A_1}} {d{\mu _{mHL}}}~\overline {g_{mHL}^{A_1}}\left( {a^*,t,{h_A}} \right)~\overline {g_{mHL}^{A_1}} \left( {b^*,t,{h_B}} \right)~\overline {g_{mHL}^{A_1}} \left( {c^*,t,{h_C}} \right) \\
&~~~~~\times~~~~~\overline {g_{mHL}^{A_1}} \left( {d^*,t,{h_D}} \right) \times g_{instanton}^{{D_4}}\left( {a,b,c,d,t} \right).\\
\end{aligned}
\end{equation}
For $A_1$, the Hall-Littlewood polynomials follow from \ref{eq:hl5.1} and can be expressed in terms of characters $[n]$ as:
\begin{equation}
\label{eq:hl5.25}
\begin{aligned}
{HL_{[n]}^{A_1}}\left( {{{\cal X}},t} \right) &={ \left\{ \begin{array}{l}
n = 0:1 + t\\
n = 1:\left[ 1 \right]\\
n \ge 2:\left[ n \right] - t\left[ {n - 2} \right]
\end{array} \right.}.\\
\end{aligned}
\end{equation}
Their generating function follows from  \ref{eq:hl5.5} and can be encoded as a highest weight generating function, using $h$ as the HL Dynkin label fugacity: 
\begin{equation}
\label{eq:hl5.26}
\begin{aligned}
g_{HL}^{A_1}\left( {{{\cal X}},t,h} \right){\rm{ }} = PE\left[ {\left[ 1 \right]h} \right]\left( {1 + t - h t\left[ 1 \right]} \right).
\end{aligned}
\end{equation}
The conjugate orthonormal Hall-Littlewood polynomials  $\overline{ {HL}_{[n]}^{A_1}}$ follow from \ref{eq:hl5.11} as:
\begin{equation}
\label{eq:hl5.27}
\begin{aligned}
{\overline {{HL} _{\left[ n \right]}^{A_1}}} \left( {{\cal X},t} \right) &={ \left\{ \begin{array}{l}
n = 0:1\\
n = 1:\left[ 1 \right]\\
n \ge 2:\left[ n \right] - t\left[ {n - 2} \right]
\end{array} \right.}.\\
\end{aligned}
\end{equation}
The generating function for the  $\overline{ {g} _{HL}^{A_1}}$ differs from \ref{eq:hl5.26} for the $ {g} _{HL}^{A_1}$ in its numerator:
\begin{equation}
\label{eq:hl5.28}
\begin{aligned}
\overline{ {g} _{HL}^{A_1}} \left( {{\cal X},t,h} \right) &= PE\left[ {\left[ 1 \right]h} \right]\left( {1 - {h^2}t} \right).\\
\end{aligned}
\end{equation}
The modified Hall-Littlewood polynomials ${mHL_{[n]}^{A_1}}$, ${\overline {{mHL}_{[n]}^{A_1}}}$ and their generating functions all differ from the above by the pre-factor, $PE[[2]t-t]$:
\begin{equation}
\label{eq:hl5.29}
\begin{aligned}
\begin{array}{l}
 {g} _{mHL}^{A_1} \left( {{\cal X},t,h} \right) = PE[[2]t-t]~{g} _{HL}^{A_1} \left( {{\cal X},t,h} \right),\\
\overline{ {g}_{mHL}^{A_1}} \left( {{\cal X},t,h} \right) =PE[[2]t-t] ~\overline {{g}_{HL}^{A_1}} \left( {{\cal X},t,h} \right).\\
\end{array}
\end{aligned}
\end{equation}
We can evaluate \ref{eq:hl5.24} by taking the conjugate generating functions from \ref{eq:hl5.29}, expanding the characters, and applying Weyl integration to obtain:
\begin{equation}
\label{eq:hl5.30}
C\left( {{h_A},{h_B},{h_C},{h_D},t} \right) = \frac{{1 - {h^2_A}{h^2_B}{h^2_C}{h^2_D}{t^4}}}{{(1 - {t^2})(1 - {h_A}{h_B}{h_C}{h_D}t)(1 - {h_A}{h_B}{h_C}{h_D}{t^2})}}.
\end{equation}
This simple HWG is of a diagonal form, in which the Dynkin label fugacities of different subgroups always appear with matching exponents. Taylor series expansion yields the explicit non-zero $C_\lambda (t)$ coefficients:
\begin{equation}
\label{eq:hl5.31}
{C_{\left[ n \right]\left[ n \right]\left[ n \right]\left[ n \right]}}\left( t \right) = \left\{ {\begin{array}{*{20}{l}}
{n = 0:{{\left( {1 - {t^2}} \right)}^{ - 1}}}\\
{n \ge 1:{t^n}{{\left( {1 - t} \right)}^{ - 1}}}.
\end{array}} \right.
\end{equation}

These can be checked by substitution back into \ref{eq:hl5.23} followed by Taylor expansion or gluing to recover the RSIMS for $D_4$. The coefficients follow a pattern related to the U(N) symmetry dressing factors discussed in Section \ref{sec:coulomb}.

\subsubsection{Branching Coefficients for RSIMS}

We can repeat the procedure described for $D_4$ for a selection of lower rank Classical and Exceptional groups. We summarise the results in Tables \ref{table17} to \ref{table22}, giving both the generating functions from which the $C_\lambda$ coefficients can be extracted, and the values for a selection of the $C_\lambda$ coefficients themselves. The denominators of the generating functions for the $C_\lambda$ express the generators of the series in terms of highest weight fugacities. The numerators of the generating functions encode a finite set of relations.

\begin{sidewaystable}[htp]
\caption{Coefficients of modified HL polynomials for A series RSIMS}
\begin{center}
\begin{tabular}{|c|c|c|}
\hline
$ Group $&$ HWG~for~C_\lambda $&$ C_\lambda~Coefficients$ \\
\hline
$ {{A_1}}$&$ 1$&$ {{C_{\left[ 0 \right]}} = 1}$ \\
\hline
$ {{A_2}}$&$ {1 - {t^2}{h_1}{h_2}}$&$ {\left\{ \begin{array}{l}
{C_{\left[ {0,0} \right]}} = 1\\
{C_{\left[ {1,1} \right]}} =  - {t^2}
\end{array} \right.}$ \\
\hline
$ {{A_3}}$&$ {\left( \begin{array}{l}
1\\
 - {t^2}\left( {1 - t} \right)h_2^2\\
 - {t^2}\left( {1 + t} \right){h_1}{h_3}\\
 + {t^3}\left( {h_1^2{h_2} + {h_2}{h_3}^2} \right)\\
 - {t^4}{h_1}^2{h_3}^2
\end{array} \right)}$&$ {\left\{ \begin{array}{l}
{C_{\left[ {0,0,0} \right]}} = 1\\
{C_{\left[ {0,2,0} \right]}} =  - {t^2}(1 - t)\\
{C_{\left[ {1,0,1} \right]}} =  - {t^2}(1 + t)\\
{C_{\left[ {0,1,2} \right]}} = {C_{\left[ {2,1,0} \right]}} = {t^3}\\
{C_{\left[ {2,0,2} \right]}} =  - {t^4}
\end{array} \right.}$ \\
\hline
$ {{A_4}}$&
${\left( \begin{array}{l}
1\\
 - \left( {{h_2}{h_3} + {h_1}{h_4}} \right){t^2}\\
 + \left( {{h_1}{h_2}^2 + {h_1}^2{h_3} - {h_1}{h_4} + {h_3}^2{h_4} + {h_2}{h_4}^2} \right){t^3}\\
 + \left( \begin{array}{l}
 - {h_1}^3{h_2} + {h_1}^2{h_3} + {h_2}{h_3} - {h_1}{h_4}\\
 - {h_1}{h_2}{h_3}{h_4} - {h_1}^2{h_4}^2 + {h_2}{h_4}^2 - {h_3}{h_4}^3
\end{array} \right){t^4}\\
 + \left( \begin{array}{l}
 - {h_1}{h_2}^2 + {h_2}{h_3} + {h_1}^3{h_3}{h_4} - 2{h_1}{h_2}{h_3}{h_4}\\
 - {h_3}^2{h_4} - {h_1}^2{h_4}^2 + {h_1}{h_2}{h_4}^3
\end{array} \right){t^5}\\
 + \left( {{h_2}^2{h_3}^2 + {h_1}^3{h_3}{h_4} - {h_1}^3{h_4}^3 + {h_1}{h_2}{h_4}^3} \right){t^6}\\
 + \left( { - {h_1}^2{h_2}{h_3}^2 + {h_1}{h_2}{h_3}{h_4} - {h_2}^2{h_3}{h_4}^2} \right){t^7}\\
 + {h_1}^2{h_2}{h_3}{h_4}^2{t^8}
\end{array} \right)}$
&$ {\left\{ \begin{array}{l}
{C_{\left[ {0,0,0,0} \right]}} = 1\\
{C_{\left[ {1,0,0,1} \right]}} =  - {t^2}\\
{C_{\left[ {0,1,1,0} \right]}} =  - {t^2}\\
 \ldots  \ldots \\
 {C_{\left[ {2,1,1,2} \right]}} =  {t^8}\\
 
\end{array} \right.}$ \\
\hline
\end{tabular}
\end{center}
\text{Some $C_\lambda$ coefficients for $A_4$ are omitted for brevity.}
\label{table17}
\end{sidewaystable}
\begin{sidewaystable}[htp]
\caption{Coefficients of modified HL polynomials for B series RSIMS}
\begin{center}
\begin{tabular}{|c|c|c|c|}

\hline
$ {Group}$&$ \begin{array}{l}
Branching\\
(Fugacities)
\end{array}$&$ {HWG~for~{C_\lambda }}$&$ {{C_\lambda}~Coefficients}$ \\
\hline
$ {{B_1}}$&$ {{A_1}}$&$ 1$&$ {{C_{\left[ 0 \right]}} = 1}$ \\
\hline
$ {{B_2}}$&$ \begin{array}{c}
{A_1} \otimes {A_1}\\
({h_A}, {h_B})
\end{array}$&$ {1 + {h_A}{h_B}t}$&$ {\left\{ \begin{array}{c}
{C_{\left[ 0 \right]\left[ 0 \right]}} = 1\\
{C_{\left[ 1 \right]\left[ 1 \right]}} = t
\end{array} \right.}$ \\
\hline

$ {{B_3}}$
&
$\begin{array}{*{20}{c}}
{{A_3}}\\
{({h_1},{h_2},{h_3})}
\end{array}$
&
$\begin{array}{l}
1 + {h_2}t - h_1^2{t^2} - h_3^2{t^2} - {h_1}{h_3}{t^2}\\
 - {h_1}{h_3}{t^3} + h_2^2{t^3} + {h_1}{h_2}{h_3}{t^4}
\end{array}$
&
$\left\{ {\begin{array}{*{20}{l}}
\begin{array}{l}
{C_{\left[ 0 \right]\left[ 0 \right]\left[ 0 \right]}} = 1\\
 \ldots 
\end{array}\\
{{C_{\left[ 1 \right]\left[ 1 \right]\left[ 1 \right]}} = {t^4}}
\end{array}} \right.$ \\
\hline

$ {{B_3}}$
&
$ \begin{array}{c}
{A_1} \otimes {A_1} \otimes {A_1}\\
({h_A}, {h_B}, {h_C})
\end{array}$
&
$\begin{array}{c}
\frac{{(1 + {h}{t^2})}}{{\left( {1 - {t^2}} \right)\left( {1 - {h}t} \right)}}\\
where~{h} \equiv {h_A}{h_B}^2{h_C}
\end{array}$
&
$ {\left\{ \begin{array}{l}
{C_{\left[ 0 \right]\left[ 0 \right]\left[ 0 \right]}} = 1/\left( {1 - {t^2}} \right)\\
{C_{\left[ n \right]\left[ {2n} \right]\left[ n \right]}} = {t^n}/\left( {1 - t} \right)
\end{array} \right.}$ \\
\hline

$ {{B_4}}$
&
$ \begin{array}{c}
{A_1} \otimes {A_1} \otimes {A_1} \otimes {A_1}\\(h_A, {h_B},{h_C},{h_D})
\end{array}$
&
$
\frac{{\left( \begin{array}{c}
1 + {h_A}{h_B}{t^2} + {h_C}{h_D}{t^2} + {h_A}{h_B}{h_C}{h_D}{t^2}\\
 - {h_A}{h_B}{h_C}{h_D}{t^3} - {h_A}{h_B}{h_C}^2{h_D}^2{t^3}\\
 - {h_A}^2{h_B}^2{h_C}{h_D}{t^3} - {h_A}^2{h_B}^2{h_C}^2{h_D}^2{t^5}
\end{array} \right)}}{{{{\left( {1 - {t^2}} \right)}^2}(1 - {h_A}{h_B}t)(1 - {h_C}{h_D}t)(1 - {h_A}{h_B}{h_C}{h_D}t)}}
$
&
$ 
\left\{ {\begin{array}{*{20}{l}}
\begin{array}{l}
{C_{\left[ 0 \right]\left[ 0 \right]\left[ 0 \right]\left[ 0 \right]}} = 1/{\left( {1 - {t^2}} \right)^2}\\
{C_{\left[ {n > 0} \right]\left[ {n } \right]\left[ 0 \right]\left[ 0 \right]}} = {t^n}/\left( {\left( {1 - t} \right)\left( {1 - {t^2}} \right)} \right)\\
{C_{\left[ n \right]\left[ n \right]\left[ m \right]\left[ m \right]}} = {t^{Max[m,n]}}/{\left( {1 - t} \right)^2}
\end{array}
\end{array}} \right.
$ \\
\hline

$ {{B_4}}$
&
$ \begin{array}{c}
{A_1} \otimes {A_3}\\(h, {h_1},{h_2},{h_3})
\end{array}$
&
$ {\frac{{\left( \begin{array}{l}
1\\
 + \left( \begin{array}{l}
 - {h^2}{h_1}^2 + {h^2}{h_2} - {h_1}{h_3}\\
 - {h^2}{h_3}^2
\end{array} \right){t^2}\\
 + \left( \begin{array}{l}
 - {h^2}{h_1}^2 + {h_2}^2 - {h_1}{h_3}\\
 + {h^2}{h_1}{h_2}{h_3} - {h^2}{h_3}^2
\end{array} \right){t^3}\\
 + \left( \begin{array}{l}
 - {h^2}{h_2}^3 + 2{h^2}{h_1}{h_2}{h_3}\\
 + {h^4}{h_1}^2{h_3}^2
\end{array} \right){t^4}\\
 + \left( \begin{array}{l}
{h^2}{h_1}{h_2}{h_3} - {h^4}{h_1}{h_2}^2{h_3}\\
 + {h^4}{h_1}^2{h_3}^2
\end{array} \right){t^5}\\
 - {h^4}{h_1}{h_2}^2{h_3}{t^6}
\end{array} \right)}}{{(1 - {t^2})(1 - {h^2}{h_2}t)}}}$
&
$ {\left\{ \begin{array}{l}
{C_{\left[ {0,0,0} \right]\left[ 0 \right]}} = 1/\left( {1 - {t^2}} \right)\\
\ldots\\
{C_{\left[ {0,n - 1,2} \right]\left[ {2n > 0} \right]}} = {t^{n + 1}}/\left( {1 - t} \right)\\
{C_{\left[ {0,n,0} \right]\left[ {2n > 0} \right]}} = {t^n}/\left( {1 - t} \right)\\
{C_{\left[ {2,n - 2,2} \right]\left[ {2n > 0} \right]}} = {t^{n + 2}}/\left( {1 - t} \right)\\
{C_{\left[ {2,n - 1,2} \right]\left[ {2n > 0} \right]}} = {t^2}/\left( {1 - t} \right)
\end{array} \right.}$ \\
\hline

\end{tabular}
\end{center}
\text{ Some $C_\lambda$ coefficients for $B_4$ are omitted for brevity.}
\label{table18}
\end{sidewaystable}
\begin{sidewaystable}[htp]
\caption{Coefficients of modified HL polynomials for C series RSIMS}
\begin{center}
\begin{tabular}{|c|c|c|c|}
\hline
$ {Group}$&
$ \begin{array}{c}
Branching\\
(Fugacities)
\end{array}$
&
$ {HWG~for~{C_\lambda}}$
&
$ {{C_\lambda}~Coefficients}$ \\
\hline
$ {{C_1}}$&$ {{A_1}}$&$ 1$&$ {{C_{[ 0 ]}} = 1}$ \\
\hline
$ {{C_2}}$&$ \begin{array}{c}
{A_1} \otimes {A_1}\\
( {{h_A},{h_B}} )
\end{array}$&$ {1 + {h_A}{h_B}t}$&
${\left\{ \begin{array}{l}
{C_{\left[ 0 \right]\left[ 0 \right]}} = 1\\
{C_{\left[ 1 \right]\left[ 1 \right]}} = t
\end{array} \right.}$ \\
\hline
$ {{C_3}}$&$ \begin{array}{l}
{A_1} \otimes {A_1}\otimes{A_1}\\
( {{h_A},{h_B},{h_C}} )
\end{array}$&$ {1  +  ( {{h_A}{h_B} +  {h_A}{h_C}  +  {h_B}{h_C}} )t}$&
${\left\{ \begin{array}{l}
{C_{\left[ 0 \right]\left[ 0 \right]\left[ 0 \right]}} = 1\\
{C_{\left[ 0 \right]\left[ 1 \right]\left[ 1 \right]}} = t\\
 \ldots 
\end{array} \right.}$
  \\
\hline
$ {{C_4}}$&$ \begin{array}{l}
{A_1}\otimes{A_1}\otimes{A_1}\otimes{A_1}\\
( {{h_A},{h_B},{h_C},{h_D}} )
\end{array}$&
${\left( \begin{array}{l}
1\\
 + \left( \begin{array}{l}
{h_A}{h_B} + {h_A}{h_C} + {h_B}{h_C}\\
 + {h_A}{h_D} + {h_B}{h_D} + {h_C}{h_D}
\end{array} \right)t\\
 +~ {h_A}{h_B}{h_C}{h_D}{t^2}
\end{array} \right)}$
&
${\left\{ \begin{array}{l}
{C_{\left[ 0 \right]\left[ 0 \right]\left[ 0 \right]\left[ 0 \right]}} = 1\\
{C_{\left[ 0 \right]\left[ 0 \right]\left[ 1 \right]\left[ 1 \right]}} = t\\
 \ldots \\
{C_{\left[ 1 \right]\left[ 1 \right]\left[ 1 \right]\left[ 1 \right]}} = t
\end{array} \right.}$
 \\
\hline
\end{tabular}
\end{center}
\text{ Some $C_\lambda$ coefficients for $C_3$ and $C_4$ are omitted for brevity.}
\label{table19}
\end{sidewaystable}
\begin{sidewaystable}[htp]
\caption{Coefficients of modified HL polynomials for D series RSIMS}
\begin{center}
\begin{tabular}{|c|c|c|c|}

\hline
$ {Group}$&$ \begin{array}{c}
Branching\\
(Fugacities)
\end{array}$&$ {HWG~for~ {C_\lambda }}$&$ {{C_\lambda}~Coefficients}$ \\
\hline

$ {{D_2}}$&$ \begin{array}{l}
{A_1} \oplus {A_1}\\
\left( {{h_A},{h_B}} \right)
\end{array}$&$ {2 - h_A^2t - h_B^2t}$&$ {\left\{ \begin{array}{l}
{C_{\left[ {0,0} \right]}} = 2\\
{C_{\left[ {2,0} \right]}} = {C_{\left[ {0,2} \right]}} =  - {t^2}
\end{array} \right.}$ \\
\hline

$ {{D_3}}$&$ \begin{array}{c}
{A_3}\\
\left( {{h_1},{h_2},{h_3}} \right)
\end{array}$&$ {\left( \begin{array}{l}
1\\
 - {t^2}\left( {1 - t} \right)h_2^2\\
 - {t^2}\left( {1 + t} \right){h_1}{h_3}\\
 + {t^3}\left( {h_1^2{h_2} + {h_2}{h_3}^2} \right)\\
 - {t^4}{h_1}^2{h_3}^2
\end{array} \right)}$&$ {\left\{ \begin{array}{l}
{C_{\left[ {0,0,0} \right]}} = 1\\
{C_{\left[ {0,2,0} \right]}} =  - {t^2}(1 - t)\\
{C_{\left[ {1,0,1} \right]}} =  - {t^2}(1 + t)\\
{C_{\left[ {0,1,2} \right]}} = {C_{\left[ {2,1,0} \right]}} = {t^3}\\
{C_{\left[ {2,0,2} \right]}} =  - {t^4}
\end{array} \right.}$ \\
\hline

$ {{D_4}}$
&
$
 \begin{array}{c}
{A_1} \otimes {A_1} \otimes {A_1} \otimes {A_1}\\
\left( {{h_A},{h_B},{h_C},{h_D}} \right)
\end{array}
$
&
$ 
\begin{array}{c}
\frac{{1 + h{t^2}}}{{(1 - {t^2})(1 - h t)}}\\
where~h \equiv {h_A}{h_B}{h_C}{h_D}
\end{array}
$
&$ {\left\{ \begin{array}{l}
{C_{\left[ 0 \right]\left[ 0 \right]\left[ 0 \right]\left[ 0 \right]}} = 1/\left( {1 - {t^2}} \right)\\
{C_{\left[ n \right]\left[ n \right]\left[ n \right]\left[ n \right]}} = {t^n}/\left( {1 - t} \right)
\end{array} \right.}$ \\
\hline

$ {{D_5}}$
&
$\begin{array}{c}
{A_3} \otimes {A_1} \otimes {A_1}\\
\left( {{h_1},{h_2},{h_3};{h_A};{h_B}} \right)
\end{array}$
&
$\frac{{\left( {\begin{array}{*{20}{l}}
1\\
{ + \left( {{h}{h_2} - {h_1}{h_3} - {h}h_1^2 - {h}h_3^2} \right){t^2}}\\
{ + \left( {{h}{h_1}{h_2}{h_3} + h_2^2 - {h_1}{h_3} - {h}h_1^2 - {h^2}h_3^2} \right){t^3}}\\
{ + \left( {2{h}{h_1}{h_2}{h_3} + h^2 h_1^2 h_3^2 - {h}h_2^3} \right){t^4}}\\
{ + \left( {{h}{h_1}{h_2}{h_3} + h^2 h_1^2 h_3^2 - h^2{h_1}h_2^2{h_3}} \right){t^5}}\\
{ - h^2{h_1}h_2^2{h_3}{t^6}}
\end{array}} \right)}}{{\begin{array}{*{20}{l}}
{(1 - {t^2})\left( {1 - h {h_2}t} \right)}\\
{where~{h} \equiv {h_A}{h_B}}
\end{array}}}$
&
${\left\{ \begin{array}{l}
{C_{\left[ {0,0,0} \right]\left[ 0 \right]\left[ 0 \right]}} = 1/\left( {1 - {t^2}} \right)\\
{C_{\left[ {0,n,0} \right]\left[ n \right]\left[ n \right]}} = {t^n}/\left( {1 - t} \right)\\
 \ldots 
\end{array} \right.}$ \\
\hline
\end{tabular}
\end{center}
\text{ $D_2$ adjoint is $[2,0] \oplus [0,2]$. Some $C_\lambda$ coefficients for $D_5$ are omitted for brevity}.
\label{table20}
\end{sidewaystable}

Naturally, the structures of the series of branching coefficients $C_\lambda$ differ between the various groups, modulo isomorphisms. Nonetheless there are a number of interesting patterns and similarities that can be observed. 
\begin{enumerate}

\item A Series. The $C_\lambda$ coefficients all constitute finite series that are symmetric under complex conjugation (reversal of Dynkin label fugacities). Since the mHL already contain symmetrisations of the adjoint by construction, the role of the $C_\lambda$ coefficients for $A_r$ is largely to recode in terms of Hall-Littlewood polynomials the class functions of characters within the ${P}_{instanton}^{A_r}$ numerators set out in Table \ref{table5}. Thus, although the coefficients for decompositions in terms of Hall-Littlewood polynomials differ from those in terms of characters, the same irreps are typically involved. Indeed a comparison of Tables \ref{table5} and \ref{table17} shows that (up to $A_3$) the Hall-Littlewood polynomial irreps match those involved in a character expansion, but that their polynomial coefficients in $t$ are considerably simpler\footnote{For $A_4$, the Hall-Littlewood irreps are a subset of those in ${P^{A_4}}_{instanton}$ (not presented herein).}.

\item B Series. With the exception of $B_3 \to A_3$, the $C_\lambda$ coefficients constitute infinite series for all mappings of rank above two. For $B_3 \to A_1 \otimes A_1 \otimes A_1 $ the generator of this infinite series is given by the $h_A {h_B}^2 h_C$ monomial corresponding to the [1][2][1] irrep, and for $B_4$ the generator is given by the ${h}^2 h_2$ monomial corresponding to the [2][0,1,0] irrep, both as identified in the branchings of the adjoint shown in Table \ref{table15}. As to be expected from the graph automorphisms in Figure \ref{branchingBCD}, the $C_\lambda$ exhibit symmetry under interchange of Dynkin fugacities ${h_A} \Leftrightarrow {h_B}$ for $B_2$, ${h_A} \Leftrightarrow {h_C}$ for $B_3  \to A_1 \otimes A_1 \otimes A_1$ and ${h_1} \Leftrightarrow {h_3}$ for $B_3 \to A_3$ and $B_4 \to A_3  \otimes A_1$.

\item C Series. In all cases, the $C_\lambda$ coefficients constitute finite series and the branching relations are completely symmetric under interchange of the $A_1$ subgroups.

\item D Series. For rank 4 and above, the $C_\lambda$ coefficients constitute an infinite series. For $D_4$ the generator of this infinite series is given by the $h_A {h_B} h_C h_D$ monomial corresponding to the [1][1][1][1] irrep and for $D_5$ the generator is given by the $h_2 {h_A} h_B$ monomial corresponding to the  [0,1,0][1][1] irrep, both as identified in the branchings of the adjoint shown in Table \ref{table15}. As to be expected from the Dynkin diagrams in Figure \ref{branchingBCD}, the $C_\lambda$ exhibit symmetry under interchange of Dynkin fugacities ${h_A} \Leftrightarrow {h_B} \Leftrightarrow {h_C} \Leftrightarrow {h_D}$ for $D_4$, and ${h_A} \Leftrightarrow {h_B}$ for $D_5$.

\end{enumerate}
%
\begin{sidewaystable}[htp]
\caption{Coefficients of modified HL polynomials for $G_2$ and $F_4$ RSIMS}
\begin{center}
\begin{tabular}{|c|c|c|c|}

\hline
$ {Group}$
&
$ \begin{array}{c}
Subgroup\\
\left( {Fugacities} \right)
\end{array}$&$ {HWG~ for~ {C_\lambda }}$&$ {{C_\lambda}~Coefficients}$ \\
\hline

$ {{G_2}}$
&
$ \begin{array}{c}{A_1} \otimes {A_1}\\
\left( {{h_A};{h_B}} \right)\end{array}$
&
$ \begin{array}{l}\frac{1 + h{t^2}}{{{{(1 - t^2)}}\left( {1 - ht} \right)}}\\
where ~h \equiv {h_A}^3{h_B}\end{array}$
&
$ {\left\{ \begin{array}{l}
{C_{\left[ 0 \right]\left[ 0 \right]}} = 1/\left( {1 - {t^2}} \right)\\
{C_{\left[ {3n} \right]\left[ n \right]}} = {t^n}/\left( {1 - t} \right)
\end{array} \right.}$ \\
\hline

$ {{G_2}}$
&
$ \begin{array}{c}{A_2}\\
\left( {{h_1},{h_2}} \right)\end{array}$
&
$1+{h_1} t+{h_2} t$
&
$\left\{ {\begin{array}{*{20}{l}}
{C_{\left[ {0,0} \right]}} = 1\\
{C_{\left[ {1,0} \right]}} = {C_{\left[ {0,1} \right]}} = t
\end{array}} \right.$ \\
\hline

$ {{F_4}}$
&
$ \begin{array}{c}
{A_1} \otimes {A_3}\\
\left( {h;{h_1},{h_2},{h_3}} \right)
\end{array}$&$ {\frac{{\left( \begin{array}{l}
1 + h{h_1}t + h{h_3}t + h{h_1}{t^2} + h{h_3}{t^2} + {h^2}{h_2}{t^2}\\
 + {h_2}{t^2} + {h^2}{h_1}{h_3}{t^2} + {h^2}{h_1}{h_3}{t^3} - {h^2}{h_2}^2{t^3}
\end{array} \right)}}{{\left( {1 - {t^2}} \right)\left( {1 - {h^2}{h_2}t} \right)}}}$&$ {\left\{ \begin{array}{l}
{C_{\left[ 0 \right]\left[ {0,0,0} \right]}} = 1/(1 - t^2)\\
 \ldots \\
\end{array} \right.}$ \\
\hline

$ {{F_4}}$
&
$ \begin{array}{c}{A_2} \otimes {A_2}\\
(h_{A1},h_{A2}; h_{B1},h_{B2})\end{array}$
&
$\begin{array}{*{20}{c}}
{\frac{{1 + {h_1}{t^2} + {h_2}{t^2} + {h_1}{t^3} + {h_2}{t^3} + {h_1}{h_2}{t^5}}}{{(1 - {t^2})( {1 - {t^3}})\left( {1 - {h_1} t} \right)\left( {1 - {h_2} t} \right)}}}\\
{where~{h_i} \equiv {h_{Ai}}{h^2_{Bi}}}
\end{array}$
&
$ {\left\{ \begin{array}{l}
{C_{\left[ {0,0} \right]\left[ {0,0} \right]}} = 1/(1 - {t^2})\left( {1 - {t^3}} \right)\\
{C_{\left[ {n,0} \right]\left[ {2n,0} \right]}} = {t^n}/\left( {1 - t} \right)\left( {1 - {t^2}} \right)\\
{C_{\left[ {0,n} \right]\left[ {0,2n} \right]}} = {t^n}/\left( {1 - t} \right)\left( {1 - {t^2}} \right)\\
{C_{\left[ {n1,n2} \right]\left[ {2n1,2n2} \right]}} = {t^{n1 + n2}}/{\left( {1 - t} \right)^2}
\end{array} \right.}$ \\
\hline

$ {{F_4}}$
&
$ \begin{array}{c}{A_1} \otimes {A_1} \otimes {A_1} \otimes {A_1}\\
(h_{A}; h_{B}; h_{C};h_{D})\end{array}$
&
$
\frac{{1 +  \ldots anti-palindrome~(1920~terms) \ldots  - {h_A}^{6}{h_B}^{6}{h_C}^{6}{h_D}^{6}{t^{23}}}}{{\left( \begin{array}{c}
{(1 - {t^2})^4}(1 - {h_A}{h_B}t)(1 - {h_A}{h_C}t)(1 - {h_B}{h_C}t)\\
 \times (1 - {h_A}{h_D}t)(1 - {h_B}{h_D}t)(1 - {h_C}{h_D}t)\\
 \times (1 - {h_A}{h_B}{h_C}{h_D}t)(1 - {h_A}^2{t^3})\\
 \times (1 - {h_B}^2{t^3})(1 - {h_C}^2{t^3})(1 - {h_D}^2{t^3})
\end{array} \right)}}
$
&
$ 
\left\{ {\begin{array}{*{20}{l}}
\begin{array}{l}
{C_{\left[ 0 \right]\left[ 0 \right]\left[ 0 \right]\left[ 0 \right]}} = 1/{\left( {1 - {t^2}} \right)^4}\\
 \ldots \\
\end{array}
\end{array}} \right.
$ \\
\hline

\end{tabular}
\end{center}
\text{ Some $C_\lambda$ coefficients for $F_4$ are omitted for brevity }
\label{table21}
\end{sidewaystable}

\begin{sidewaystable}[htp]
\caption{Coefficients of modified HL polynomials for E series RSIMS}
\begin{center}
\begin{tabular}{|c|c|c|c|}

\hline
$ {Group}$
&
$ \begin{array}{c}
Branching\\
\left( {Fugacities} \right)
\end{array}$&$ {HWG~ for~ {C_\lambda }}$&$ {{C_\lambda}~Coefficients}$ \\

\hline
$ {{E_6}}$
&
$ \begin{array}{c}
{A_2} \otimes {A_2} \otimes {A_2}\\
({h_{Ai};h_{Bi};h_{Ci})}\\
where~i = \left\{ {1,2} \right\}
\end{array}$
&$ \begin{array}{c}
\frac{{1 + {h_1}{t^2} + {h_2}{t^2} + {h_1}{t^3} + {h_2}{t^3} + {h_1}{h_2}{t^5}}}{{\left( {1 - {t^2}} \right)\left( {1 - {t^3}} \right)(1 - {h_1}t)(1 - {h_2}t)}}\\
where ~{h_i} \equiv {h_{Ai}}{h_{Bi}}{h_{Ci}}
\end{array}$&
$ {\left\{ \begin{array}{l}
{C_{\left[ {0,0} \right]\left[ {0,0} \right]\left[ {0,0} \right]}} = 1/(1 - {t^2})\left( {1 - {t^3}} \right)\\
{C_{\left[ {0,n} \right]\left[ {0,n} \right]\left[ {0,n} \right]}} = {t^n}/\left( {1 - t} \right)\left( {1 - {t^2}} \right)\\
{C_{\left[ {n,0} \right]\left[ {n,0} \right]\left[ {n,0} \right]}} = {t^n}/\left( {1 - t} \right)\left( {1 - {t^2}} \right)\\
{C_{\left[ {n1,n2} \right]\left[ {n1,n2} \right]\left[ {n1,n2} \right]}} = {t^{n1 + n2}}/{\left( {1 - t} \right)^2}
\end{array} \right.}$ \\
\hline

$ {{E_6}}$
&
$ \begin{array}{c}
{A_5} \otimes {A_1} \\
({h_1,h_2,h_3,h_4,h_5;h)}\\
\end{array}$
&
$ 
\frac{{\left( {1 +  \ldots 250~terms \ldots  + {h^4}{h_1}^2{h_2}{h_3}^2{h_4}{h_5}^2{t^{12}}} \right)}}{{(1 - {t^2})(1 - h{h_3}t)}}
$
&
$
\left\{ {\begin{array}{*{20}{l}}
\begin{array}{l}
{C_{\left[ {0,0,0,0,0} \right]\left[ 0 \right]}} = 1/\left( {1 - {t^2}} \right)\\
 \ldots 
\end{array}
\end{array}} \right.
$ \\
\hline

\hline
$ {{E_7}}$
&
$\begin{array}{*{20}{c}}
{{A_3} \otimes {A_3} \otimes {A_1}}\\
{({h_{Ai}};{h_{Bi}};h)}\\
{where~ i = \left\{ {1,2,3} \right\}}\\
\end{array}$
&
$\begin{array}{c}
\frac{{{C_{E7}}\left( {{h_i},h,t} \right)}}{{{{(1 - {t^2})}^2}(1 - {t^3})(1 - {t^4})(1 - h{h_1}t)(1 - {h_2}t)(1 - h{h_3}t)}}\\
 \times \frac{1}{{(1 - {h^2}{t^2})(1 - {h^2}{h_2}{t^2})(1 - {h_1}{h_3}{t^2})(1 - {h_1}^2{t^3})(1 - {h_3}^2{t^3})}}\\
where~{h_i} \equiv {h_{Ai}}{h_{Bi}}
\end{array}$
&
$
\left\{ {\begin{array}{*{20}{l}}
\begin{array}{l}
{C_{\left[ {0,0,0} \right]\left[ {0,0,0} \right]\left[ 0 \right]}} = \frac{1}{{{{(1 - {t^2})}^2}(1 - {t^3})(1 - {t^4})}}\\
 \ldots 
\end{array}
\end{array}} \right.
$ \\
\hline
$ {{E_8}}$
&
$ {{A_5} \otimes {A_2} \otimes {A_1}}$&$ {to~be~calculated}$&$ {to~be~calculated}$ \\
\hline
\end{tabular}\\
\end{center}
\text{ Some $C_\lambda$ coefficients for $E_7$ and $E_8$ are omitted for brevity. }\\
\text{$C_{E7}$ consists of 704 monomials and is shown in Appendix 3}
\label{table22}
\end{sidewaystable}
The $C_\lambda$ coefficients for Exceptional groups do not fall into any simple pattern, but some categories can be identified in Tables \ref{table21} and \ref{table22}:
\begin{enumerate}
\item Finite series. For $G_2 \to A_2$, the series of coefficients is finite. 
\item $G_2, B_3, D_4 \to n A_1$  family. The $C_\lambda$ for $G_2 \to A_1 \otimes A_1$ form a complete intersection, which has a generator given by the ${h_A}^3 {h_B}$ monomial corresponding to the [3][1] irrep. Interestingly, the generating functions for $G_2, B_3, D_4 \to n A_1$  differ only in the composition of their respective monomials ${h_A}^3 h_B$, $h_A {h_B}^2 h_C$ and $h_A {h_B} h_C h_D$. The reasons can be traced to the folding relationships between the extended Dynkin diagrams of these groups.
\item $ F_4, E_6 \to n A_2$ family. For $F_4$ to ${A_2} \otimes {A_2}$, the generators are given by the $h_{A1}{h_{B1}}^2$ and $h_{A2}{h_{B2}}^2$ monomials corresponding, respectively, to the [0,1][0,2] and [1,0][2,0] irreps. For $E_6$ to ${A_2} \otimes {A_2}  \otimes {A_2}$, the generators are given by the  $h_{A1} h_{B1}h_{C1}$ and $h_{A2} h_{B2}h_{C2}$ monomials corresponding, respectively, to the [1,0][1,0][1,0] and [0,1][0,1][0,1] irreps and the $C_\lambda$ coefficients are invariant under complex conjugation and under exchange of subgroups ${h_A} \Leftrightarrow {h_B} \Leftrightarrow {h_C}$. Interestingly, the structure of the generating functions for $F_4$ and $E_6$ is the same, differing only by their respective monomials ${h_{Ai}} {h_{Bi}}^2$ and ${h_{Ai}} {h_{Bi}} {h_{Ci}}$. The source can be traced to the folding relationship between the extended Dynkin diagrams of these two groups. Even though the generating function for the $C_\lambda$ is not a complete intersection, the $C_\lambda$ coefficients form a simple pattern.
\end{enumerate}
In the case of the other Exceptional group decompositions, the HWGs typically have complicated numerators. 

Interestingly, the denominators of the $C_\lambda$ coefficients for all groups appear to take a simple form determined by the zeros of the Dynkin labels in a similar manner to the $P_{U(N)}^G$ factors encountered in the Coulomb branch monopole construction for RSIMS.

\FloatBarrier

In \cite{Gadde:2011uv}, it is conjectured that, whenever three modified Hall-Littlewood polynomials of the A series are combined by a three punctured sphere, the $C_{\lambda ,\mu  ,\sigma }(t)$  branching coefficients should follow a symmetric diagonal pattern, such that they are only non-zero when the partitions $\{\lambda ,\mu ,\sigma \}$ are the same for each subgroup. This is exemplified by the $C_\lambda$ monomials for $E_6$ which take the form ${h_{Ai}} {h_{Bi}} {h_{Ci}}$. For the highly symmetric root system of $D_4$, where the branching occurs symmetrically into four $A_1$ subgroups, an extension of this symmetric diagonal pattern to four punctures applies. However, the structures of the monomials for non-simply laced groups such as $B_3$ are more subtle, involving different weights, and therefore lie outside a symmetric diagonal ansatz. This is also the case for many of the other mappings we have studied.

The $C_\lambda$ coefficients for $E_6$ match those in \cite{Gadde:2011uv}, when adjusted for normalisation of the modified Hall-Littlewood polynomials. However, the patterns of the $C_\lambda$ coefficients for $E_7 \to A_3 \otimes A_3 \otimes A_1$ and, presumably, $E_8 \to A_5 \otimes A_2 \otimes A_1$, differ markedly from \cite{Gadde:2011uv}, even though the resulting RSIMS are the same. This is because our approach in this Section has been to decompose RSIMS in terms of modified Hall-Littlewood polynomials as defined by \ref{eq:hl5.19}. On the other hand, \cite{Gadde:2011uv} applies a {\it non-maximal puncture} methodology when the subgroups are of different rank. This further modifies the mHL into a set of non-orthogonal functions that cannot be deployed as a basis. The non-maximally punctured mHL constructions for $E_7$ and $E_8$ in \cite{Gadde:2011uv} do, however, follow from the monopole construction adapted to star shaped quivers, as discussed in Section \ref{sec:coulomb}, by the gauge choices of $q_{3,4}=0$ and $q_{3,6}=0$ for $E_7$ and $E_8$ respectively.

Generally, the decomposition of RSIMS using mHL polynomials leads to HWGs with a small number of generators, as can be seen from Table \ref{table23}. This arises because the mHL polynomials contain embedded generators equal in number to the roots of the product group.\footnote{Recall that in the case of HWGs built on characters of representations, the number of embedded generators is limited by the degree of the dimensional polynomial of the group, which equals the number of positive roots.} The difference between the Hilbert series dimension and the number of HWG generators plus the mHL dimension is balanced by the constraints or relations, if any, that follow from the HWG numerators. We can analyse these in terms of (a) the simple number of relations, calculated by setting all the HWG fugacities $h$ to unity, and (b) the effect of constraints due to the precise structures of the HWG numerators, taking into account differences between the $h$ fugacities. We calculate the impact of these latter constraints by difference in Table \ref{table23} and observe that they only arise for some HWG numerators that have relations involving subgroups containing $A_2$ or higher rank groups.

\begin{table}
\caption{Dimensions of RSIMS Hilbert Series and HWGs from A Series mHLs}
\begin{center}
\begin{tabular}{|c|c|c|c c|c|c|}
\hline
$ {Group}$
&$\begin{array}{c}HS\\Dimension\\\ \sum {a,b,c,d} \end{array}$
&$Subgroup$
&$\begin{array}{c}HWG\\Generators\\(a)\end{array}$
&$\begin{array}{c}HWG\\Relations\\(b)\end{array}$
&$\begin{array}{c}\\Constraints\\(c)\end{array}$
&$\begin{array}{c}mHL\\Dimension\\(d)\end{array}$ \\

\hline
$ {{A_1}}$&$ 2$&${A_1}$&$0$&$0 $&$0 $&$2 $ \\
\hline
$ {{A_2}}$&$4 $&${A_2}$&$0$&$-1 $&$-1 $&$6 $ \\
\hline
$ {{A_3}}$&$6 $&${A_3}$&$0$&$-1 $&$-5  $&$12$ \\
\hline
$ {{A_4}}$&$8 $&${A_4}$&$0$&$-2 $&$-10  $&$20$ \\
\hline
$ {{B_3}}$&$8$&${A_3}$&$0$&$-1 $&$-3 $&$12$ \\

$ {{}}$&$ $&${A_1^{ \otimes 3}}$&$2$&$0 $&$0 $&$6 $ \\

\hline
$ {{B_4}}$&$12 $&${A_3 \otimes A_1}$&$2$&$-2 $&$-2  $&$14 $ \\

$ {{}}$&$ $&${A_1^{ \otimes 4}}$&$5$&$-1 $&$0  $&$8$ \\
\hline
$ {{C_2}}$&$4 $&${A_1^{ \otimes 2}}$&$0$&$0 $&$0 $&$4 $ \\
\hline
$ {{C_3}}$&$6 $&${A_1^{ \otimes 3}}$&$0$&$0 $&$0  $&$6 $ \\
\hline
$ {{C_4}}$&$8 $&${A_1^{ \otimes 4}}$&$0$&$0 $&$0 $&$8 $ \\
\hline

$ {{D_4}}$&$10 $&${A_1^{ \otimes 4}}$&$2$&$ 0$&$ 0 $&$8 $ \\
\hline
$ {{D_5}}$&$14 $&${A_3 \otimes {A_1^{ \otimes 2}}}$&$2$&$-2 $&$-2  $&$16$ \\
\hline
$ {{E_6}}$&$22$&$ {A_2^{ \otimes 3}}$&$4$&$0 $&$0 $&$18$ \\

$ {{}}$&$$&$ {A_5} \otimes {A_1} $&$2$&$-2 $&$-8 $&$30$ \\
\hline
$ {{E_7}}$&$34$&$ {{A_3^{ \otimes 2}} \otimes {A_1}} $&$ 12$&$-4 $&$0 $ &$26 $\\
\hline
$ {{E_8}}$&$58$&$ {{A_5} \otimes {A_2} \otimes {A_1}} $&$\ge 20$&$? $&$?  $ &$38 $\\
\hline
$ {{F_4}}$&$16$&$ {{A_3} \otimes {A_1}}$&$2$&$0 $&$0 $&$14$ \\

$ {{}}$&$$&${A_2^{ \otimes 2}}$&$4$&$0 $&$0 $&$12$ \\

$ {{}}$&$$&$ {A_1^{ \otimes 4}}$&$15$&$-7 $&$0 $&$8$ \\
\hline
$ {{G_2}}$&$6$&${A_2}$&$0$&$0 $&$0 $&$6$ \\

$ {{}}$&$$&${A_1^{ \otimes 2}}$&$2$&$0 $&$0 $&$4$ \\
\hline
\end{tabular}
\end{center}
\text{(a) Number of poles in denominator of HWG determined by setting $h$ fugacities to 1.}\\
\text{(b) Number of poles in numerator of HWG determined by setting $h$ fugacities to 1.}\\
\text{(c) Hidden constraints on HWG/mHL lattice calculated by difference.}\\
\text{(d) Dimension of mHL polynomial equals number of roots of sub-group.}
\label{table23}
\end{table}

\FloatBarrier
\section{RSIMS from Higgs Branches via Product/Factor Groups}
\label{sec:Higgs}

\subsection{Weyl Integration/Molien Series Construction for Classical Groups}

A quite different set of constructions for the Hilbert series of the moduli spaces of instantons has been studied using the Higgs branch of SUSY quiver gauge theories \cite{Benvenuti:2010pq}. These Hilbert series enumerate the gauge invariant objects (``GIOs") of fields transforming in particular representations of Classical product groups, described by their characters. Before proceeding to discuss their field theoretic interpretations, it is useful to summarise the generating functions and the product group structures. These are set out in Table \ref{table24}, where we focus once again on RSIMS.

\begin{sidewaystable}
\caption{RSIMS Generating Functions from GIOs of Product Groups}
\begin{center}
\begin{tabular}{|c|c|c|c|c|}
\hline
$ {Group}$&$Adjoint$&$ \begin{array}{c}Instanton\\HWG\end{array}$&$ \begin{array}{c}Product\\Group\end{array}$&$ {Generating~Function}$ \\
\hline
$ {{A_1} \cong {B_1} \cong {C_1}}$&$ {\left[ 2 \right]}$&$
 {{m^2}{t^2}}$&$ {{A_1} \times U\left( 1 \right)}$&$ {\oint_{U(1)} {d\mu PE\left[ {\left[ 1 \right]q + \left[ 1 \right]{q^{ - 1}},t} \right]PE\left[ { - 1,{t^2}} \right]} } $ \\
$ {{A_{r \ge 2}}}$&$ {\left[ {1, \ldots 1} \right]}$&$ {{m_1}{m_r}{t^2}}$&$ {{A_r} \times U\left( 1 \right)}$&$ {\oint_{U(1)} {d\mu PE\left[ {\left[ {1,0, \ldots } \right]q + \left[ {0, \ldots ,1} \right]{q^{ - 1}},t} \right]PE\left[ { - 1,{t^2}} \right]} }$ \\
\hline
${{B_2} \cong {C_2}}$&$ {\left[ {0,2} \right]}$&$ {{m_2}^2{t^2}}$&$ {{B_2} \times {C_1}}$&$ {\oint_{C_1} {d\mu PE\left[ {\left[ {1,0} \right]\left[ 1 \right] t} \right]PE\left[ { - \left[ 2 \right],{t^2}} \right]} }$ \\
$ {{B_{r \ge 3}}}$&$ {\left[ {0,1, \ldots 0} \right]}$&$ {{m_2}{t^2}}$&$ {{B_r} \times {C_1}}$&$ {\oint_{C_1} {d\mu PE\left[ {\left[ {1,0, \ldots 0} \right]\left[ 1 \right] t} \right]PE\left[ { - \left[ 2 \right],{t^2}} \right]} }$ \\
\hline
${{C_1} \cong {A_1} \cong {B_1}}$&$ {\left[ 2 \right]}$&$ {{m^2}{t^2}}$&$ {{C_1}\times {O(1)}}$&$ {\frac{1}{2}\left( {PE\left[ {\left[ 1 \right],t} \right] + PE\left[ {\left[ 1 \right], - t} \right]} \right)}$ \\
$ {{C_{r \ge 2}}}$&$ {\left[ {2,0, \ldots } \right]}$&$ {{m_1}^2{t^2}}$&$ {{C_r}\times {O(1)}}$&$ {\frac{1}{2}\left( {PE\left[ {\left[ {1,0, \ldots } \right],t} \right] + PE\left[ {\left[ {1,0, \ldots } \right], - t} \right]} \right)}$ \\
\hline
$ {{D_2}}$&$ {\left[ {2,0} \right] \oplus [0,2]}$&$ {m_1^2{t^2} + m_2^2{t^2}}$&$ {{D_2} \times {C_1}}$&$ {\oint_{C_1} {d\mu PE\left[ {\left[ {1,1} \right]\left[ 1 \right],t} \right]PE\left[ { - \left[ 2 \right],{t^2}} \right]} }$ \\
$ {{D_3}}$&$ {\left[ {0,1,1} \right]}$&$ {{m_2}{m_3}{t^2}}$&$ {{D_3} \times {C_1}}$&$ {\oint_{C_1} {d\mu PE\left[ {\left[ {1,0,0} \right]\left[ 1 \right],t} \right]PE\left[ { - \left[ 2 \right],{t^2}} \right]} }$ \\
$ {{D_{r \ge 4}}}$&$ {\left[ {0,1, \ldots ,0} \right]}$&$ {{m_2}{t^2}}$&$ {{D_r} \times {C_1}}$&$ {\oint_{C_1} {d\mu PE\left[ {\left[ {1,0,0,\ldots,0} \right]\left[ 1 \right],t} \right]PE\left[ { - \left[ 2 \right],{t^2}} \right]} }$ \\
\hline
\end{tabular}
\end{center}
\label{table24}
\end{sidewaystable}

In all cases, the RSIMS are constructed from one or more basic representations of the Yang-Mills symmetry group $G$ through symmetrisation or anti-symmetrisation, followed by projection of the GIOs (or singlets) of the quiver gauge group through Weyl integration or a Molien average. Different quiver gauge groups are required to yield the RSIMS, depending on the Yang-Mills group. Some balancing HyperK\text{\" a}hler quotient terms may also be required to remove unwanted irreps. It is straightforward to verify that evaluation of the contour integrals gives precisely the constructions for RSIMS shown in Table \ref{table5}.

In the case of $G$ equals $SU(N)$, an RSIMS is built by taking the PE of (i) a quark transforming in a product group comprising the $SU(N)$ fundamental and a $U(1)$ representation with (ii) an antiquark transforming in the $SU(N)$ anti fundamental and a conjugate $U(1)$ representation. The Weyl integral projects out singlets of the $U(1)$ quiver gauge group and thus eliminates all tensor products other than those containing equal numbers of quarks and antiquarks. These tensor products all transform in some symmetrisation of the adjoint representation of $G$, or as singlets. The balancing factor of $(1-t^2)$ takes a quotient which eliminates the singlets resulting from the quark-antiquark tensor products, leaving only the adjoint representation and its symmetrisations.

In the case of $G$ equals $SO(n)$, the adjoint is formed from the vector representation by anti-symmetrisation. The chosen product group representation is built from the vector of the $SO(n)$ series and a quiver gauge field transforming in the $C_1$ fundamental. The series generated by the PE contains representations which have the same symmetry properties with respect to both parts of the product group. Thus, anti-symmetrisations of the $SO(n)$ vector representation are coupled to the antisymmetric rank two invariant tensor of the symplectic $C_1$ group. Projecting out these antisymmetric $C_1$ invariants by Weyl integration therefore selects objects transforming in the adjoint representation of the $SO(n)$ instanton symmetry group. The quotient term $PE[-[2],t^2]$ is necessary since the generating function otherwise also generates singlets of the quiver gauge group that are not part of the RSIMS.

\FloatBarrier

In the case of $USp(2n)$ Yang-Mills groups, the adjoint is formed from the fundamental of the group by symmetrisation. In this case the chosen quiver gauge group is the discrete $O(1)$ group, which is isomorphic to $Z_2$. The singlets of the quiver gauge group are obtained from a Molien sum \cite{Benvenuti:2006qr}, which replaces Weyl integration over a continuous group by an average over a discrete group.

It is interesting to note that the various isomorphisms between Classical groups give rise to alternative possible product group and quiver gauge group choices for the construction of instantons for $A_1, B_1, C_1, B_2, C_2, A_3$ and $D_3$ Yang-Mills symmetry groups. Specifically, we can use the isomorphisms to construct instanton moduli spaces from the spinor as well as vector representations of $B_2$ and $D_3$ instanton symmetry groups.

Constructions of this type are not known for cases where the Yang-Mills group is an Exceptional group; while the adjoint of an Exceptional group is formed by antisymmetrisation of the fundamental representation, many other irreps are generated in addition and no simple quotient has yet been identified for their exact cancellation.

\subsection{Higgs Branch Quiver Theories}
We now turn briefly to the field theoretic interpretation of these product group constructions of RSIMS. These theories arise on the Higgs branches of various SUSY quiver theories that involve fields transforming in both quiver gauge and instanton Yang-Mills Classical group representations.
\begin{table}
\caption{Field Content of Higgs Branch Quiver Theories for RSIMS}
\begin{center}
\begin{tabular}{|c|c|c|c|c|}
\hline
$ {Quiver~Theory}$&$ {Fields}$&$ \begin{array}{c}Quiver~Gauge\\Charges\end{array}$&$ \begin{array}{c}Yang~Mills\\ Irreps \end{array}$&$ {Superpotential}$ \\
\hline
$ {SU\left(N \right) \times U\left(1 \right) \times {\bar U}{{\left( 1 \right)}_{gauge}}}$&$ \begin{array}{c}\Phi \\{X_{12}}\\{X_{21}}\end{array}$&$ \begin{array}{c}1\\1/q\\q\end{array}$&$ \begin{array}{l}
\left[ {0, \ldots 0} \right] \\
\left[ {1, \ldots 0} \right] \\
\left[ {0, \ldots 1} \right] \\
\end{array}$&$ {Tr\left( {{X_{12}}\Phi {X_{21}}} \right)}$ \\
\hline
$ {SO\left(n \right) \times USp{{\left(2 \right)}_{gauge}}}$&$ \begin{array}{c}S\\Q\end{array}$&$ \begin{array}{c}
\left[ 2 \right]\\
\left[ 1 \right]
\end{array}$&$ \begin{array}{c}
\left[ {0, \ldots 0} \right]\\
\left[ {1, \ldots 0} \right]
\end{array}$&$ {Tr\left( {{Q_a}{\varepsilon ^{ab}}{S_{bc}}{\varepsilon ^{cd}}{Q_d}} \right)}$ \\
\hline
$ {USp\left( {2n} \right) \times O{{\left(1\right)}_{gauge}}}$&$ \begin{array}{c}A\\Q
\end{array}$&$ \begin{array}{c}1\\ \pm 1
\end{array}$&$ \begin{array}{c}
\left[ {0, \ldots 0} \right]\\
\left[ {1, \ldots 0} \right]
\end{array}$&$ {Tr\left( {QAQ} \right)}$ \\
\hline
\end{tabular}
\end{center}
\label{table25}
\end{table}
The RSIMS are created from the product groups shown in Table \ref{table24} arise when the fields in Table \ref{table25} are symmetrised using the PE in the background of a superpotential. The F-term vacuum constraints that result from the superpotentials shown give rise to relations that correspond exactly to the balancing terms in Table \ref{table24}. Only the quiver theories with $SU(N)$ or $SO(n)$ instanton symmetry groups have such non-trivial balancing terms. In the case of the $USp(2n)$ instanton symmetry groups, the F-term vacuum constraints simply cause the fields to vanish and do not give rise to such relations.

As elaborated in \cite{Benvenuti:2010pq, Nekrasov:2004vw} these quiver theories arise on systems of Dp branes against a background of Dp+4 branes in type II string theories. Specifically taking p=3, we obtain a 3+1 dimensional space-time with ${\cal N}=2$ SUSY, spanned by the D3 branes. 

The instantons can be assigned positions on the transverse directions on the D7 branes parameterised using ${\mathbb C}^2$. When there is only one instanton, the fields specifying the global position of the instanton decouple from the quiver gauge and instanton Yang-Mills fields and this gives rise to an RSIMS that is determined solely by the quiver gauge and Yang-Mills field representations. The brane construction corresponding to the unitary theories is straightforward, however the orthogonal and symplectic theories require the use of orientifold planes \cite{Benvenuti:2010pq}.

\begin{figure}[htbp]
\begin{center}
\includegraphics[scale=0.60]{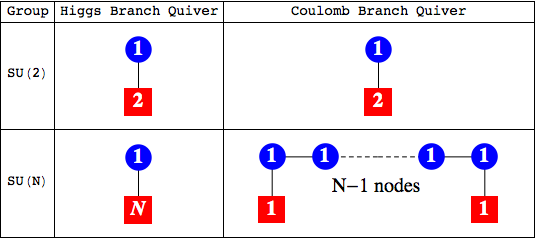}\\
\caption[Quiver diagrams for Higgs and Coulomb branch A series RSIMS.]{Quiver diagrams for Higgs and Coulomb branch A series RSIMS. Blue nodes denote $U(1)$ gauge groups. Red nodes denote flavours. The quiver for SU(2) is self-dual under mirror symmetry. The Coulomb branch quivers correspond to extended Dynkin diagrams when the flavour nodes, which have zero monopole charges, are identified.}
\label{mirrors}
\end{center}
\end{figure}

The Higgs and Coulomb branch quiver theories for A series RSIMS are related by mirror symmetry \cite{Gaiotto:2008ak, Hanany:2011tt}. Without digressing further on this important topic, we summarise in Figure \ref{mirrors} the Higgs branch and Coulomb branch quiver theories corresponding to the A series RSIMS, which are mirror to each other.
\FloatBarrier
\section{Discussion and Conclusions}
\label{sec:conclusions}

The construction of RSIMS using group theoretic methods based on the Weyl Character Formula is, in principle, straightforward for both Classical and Exceptional groups \footnote{subject only to computational challenges for higher rank groups}. As discussed in Section \ref{sec:plethystics}, the single instanton moduli space constructions for both Classical and Exceptional groups given in \cite{Keller:2011ek} can be counted within this category, being derivable from the Weyl Character Formula. What is remarkable is that these instanton moduli spaces can also be obtained by four further quite different methods, three of which have precise interpretations within SUSY quiver gauge theories.

\begin{enumerate}

\item Coulomb branch RSIMS constructions under ${\cal N}=4$ SUSY in 2+1 dimensions were given in \cite{Cremonesi:2013lqa} for simply laced groups and in \cite{Cremonesi:2014xha} it was shown how these constructions can be extended to non-simply laced groups. These quiver gauge theory constructions discussed in Section \ref{sec:coulomb} describe a product group of $U(N)$ {\it monopole} operators labelling points in the root lattice of the Classical or Exceptional group.

\item Higgs branch RSIMS constructions under ${\cal N}=2$ SUSY in 3+1 dimensions are given in Table \ref{table25} \cite{Benvenuti:2010pq}. These quiver gauge theory constructions discussed in Section \ref{sec:Higgs} build instanton moduli spaces as the GIOs of symmetrisations of (the characters of) chiral scalar fields transforming under various product group representations. They are only known for Classical instanton symmetry groups.

\item We have shown in Section \ref{sec:weylbranch} how it is possible to use mappings the weight space of any Classical or Exceptional group and its semi-simple subgroups to deconstruct an RSIMS in terms of the moduli spaces of its subgroup irreps. In Section \ref{sec:weylbranch} we focused on maximal semi-simple subgroups reached via a single elementary transformation. In the case of all Classical and some Exceptional groups, such mappings lead to simple HWGs in terms of subgroup irreps, whose moduli spaces are complete intersections of dimension six or less. 

\item In Section \ref{sec:mHL} we have shown how it is possible to extend the subgroup decomposition approach to utilise A series modified Hall-Littlewood polynomials, in place of the characters of representations. These correspond to $T(SU(N))$ Coulomb branch quiver theories in the presence of background charges. This method leads to an interesting simplification of the HWGs for certain families of RSIMS deconstructions, as discussed further below.

\end{enumerate}

Analysis confirms, on a case by case basis, the identity of the refined Hilbert series (or character expansions) resulting from the different constructions. The methods all lead to identical moduli spaces.

The relationship between the RSIMS of a group and its subgroup moduli spaces is of particular interest. As shown in Section \ref{sec:weylbranch}, in the case of a pair of maximal semi-simple subgroups of BCD Classical groups\footnote{Assuming minimum ranks of 2 and 3 respectively for any B and D series subgroups} reached via a single elementary transformation, we have the simple schema for an RSIMS deconstruction into a subgroup HWG:
\begin{equation}
\label{eq:disc1}
\begin{aligned}
g_{instanton}^{B~or~D} & \to PE\left[ {\left( {{{\theta  + \theta'  + v}} \otimes {{v'}}} \right)t + \left( {1 + {{g}} + {{g'}} + {{v}} \otimes {{v'}}} \right){t^2} - g \otimes g' {t^4}} \right],\\
g_{instanton}^{C} & \to PE\left[ {\left( \theta  + \theta'  + v \otimes v' \right)t - g \otimes g' {t^2}} \right],
\end{aligned}
\end{equation}
where $\theta, v$ and $g$ refer to adjoint, vector and graviton (symmetrised vector) representations, respectively.

These Classical group RSIMS split into subgroup moduli spaces with their HWGs defined, at order $t$, by the adjoint representations of subgroups together with the vector representation of the product group formed by the sub-groups. The HWGs have the same form for all the orthogonal and symplectic groups studied (providing the component subgroups have sufficient rank) and we conjecture that this remains so for higher rank parent groups. These HWGs are all of dimension six or less.

In the case of Exceptional groups, not all mappings to maximal semi-simple subgroups lead to HWGs of low dimension, and the dimensional analysis in Section \ref{sec:weylbranch} shows that this results from the low degrees of the dimensional polynomials of the subgroups, relative to the dimensions of the Hilbert series for the parent group RSIMS.

A sequence of elementary transformations to a regular semi-simple subgroup always leads to a decomposition of the adjoint representation of the parent that includes the adjoint representations of its subgroups \cite{Fuchs:1997bb}. This makes it possible to find HWGs of low dimension utilising modified Hall-Littlewood polynomials. These incorporate the plethystic function $PE[(adjoint-rank) t]$ in their construction and, as shown in Section \ref{sec:mHL}, have dimensional polynomials with a higher degree than those of corresponding representations. Specifically, the degree of the dimensional polynomial of an mHL polynomial is bounded by the number of roots of the subgroup, rather than the number of positive roots, as in the case of a representation based on characters, leading to a factor of two difference.

The resulting low dimensions of the the HWGs built on mHL polynomials leads, in some cases, to particularly simple decompositions of RSIMS into A series subgroups. In particular, the HWGs for A and C series groups (and their isomorphisms) and those for mappings to a single A series subgroup (such as $G_2 \to A_2, B_3 \to A_3, [E_7 \to A_7, E_8 \to A_8]$) are given by finite series of mHL polynomials. There are also two families, $G_2,B_3,D_4 \to {A_1}^r $ and $F_4, E_6 \to  {A_2}^{r/2} $ that have simple HWGs of dimension 2 and 4 respectively, as shown in Table \ref{table26}.

\begin{table}
\caption{RSIMS HWGs from $T(SU(N))$ families}
\begin{center}
\begin{tabular}{|c|c|c|}
\hline
$ {Family}$&$ {HWG~for~{C_\lambda }}$&$ {}$ \\
\hline
$ {{G_2},{B_3},{D_4} \to {A_1}^{\otimes r}}$&$ {\frac{{1 - {h^2}{t^4}}}{{\left( {1 - {t^2}} \right)\left( {1 - ht} \right)\left( {1 - h{t^2}} \right)}}}$&$ {\left\{ \begin{array}{l}
{G_2}:h \equiv h_A^3{h_B}\\
{B_3}:h \equiv {h_A}h_B^2{h_C}\\
{D_4}:h \equiv {h_A}{h_B}{h_C}{h_D}
\end{array} \right.}$ \\
\hline
$ {{F_4},{E_6} \to {A_2}^{\otimes r/2}}$&$ {\begin{array}{*{20}{c}}
{\frac{{1 + {h_1}{t^2} + {h_2}{t^2} + {h_1}{t^3} + {h_2}{t^3} + {h_1}{h_2}{t^5}}}{{(1 - {t^2})(1 - {t^3})\left( {1 - {h_1}t} \right)\left( {1 - {h_2}t} \right)}}}
\end{array}}$&$ {\left\{ \begin{array}{l}
{F_4}:{h_i} \equiv {h_{Ai}}{h^2_{Bi}}\\
{E_6}:{h_i} \equiv {h_{Ai}}{h_{Bi}}{h_{Ci}}
\end{array} \right.}$ \\
\hline
\end{tabular}
\end{center}
\label{table26}
\end{table}

\begin{figure}[htbp]
\begin{center}
\includegraphics[scale=0.5]{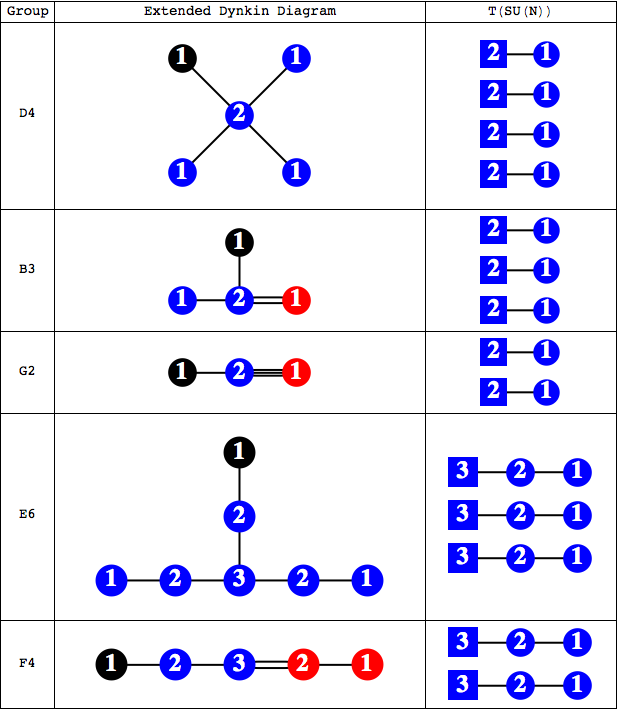}\\
\caption[Quivers for RSIMS deconstructions to $T(SU(N))$ families.]{Quivers for RSIMS deconstructions to $T(SU(N))$ families. The Extended Dynkin diagrams are labelled as follows: blue nodes denote long roots, red nodes denote short roots, black nodes denote the long roots added in the affine construction and the dual Coxeter numbers of each node are shown. $T(SU(N))$ square quiver nodes are labelled by the number $N$ of flavours and round nodes by $U(N)$ gauge symmetries. The RSIMS deconstruction maps the central node of the Extended Dynkin diagram to the $T(SU(N))$ flavour nodes.}
\label{fig:tsunfamilies}
\end{center}
\end{figure}

These decompositions of RSIMS in terms of mHL polynomials reflect structural relationships between Coulomb branch quiver theories for extended Dynkin diagrams and those for $T(SU(N))$. We indicate in Figure \ref{fig:tsunfamilies} the quiver diagrams involved in these family relationships. The extended Dynkin diagrams can be constructed by identifying the flavour nodes of the $T(SU(N))$ quivers; this construction is provided algebraically by the $C_\lambda$ coefficients, obtained from the HWGs. As noted in Sections \ref{sec:coulomb} and \ref{sec:mHL}, the Coulomb branch constructions map directly to the mHL constructions under a gauge choice that sets the lowest $U(1)$ monopole charge of the central node of the $D_4$ or $E_6$ extended Dynkin diagram to zero.

It is clear that quiver diagrams play a profound role in encoding precise relationships between the root spaces of Classical and Exceptional groups and the Coulomb branches of SUSY gauge theories with unitary symmetry groups. Although much work remains to be done to provide a complete account, we can identify some key relationships.

Firstly, the quiver diagrams for all the affine Dynkin diagrams and $T(SU(N))$ theories are balanced, as defined in Section \ref{sec:coulomb}. This ensures that any integer assignment of charges to nodes under the Coulomb branch construction leads to integer conformal dimension, as is necessary for conformal dimension to map to shifts around the root lattice of a group. When the extended Dynkin diagram is taken as the quiver, conformal dimension turns out to be $1$ for those field configurations corresponding to the roots of a group; interestingly, this applies equally for non-simply laced groups, notwithstanding the presence of roots of different lengths \footnote{Providing the prescription for conformal dimension in Section \ref{sec:coulomb} based on the off-diagonal entries in the Cartan matrix is adhered to}. Thus, in the RSIMS construction, conformal dimension increases by $1$ for each new set of dominant weights and orbits introduced by each symmetrisation of the adjoint. In SUSY field theory, conformal dimension corresponds to the R-charge of fields within multiplets.  Since conformal dimension defines a foliation of the root space of a Classical or Exceptional group, the R-charges can be viewed as corresponding to sets of adjacent orbits or {\it shells} in root space.

Secondly, each Coulomb branch monopole construction also depends crucially on its $U(N)$ symmetries, which correspond to key group theoretic parameters. In the case of the RSIMS construction, these $U(N)$ symmetries match the dual Coxeter numbers of the nodes in the extended Dynkin diagram. We have shown how the matching of extended Dynkin diagram $U(N)$ symmetries to those of $T(SU(N))$, leads to simple HWGs for RSIMS in terms of mHL polynomials. In cases where the fit between $T(SU(N))$ structures and the extended Dynkin diagram is not so good, we can still deconstruct an RSIMS in terms of mHL polynomials, but at an increase in the complexity of the relations encoded in the HWG numerators.

\paragraph{Conclusion}
A wide variety of methods can be deployed to construct and deconstruct the single instanton moduli spaces of any Classical or Exceptional group. We have shown how generating functions for characters and Hall-Littlewood polynomials, and the related modified Hall-Littlewood polynomials, can be used to give efficient decompositions of RSIMS in terms of HWGs that draw on their semi-simple subgroups. These decompositions are faithful and the original series can be recovered by recombining characters, or mHL polynomials as appropriate, with the $C_\lambda$ series of coefficients generated by the HWGs. In many cases, the $C_\lambda$ coefficients depend in a simple way on the Dynkin labels of the subgroup representations or mHL polynomials. While these calculations can be implemented in a purely algebraic manner, relationships between SUSY quiver theories play a valuable role in guiding the identification of constructions that lead to simple HWGs; conversely the relationships between moduli spaces that we have identified translate to precise relationships between different SUSY quiver theories.

\paragraph{Further Work}
It could be interesting to extend our use of generating functions for Hall-Littlewood polynomials to the moduli spaces of non-unitary groups and to explore the circumstances under which these provide simple deconstructions of such spaces. 

We have identified two classes of star shaped $T(SU(N))$ quiver theories involving $E_6$ to $A_2 ^{\otimes 3} $  and $D_4$ to $A_1 ^{\otimes 4} $, such that each class is defined by a common HWG when deconstructed into mHL polynomials. If such families are generalised to include theories with more and/or higher rank limbs, can we formulate the principles and find the HWGs for their deconstruction into simpler components?

It may also be interesting to explore whether these approaches can be used to obtain simple descriptions of other moduli spaces of physical interest, including multiple instanton moduli spaces.

\FloatBarrier

\paragraph{Acknowledgements}
Rudolph Kalveks expresses his gratitude to Andrew Thomson, Imperial College for many valuable discussions.


\section{Appendices}
\subsection{Appendix 1: $P^G_{instanton}$ for Low Rank Classical Groups}
\begin{center}
\includegraphics[scale=1]{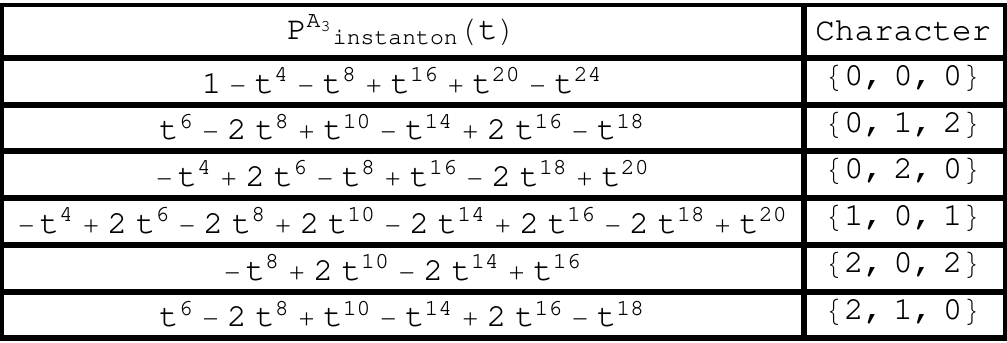}\\
\end{center}
\begin{center}
\includegraphics[scale=1]{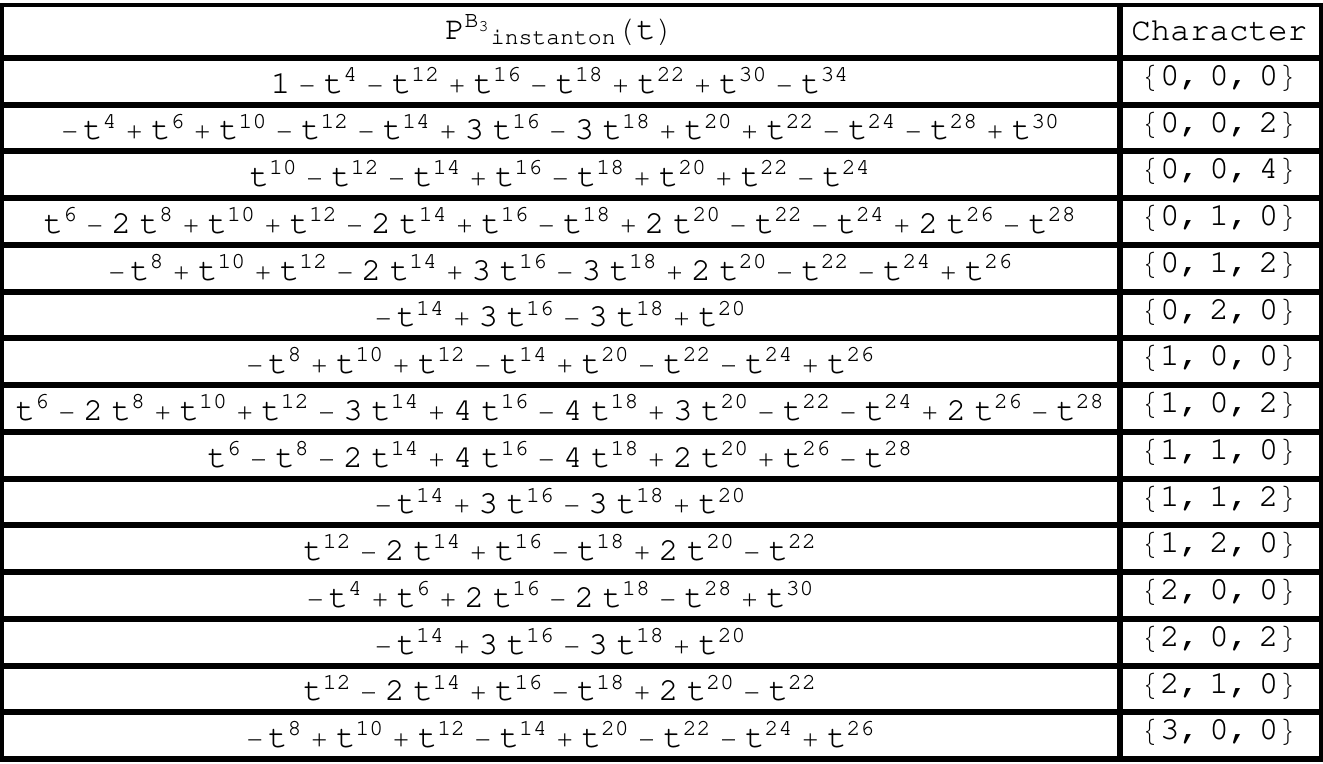}\\
\end{center}
\begin{center}
\includegraphics[scale=1]{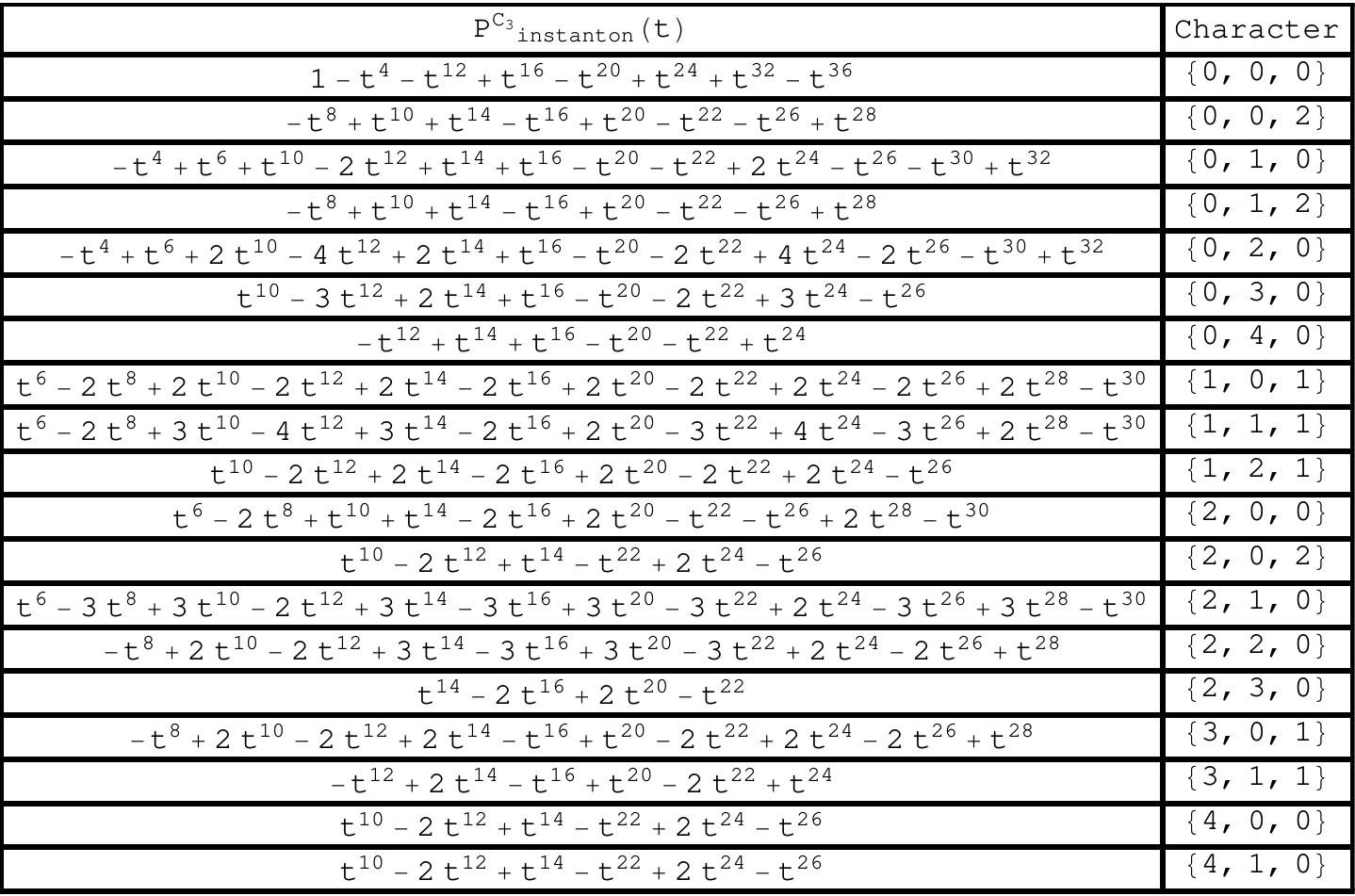}\\
\end{center}
\begin{center}
\includegraphics[scale=1]{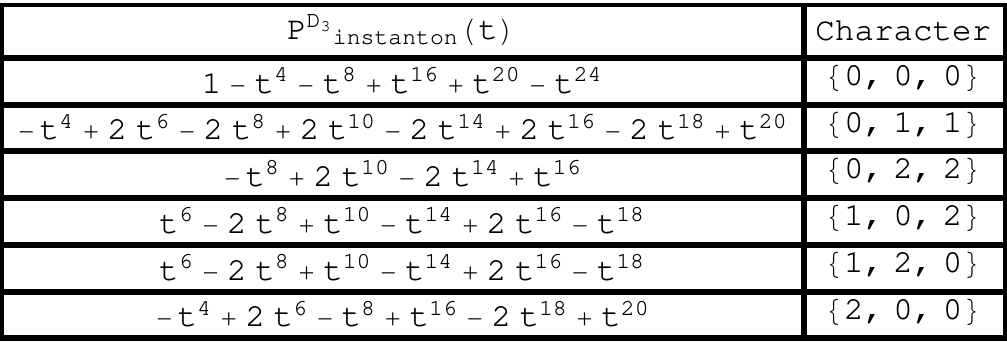}\\
\end{center}
\begin{center}
\includegraphics[scale=1]{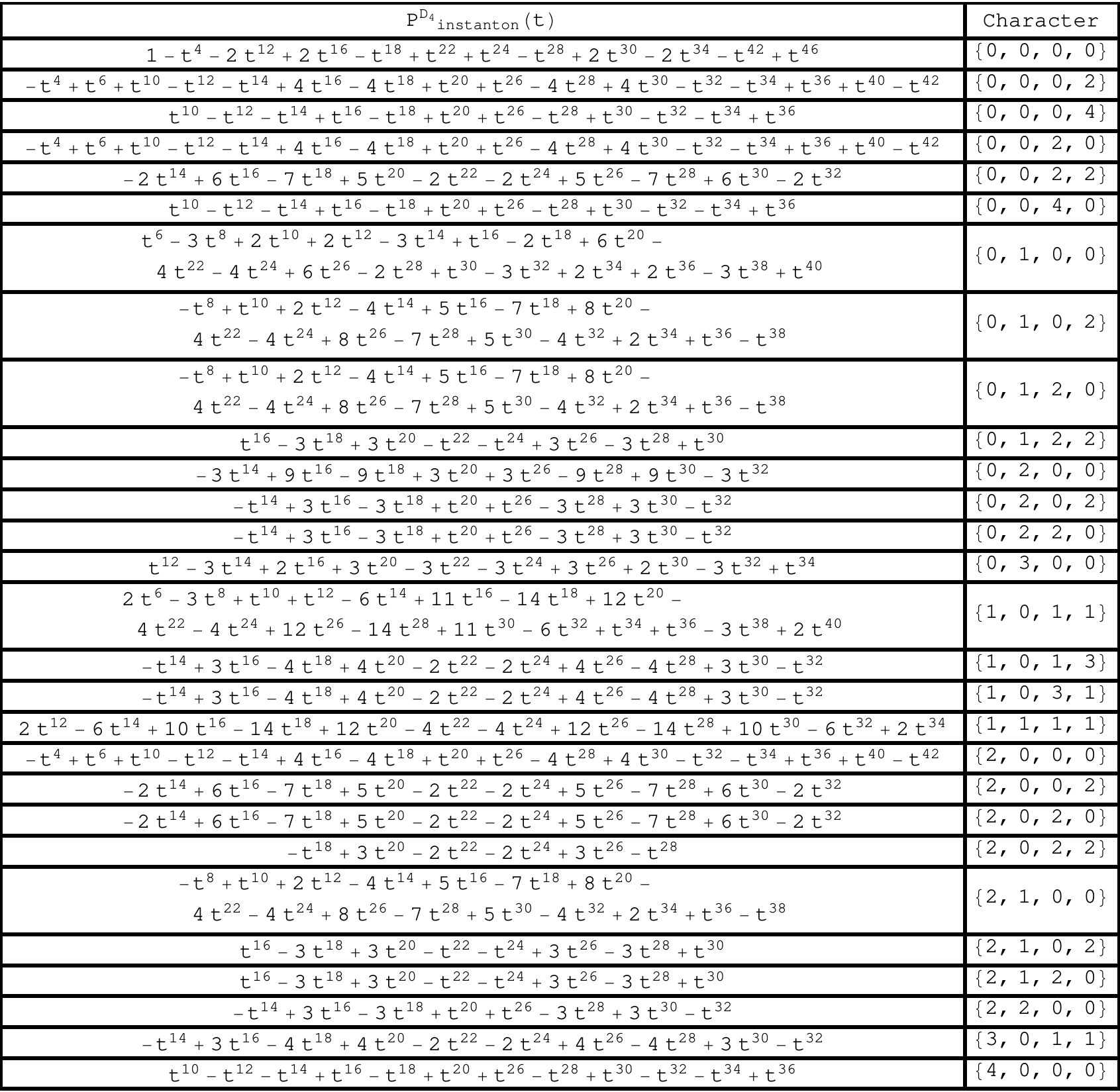}\\
\end{center}
\clearpage

\subsection{Appendix 2: $P^{F_4}_{instanton}$}
\begin{center}
\includegraphics[scale=1]{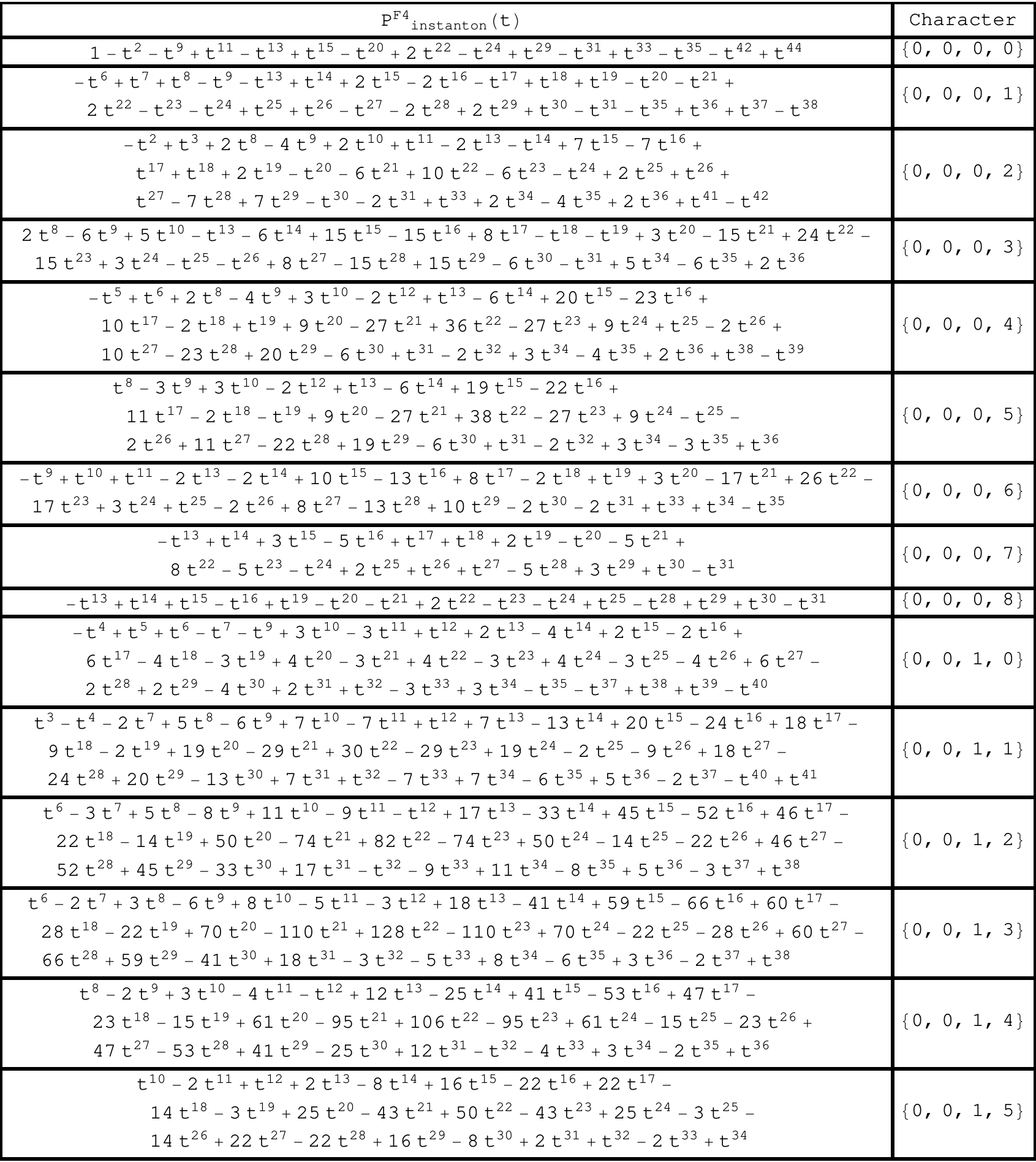}\\
\end{center}
\begin{center}
\includegraphics[scale=1]{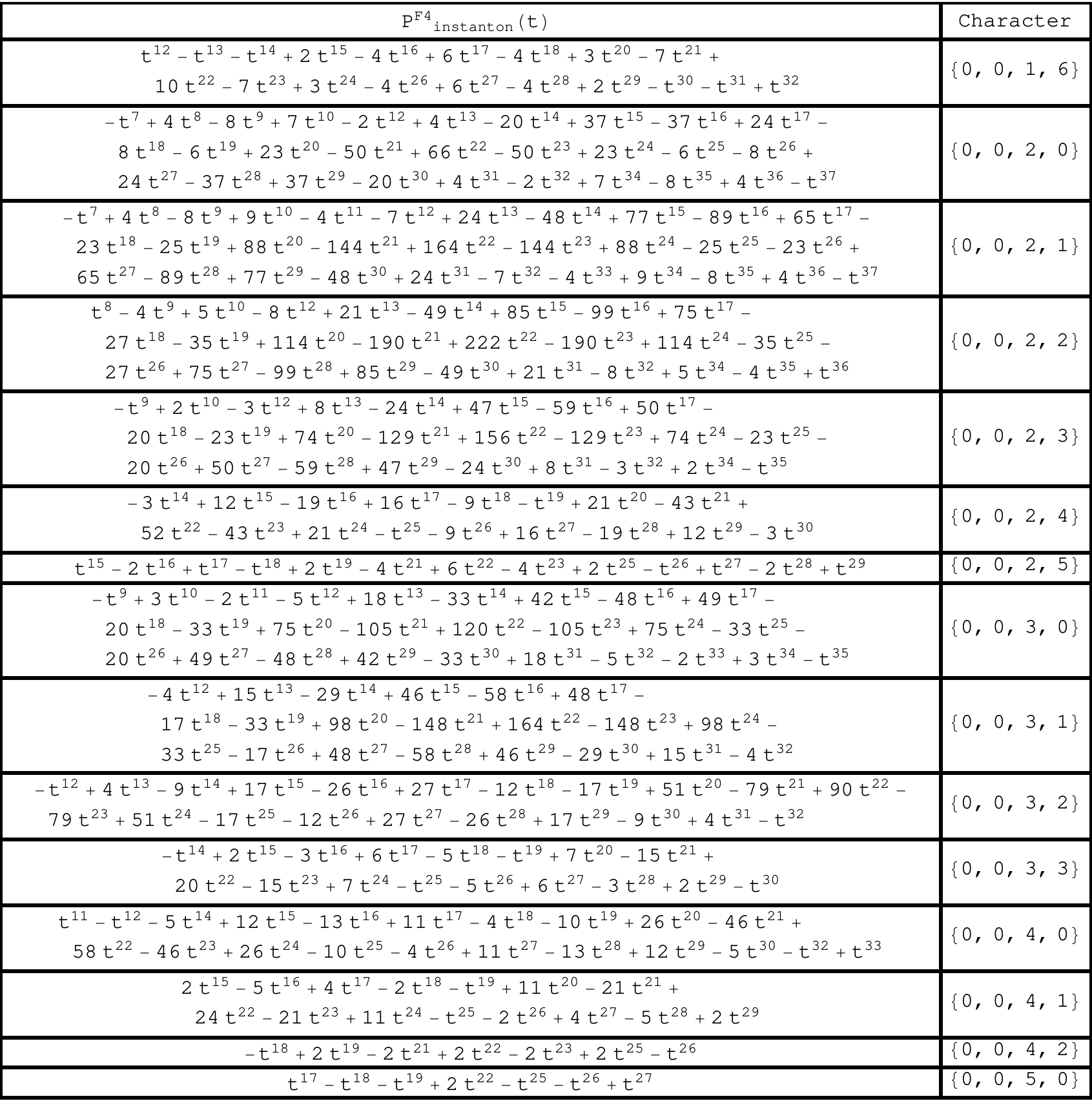}\\
\end{center}
\begin{center}
\includegraphics[scale=1]{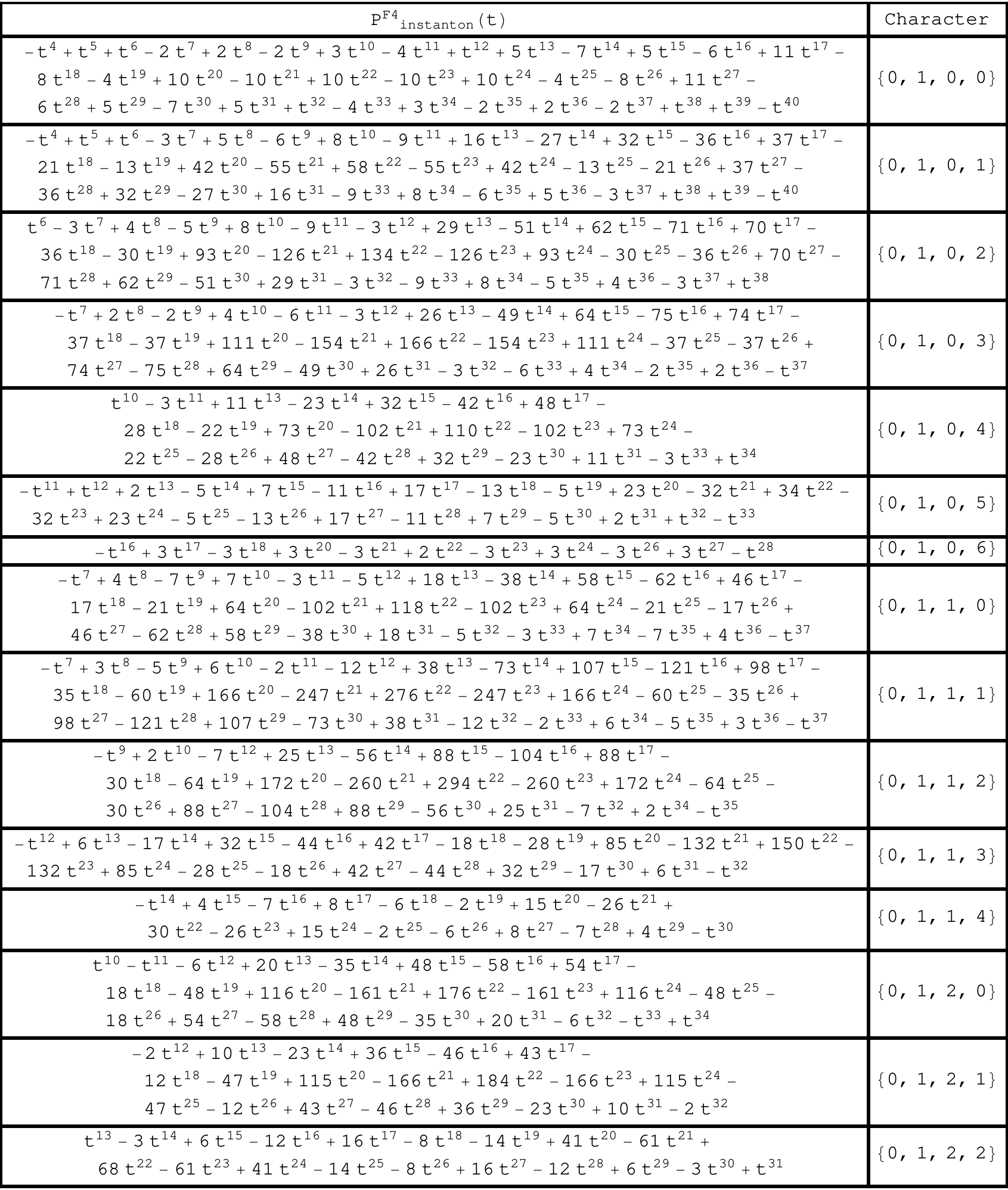}\\
\end{center}
\begin{center}
\includegraphics[scale=1]{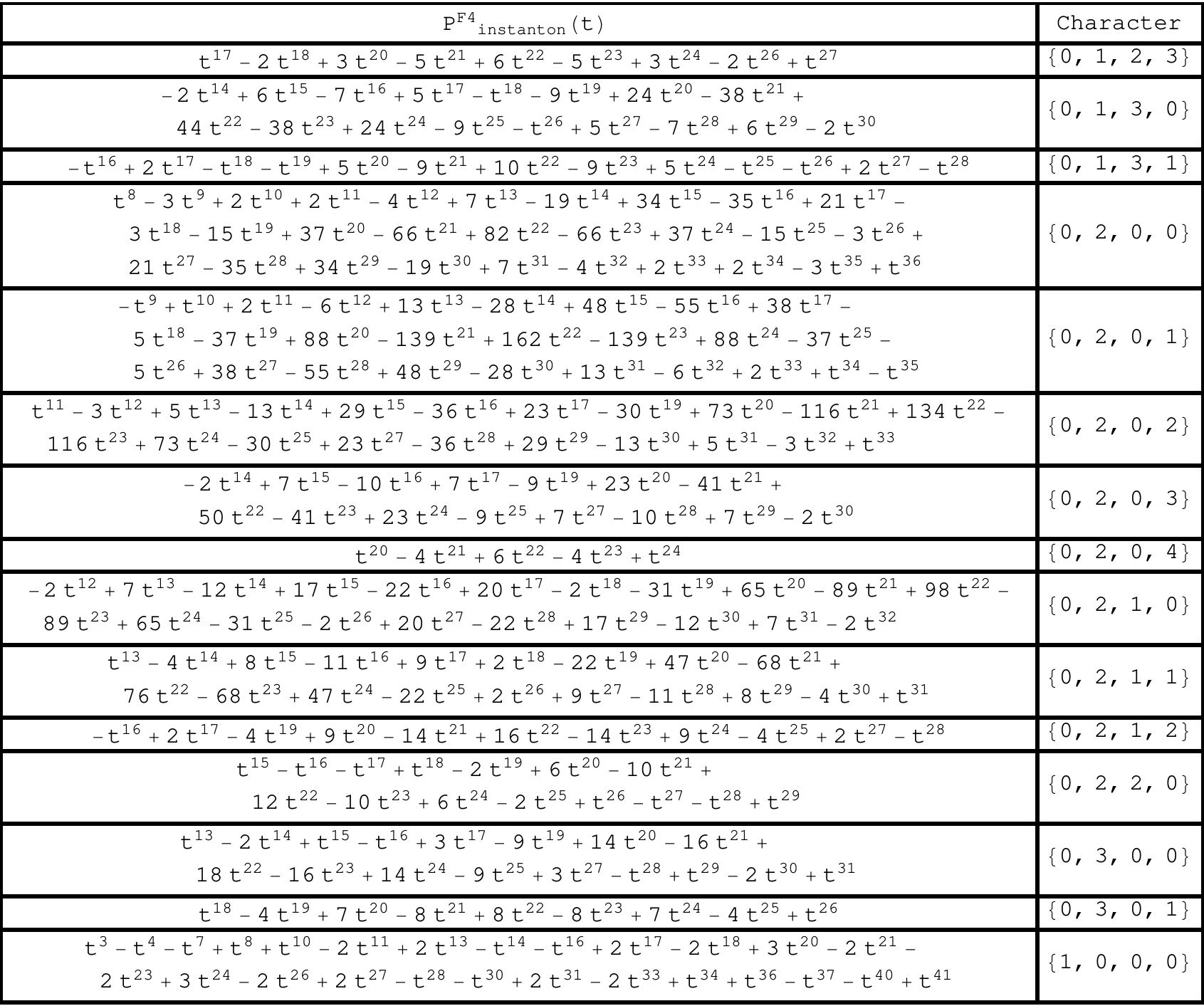}\\
\end{center}
\begin{center}
\includegraphics[scale=1]{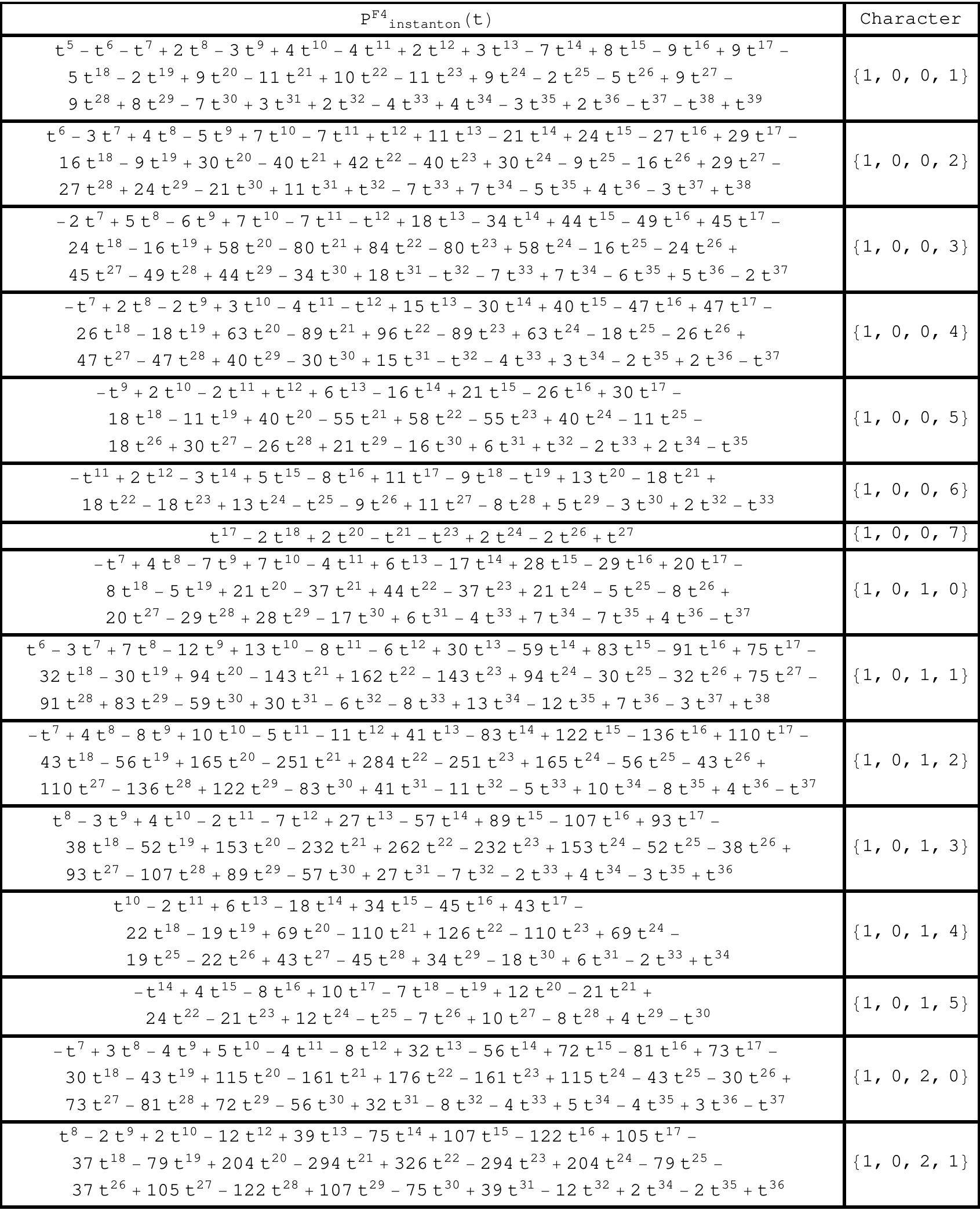}\\
\end{center}
\begin{center}
\includegraphics[scale=1]{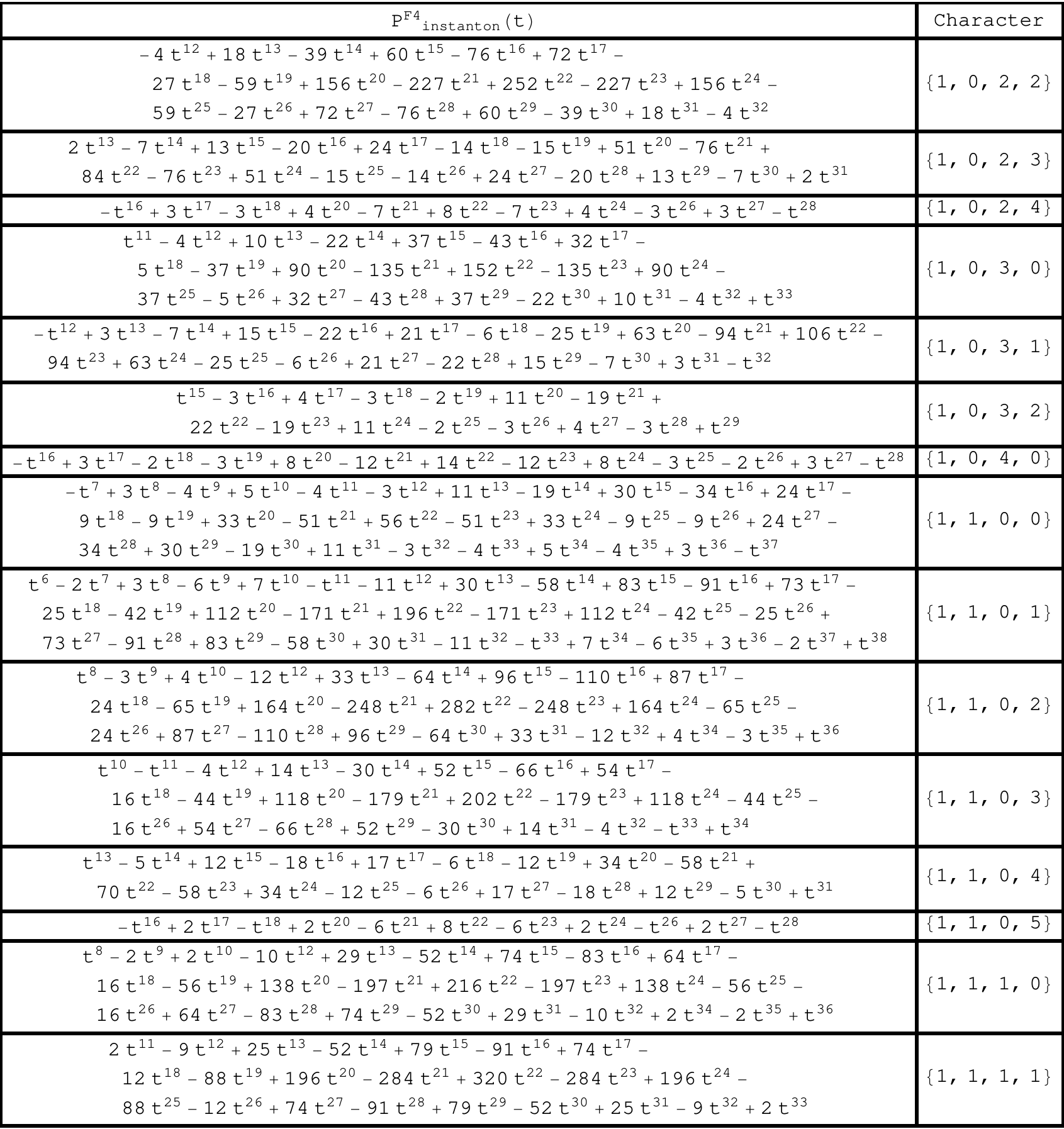}\\
\end{center}
\begin{center}
\includegraphics[scale=1]{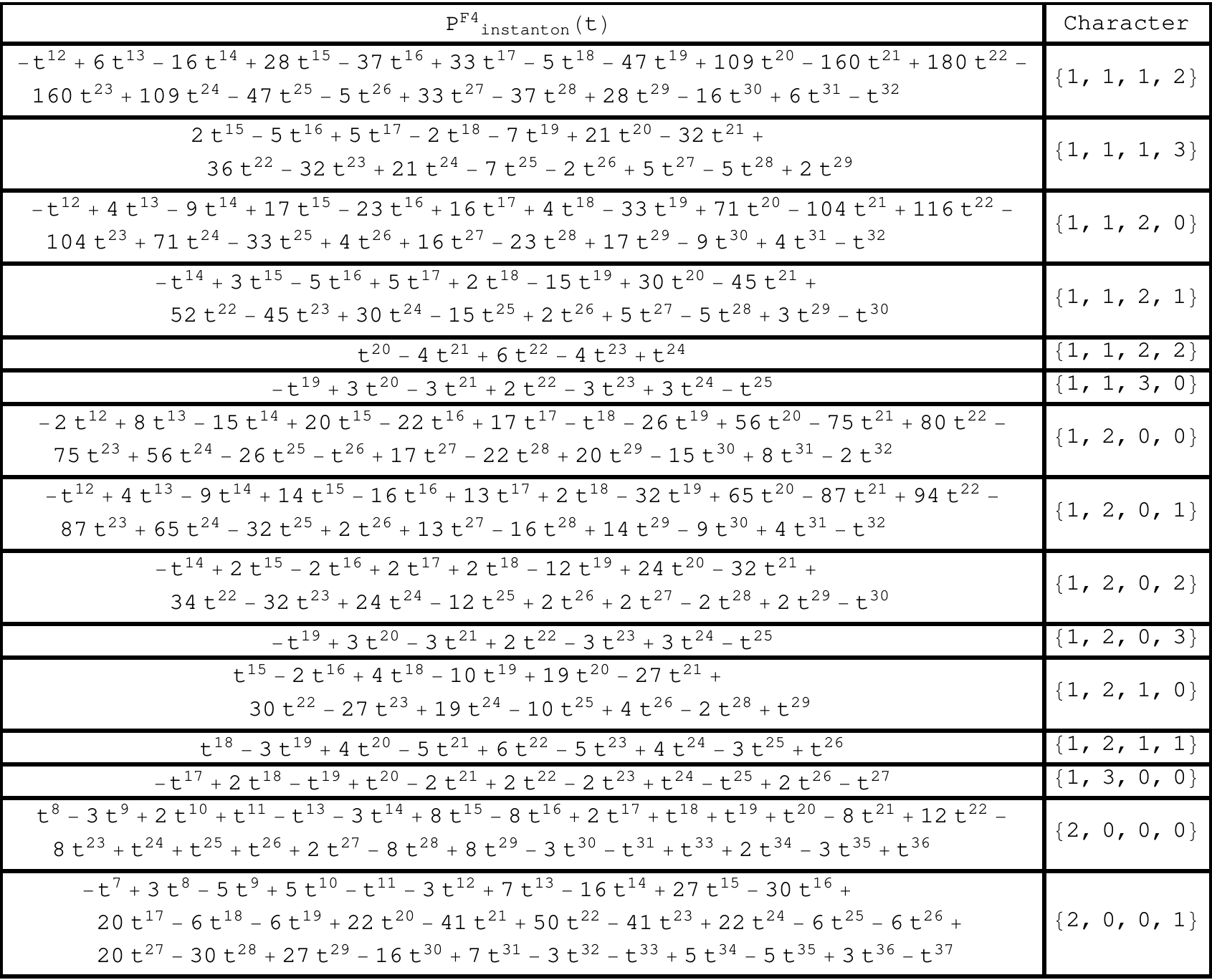}\\
\end{center}
\begin{center}
\includegraphics[scale=1]{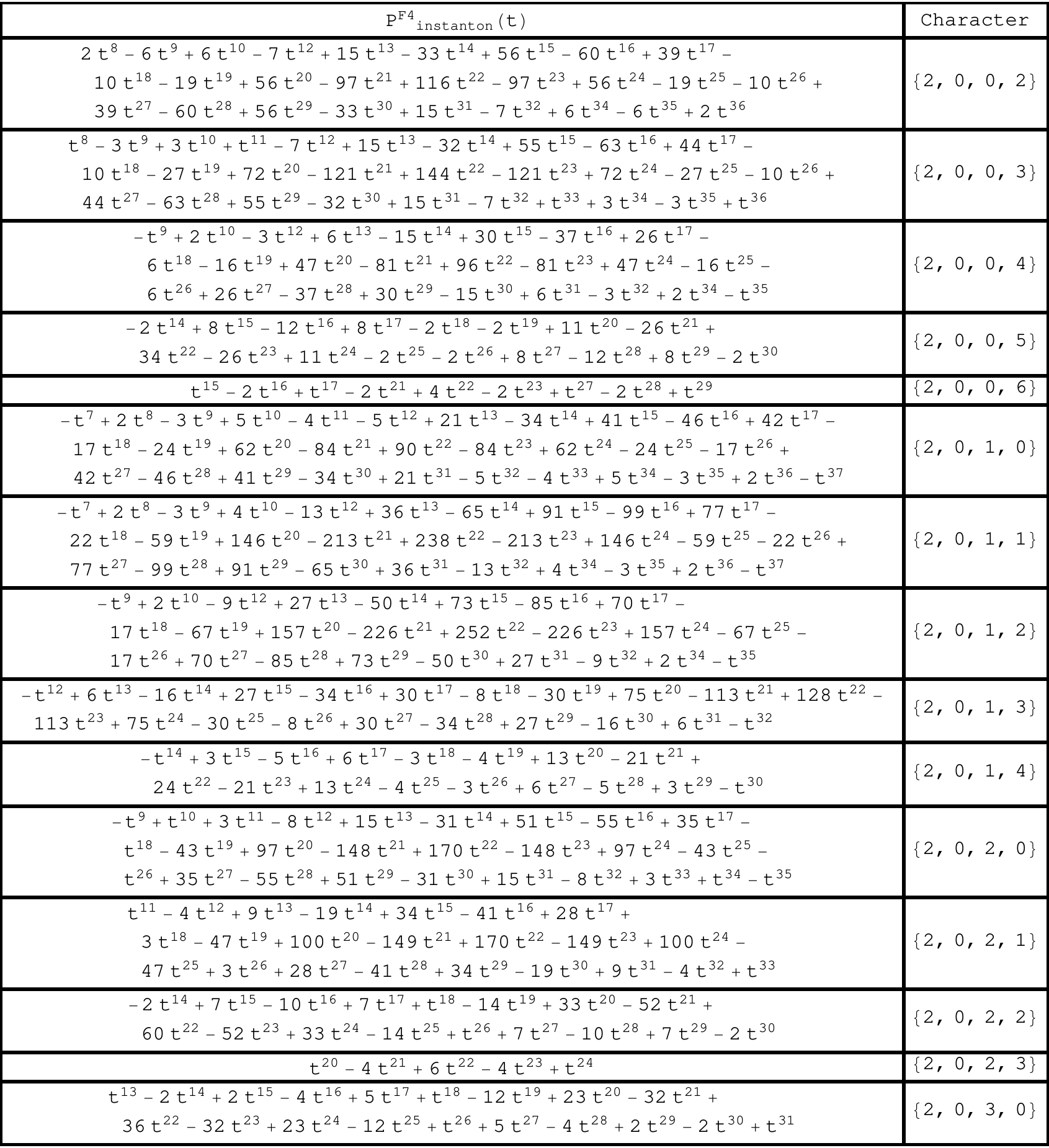}\\
\end{center}
\begin{center}
\includegraphics[scale=1]{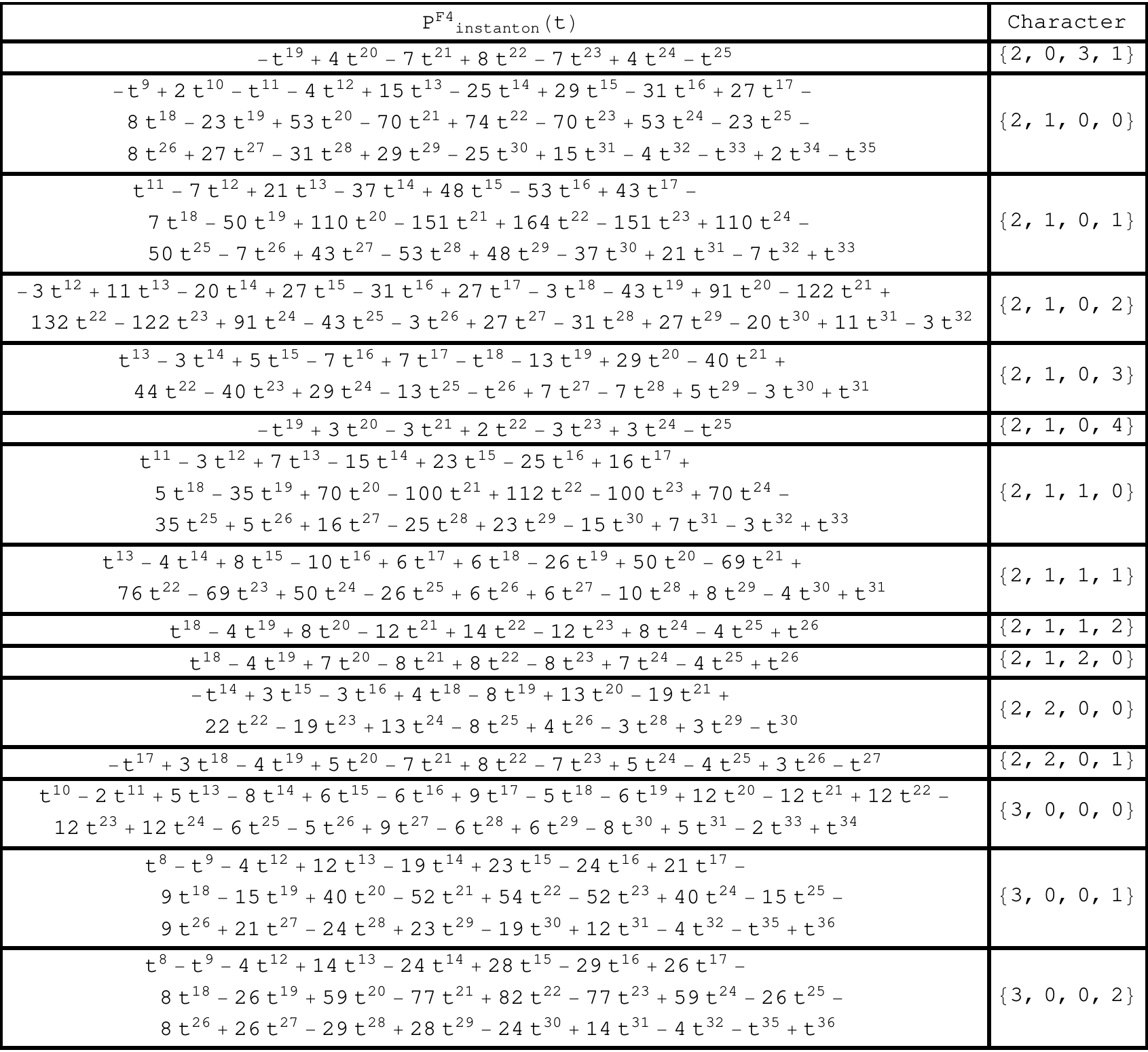}\\
\end{center}
\begin{center}
\includegraphics[scale=1]{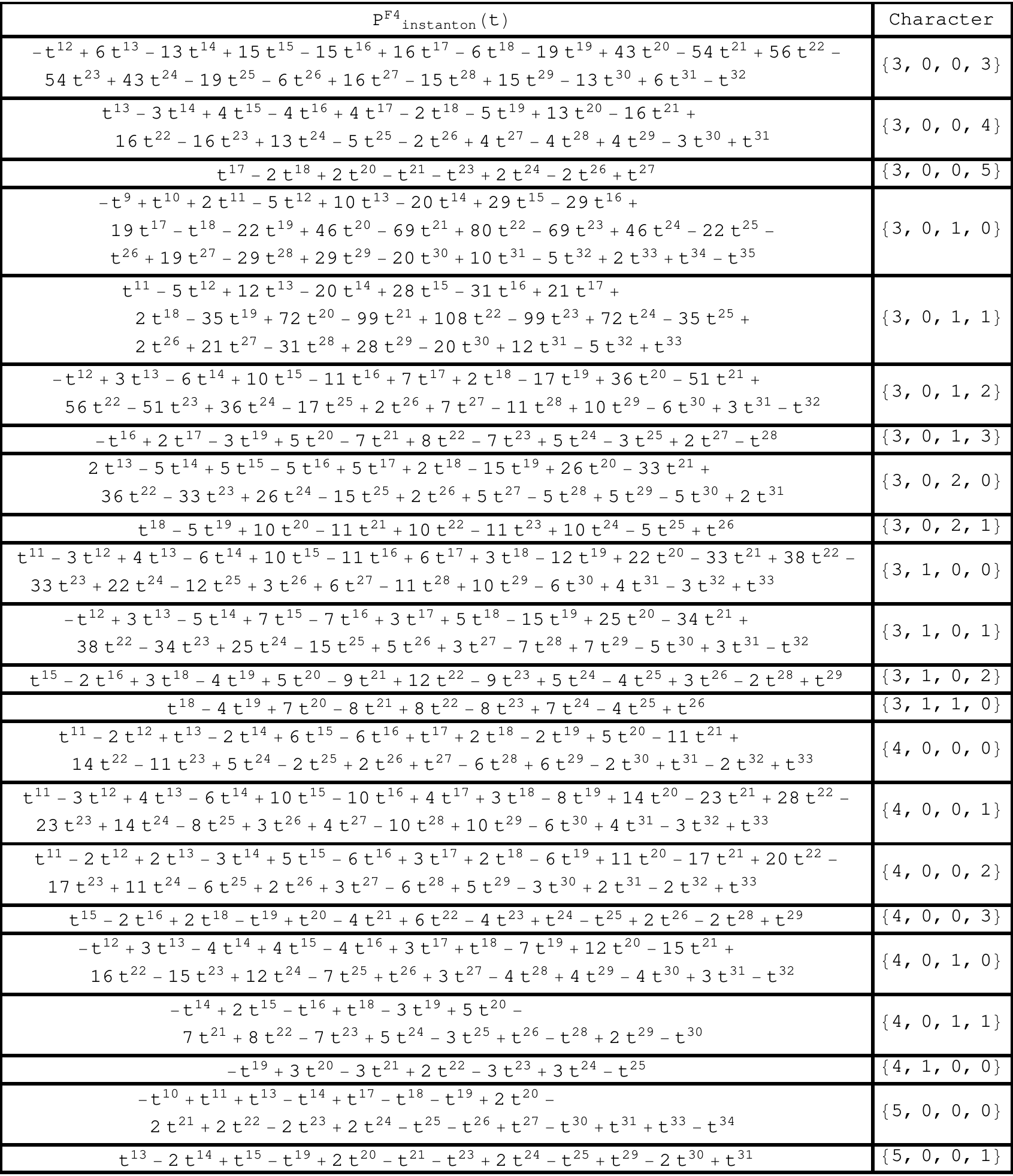}\\
\end{center}
\clearpage
\subsection{Appendix 3: $C_{E_7}$}
\begin{center}
\includegraphics[scale=1]{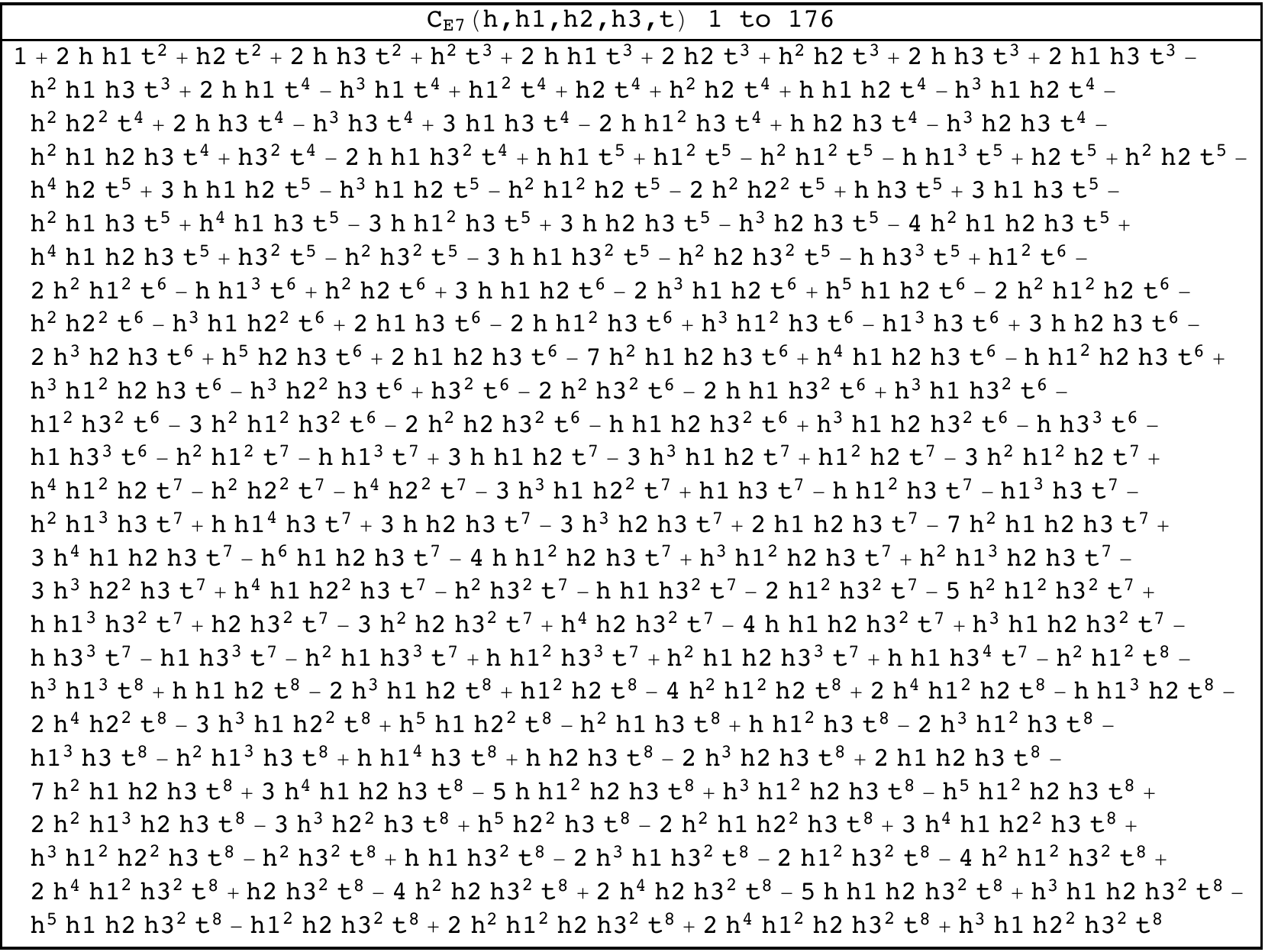}\\
\end{center}
\begin{center}
\includegraphics[scale=1]{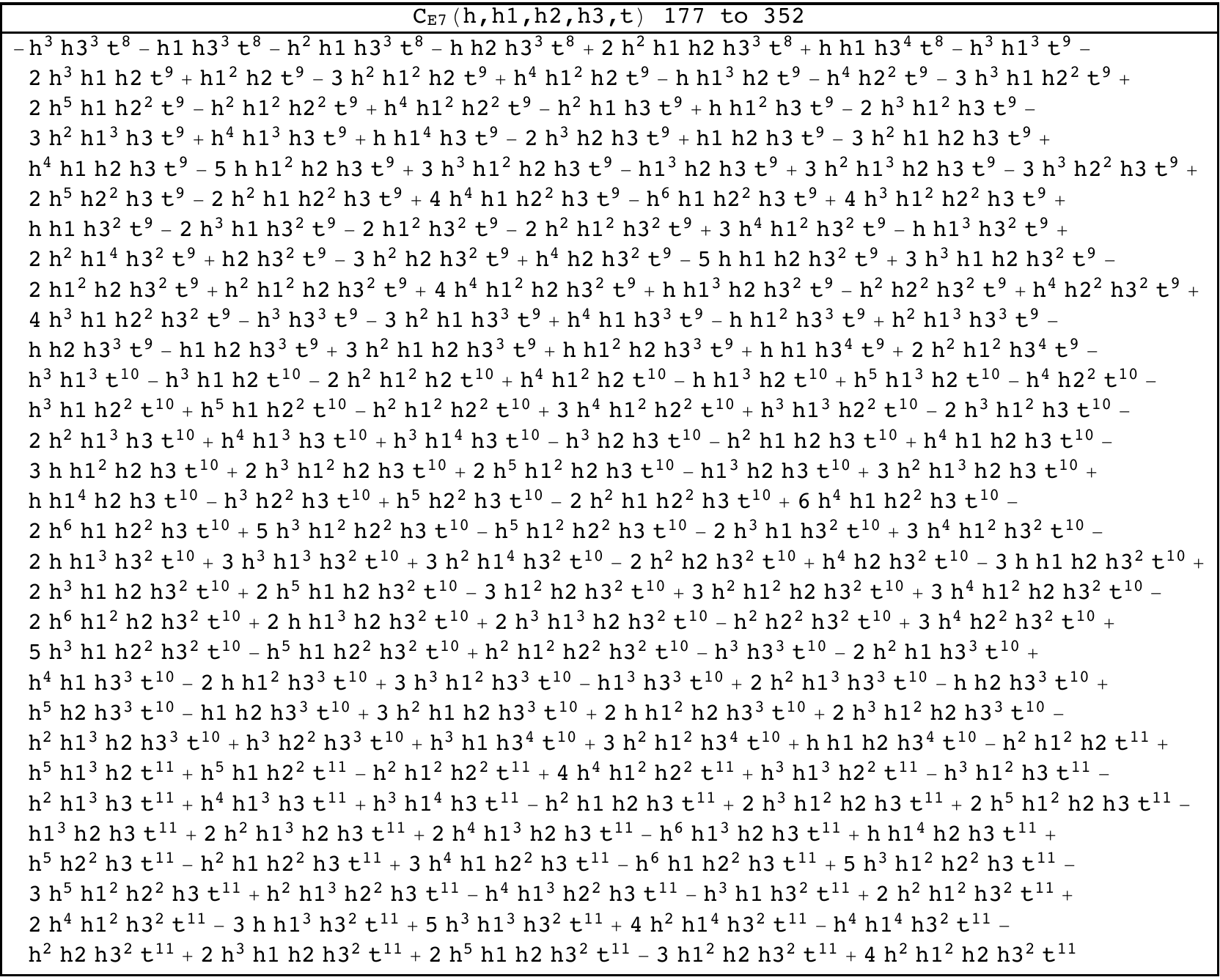}\\
\end{center}
\begin{center}
\includegraphics[scale=1]{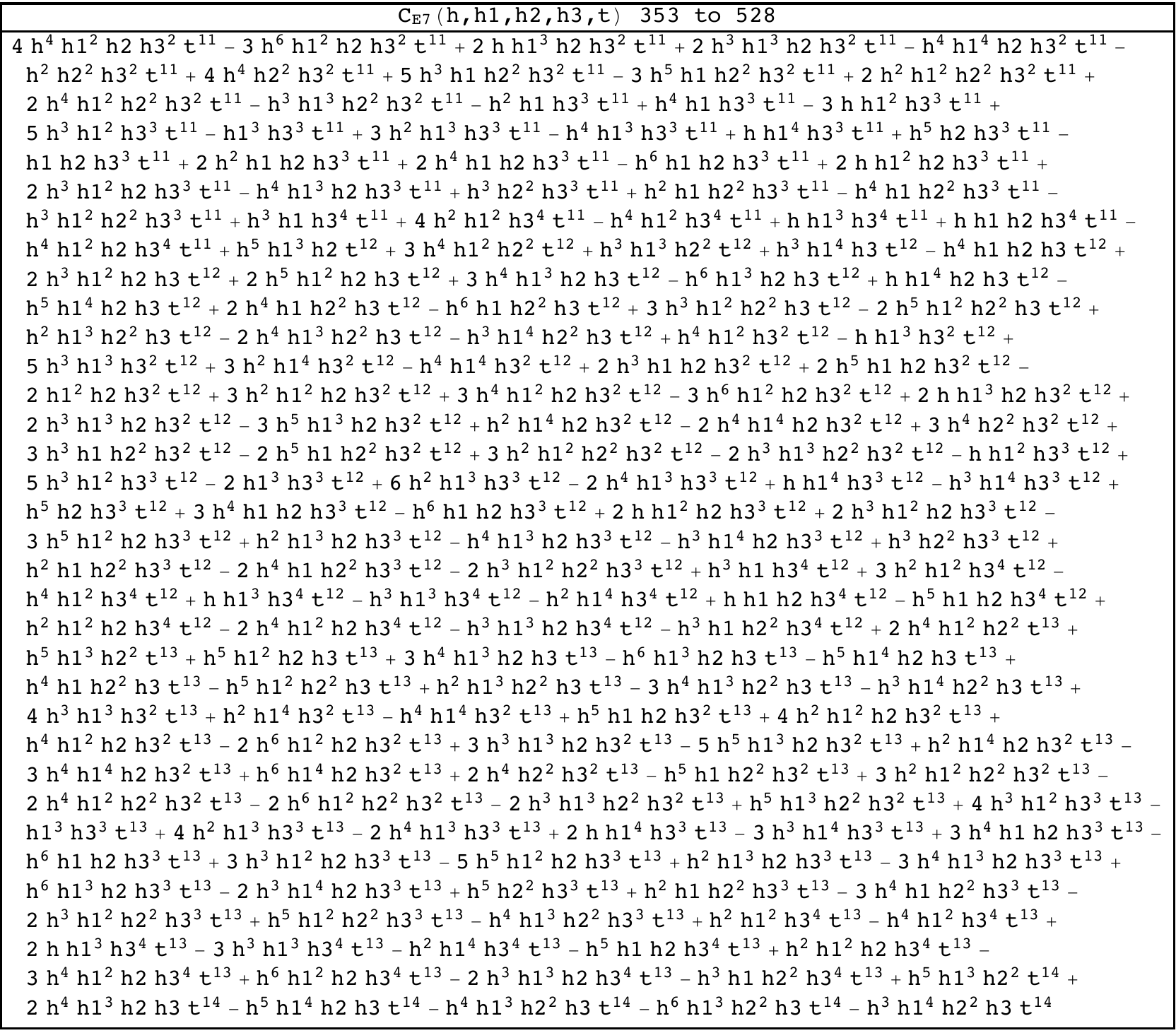}\\
\end{center}
\begin{center}
\includegraphics[scale=1]{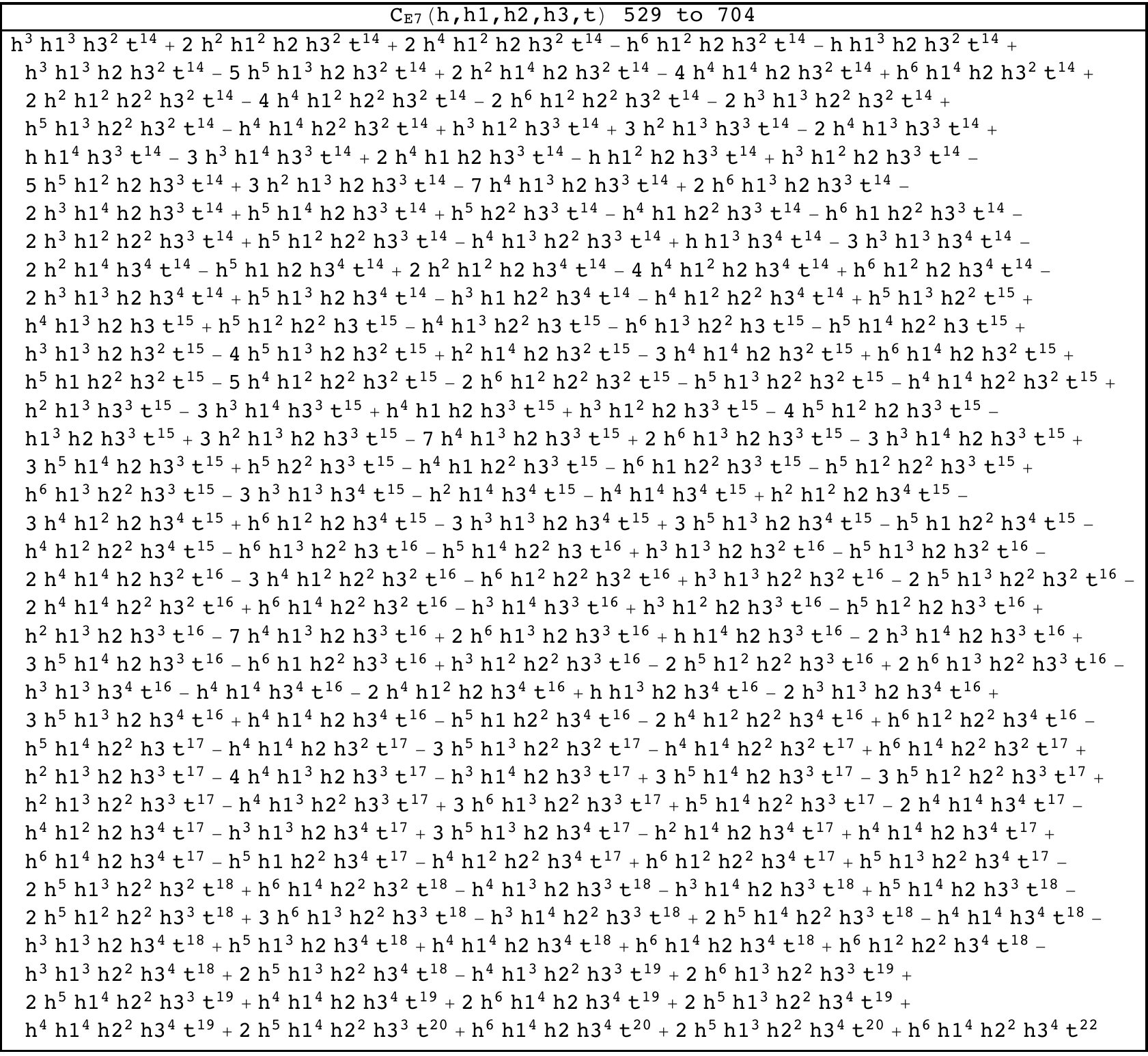}\\
\end{center}
\clearpage

\bibliographystyle{JHEP}
\bibliography{RJKBibLib}


\end{document}